\pdfoutput=1

\documentclass[sigplan,nonacm]{acmart}

\usepackage[]{hyperref}
\usepackage{xcolor}
\providecommand{\varnothing}{\emptyset}

\allowdisplaybreaks[1]

\setcopyright{none}

\usepackage{macros}

\makeatletter
\let\ACM@orig@mkabstract\@mkabstract
\def\@mkabstract{\begingroup\ACM@orig@mkabstract\endgroup}
\makeatother

\begin{document}

\title[Analog Quantum Experiments Meet Automatic Differentiation -
from Programs to Atoms]{Analog Quantum Experiments Meet Automatic
Differentiation - from Programs to Atoms}



\author[Y. Shi]{Yue Shi}
\affiliation{
  \department{Department of Physics}
  \institution{Princeton University}
  \country{United States}
}
\email{YS2251@princeton.edu}

\author[P. Liu]{Pengyu Liu}
\affiliation{
  \department{Department of Computer Science}
  \institution{Carnegie Mellon University}
  \country{United States}
}
\email{pengyul@andrew.cmu.edu}

\author[Y. Peng]{Yuxiang Peng}
\affiliation{
  \department{Department of Computer Science}
  \institution{Purdue University}
  \country{United States}
}
\email{yxpeng@purdue.edu}


\begin{abstract}

Gradient-based optimization of analog quantum hardware poses a
concrete programming problem: given an executable experiment
$P(\theta)$ that returns only measurement samples, what executable
experiment should the system run to obtain
$(\partial/\partial\theta_\ell)\llbracket P(\theta)\rrbracket$?
Conventional automatic differentiation cannot propagate through
the inaccessible quantum state, while finite differences require a noise-sensitive step size and analytic gradient identities cover only
parts of an experiment.

We present DiffSimuQ, which shows that a broad class of open-loop
analog quantum experiments is closed under automatic differentiation.
A compositional transformation maps $P(\theta)$ to a derivative
program $\mathcal D_\ell[P(\theta)]$ in the same language whose expected
outcome is exactly the derivative. Only continuous evolution requires
a strategy-specific rule, and we give two realizations. Our Nyquist shift
rule (NSR) applies to general time-dependent evolution: we prove that
the observable response to shifting the Hamiltonian along its tangent
is band-limited even for non-commuting controls, enabling exact
reconstruction from shifted executions. For involutory generators, a parameter-shift realization (PSR) inserts a parameter-independent operation at a sampled time. 

Because derivative programs remain ordinary programs, source and
derivative experiments share one compiler, which reuses the source
realization, lowers branches to device-ready neutral-atom controls,
and selects between NSR and PSR based on estimated cost. Under
shift-independent time-local Lindblad noise, the exact derivative
programs differentiate the objective realized by the device, with
hardware realization errors treated separately. In machine-level
noisy emulation, NSR and PSR avoid the finite-difference error floor
and improve closed-loop optimization; they win in complementary cost
regimes, while each derivative branch at 1000 qubits add at most $0.3\%$
to source compilation time.
\end{abstract}

\newif\iftechrep\techreptrue 


\maketitle

\section{Introduction}
\label{sec:intro}

Many quantum workloads share the same control loop: a program sets Hamiltonian controls $\theta$, runs an
experiment, and produces measurement samples whose average estimates
a scalar objective $C(\theta)$. An optimizer then asks for
$\partial C/\partial\theta_\ell$. Calibration, state preparation,
quantum control, Hamiltonian learning, analog simulation, and
variational algorithms all have this form
\cite{glaser2015training,georgescu2014quantum,cerezo2021variational}.
Classical automatic differentiation (AD) answers it by transforming a program into another
program that computes its derivative \cite{baydin2018automatic}.
For a physical quantum experiment, the concrete question is therefore:
\emph{what new experiment should the system run to obtain the
derivative?}

The difficulty is that a quantum device exposes measurement samples,
not the execution trace required by conventional AD. Its evolving
state cannot be inspected, and measurement is stochastic and
destructive, so the derivative cannot be backpropagated through one
execution. Existing alternatives answer only parts of the question.
A classical simulator restores an accessible state but loses
scalability. Finite differences rerun the experiment at nearby
controls, but at every finite step incur truncation error; shrinking
the step amplifies sampling and control-setpoint error, even if the
step is re-tuned as the execution budget grows. Analytic quantum
gradient identities remove this step for particular evolutions
\cite{schuld2019evaluating,banchi2021measuring,theis2023proper},
but do not by themselves differentiate the surrounding sampling,
measurement, readout, and weighted aggregation or produce a complete
program that a compiler can lower to hardware. The missing abstraction
is an experiment-level analogue of classical AD.

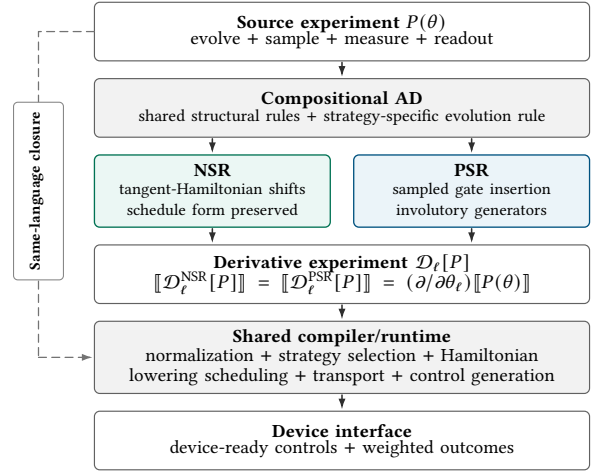
\begin{figure}[t]
  \centering
  \begin{tikzpicture}

  \definecolor{NSRGreen}{HTML}{009E73}
  \definecolor{PSRBlue}{HTML}{0072B2}

  \tikzset{%
    >={Latex[length=1.45mm,width=.95mm]},%
    font=\scriptsize,%
    flow/.style={->,draw=black!82,line width=.58pt,shorten <=.60pt,shorten >=.60pt,line cap=round,line join=round},%
    bypass/.style={draw=black!55,densely dashed,line width=.48pt,line cap=round,line join=round},%
    bypass arrow/.style={->,draw=black!55,densely dashed,line width=.48pt,shorten >=.60pt,line cap=round,line join=round},%
    box/.style={draw=black!62,line width=.44pt,rounded corners=2pt,align=center,outer sep=0pt,inner xsep=3.3pt,inner ysep=1.2pt},%
    wide/.style={box,minimum width=6.55cm,text width=6.02cm},%
    method/.style={box,minimum width=3.12cm,text width=2.70cm},%
    twoline/.style={minimum height=7.75mm},%
    threeline/.style={minimum height=9.55mm},%
    transform/.style={wide,fill=black!5},%
    nsrbox/.style={method,threeline,fill=NSRGreen!8,draw=NSRGreen!72!black,line width=.50pt},%
    psrbox/.style={method,threeline,fill=PSRBlue!8,draw=PSRBlue!72!black,line width=.50pt},%
    sidebox/.style={draw=black!48,line width=.40pt,rounded corners=1.6pt,fill=white,minimum width=6.6mm,minimum height=23.5mm,inner sep=1.1pt,align=center,outer sep=0pt}%
  }

  \def\layergap{2.35mm}

  \node[
    wide,
    twoline,
    fill=white
  ] (src) {
    \textbf{Source experiment} $P(\theta)$\\[-.66mm]
    evolve $+$ sample $+$ measure $+$ readout
  };

  \node[
    transform,
    twoline,
    anchor=north
  ] (ad) at ([yshift=-\layergap]src.south) {
    \textbf{Compositional AD}\\[-.60mm]
    {\fontsize{6.35}{6.85}\selectfont
      shared structural rules $+$ strategy-specific evolution rule}
  };

  \node[
    nsrbox,
    anchor=north west
  ] (nsr) at ([yshift=-\layergap]ad.south west) {
    \textbf{NSR}\\[-.40mm]
    {\fontsize{6.15}{6.75}\selectfont
      tangent-Hamiltonian shifts}\\[-.08mm]
    {\fontsize{6.15}{6.75}\selectfont
      schedule form preserved}
  };

  \node[
    psrbox,
    anchor=north east
  ] (psr) at ([yshift=-\layergap]ad.south east) {
    \textbf{PSR}\\[-.40mm]
    {\fontsize{6.15}{6.75}\selectfont
      sampled gate insertion}\\[-.08mm]
    {\fontsize{6.15}{6.75}\selectfont
      involutory generators}
  };

  \node[
    wide,
    twoline,
    fill=white,
    anchor=north
  ] (deriv) at
    ([yshift=-\layergap]$(nsr.south)!0.5!(psr.south)$) {
    \textbf{Derivative experiment} $\mathcal{D}_\ell[P]$\\[-.42mm]
    $\llbracket \mathcal{D}_\ell^{\mathrm{NSR}}[P]\rrbracket
      =
      \llbracket \mathcal{D}_\ell^{\mathrm{PSR}}[P]\rrbracket
      =
      (\partial/\partial\theta_\ell)
      \llbracket P(\theta)\rrbracket$
  };

  \node[
    transform,
    threeline,
    anchor=north
  ] (compiler) at ([yshift=-\layergap]deriv.south) {
    \textbf{Shared compiler/runtime}\\[-.40mm]
    normalization $+$ strategy selection $+$ Hamiltonian \\[-.08mm]
    lowering scheduling $+$ transport $+$ control generation
  };

  \node[
    wide,
    twoline,
    fill=white,
    anchor=north
  ] (device) at ([yshift=-\layergap]compiler.south) {
    \textbf{Device interface}\\[-.66mm]
    device-ready controls $+$ weighted outcomes
  };

  \draw[flow]
    (src.south) -- (ad.north);

  \draw[flow]
    ([xshift=-1.715cm]ad.south) -- (nsr.north);

  \draw[flow]
    ([xshift=1.715cm]ad.south) -- (psr.north);

  \draw[flow]
    (nsr.south) -- ([xshift=-1.715cm]deriv.north);

  \draw[flow]
    (psr.south) -- ([xshift=1.715cm]deriv.north);

  \draw[flow]
    (deriv.south) -- (compiler.north);

  \draw[flow]
    (compiler.south) -- (device.north);

  \coordinate (rail) at ([xshift=-7.4mm]src.west);

  \draw[bypass]
    (src.west)
    -- (rail)
    -- (rail |- compiler.west);

  \draw[bypass arrow]
    (rail |- compiler.west)
    -- (compiler.west);

  \node[sidebox] (closure)
    at ($(rail |- ad.north)!0.50!(rail |- deriv.south)$) {
    \rotatebox{90}{
      \fontsize{5.75}{6.25}\selectfont
      \textbf{Same-language closure}
    }
  };

\end{tikzpicture}
  \caption{Overview of DiffSimuQ. A source experiment is transformed
    compositionally into a derivative program in the same language, realized either by NSR or by PSR, and sent through the same compilation stack.}
  \label{fig:overview}
\end{figure}

A satisfactory answer should satisfy three requirements. First, it
should be \emph{compositional}: differentiating a complete experiment
should follow from local rules for its sampling, evolution,
measurement, and classical aggregation, rather than from a new
derivation for every workload. Second, it should be \emph{closed}: the
derivative should itself be an experiment program in the same language,
so that it can reuse the source program's runtime and compiler. Third, it should be \emph{realizable}: generated branches should admit
a compiler path to physical controls, with hardware feasibility and
any realization error made explicit rather than folded into the
derivative semantics. Together, these
requirements turn gradient estimation from a numerical recipe into a
programming and compilation abstraction.

We present DiffSimuQ, which provides this abstraction for a broad
class of open-loop analog quantum experiments. Its language represents
classical sampling, continuous-time evolution, measurement, readout,
and weighted families. A program $P(\theta)$ denotes an expected real
value, and DiffSimuQ transforms it according to
\[
  P(\theta)\longmapsto \mathcal D_\ell[P(\theta)],
  \qquad
  \llbracket\mathcal D_\ell[P(\theta)]\rrbracket
  =
  \tfrac{\partial}{\partial\theta_\ell}
  \llbracket P(\theta)\rrbracket .
\]
The equality is semantic and exact: $\mathcal D_\ell[P(\theta)]$ is itself an executable
program whose sample mean estimates the derivative without a tunable
differentiation step.
Our current scope is open-loop experimentation
with parameter-independent sampling and evolution boundaries;
mid-circuit measurement and real-time feedback remain outside the
language.

The transformation follows program structure. Readout and family
weights use ordinary differentiation, classical sampling is preserved,
and sequencing propagates the derivative compositionally. Only
continuous evolution is left as an interface to the underlying
physics. We prove a local-to-global result: any evolution rule
satisfying this local interface induces a sound derivative
transformation for the complete experiment. Thus the experiment-level derivative is defined once at the language level, while physical rules independently instantiate the evolution interface.

We instantiate this evolution derivative interface in two ways. The 
first is a new \emph{Nyquist shift rule} (NSR), applicable to general
time-dependent evolution. For an evolution under $H(\theta,t)$, let
$\dot H_\ell=\partial H/\partial\theta_\ell$ and consider the
executable family $H(\theta,t)+s\dot H_\ell(\theta,t)$. We prove that
the resulting observable response $g(s)$ is band-limited in the shift
$s$, even when the controls vary in time and the Hamiltonians do not
commute across controls or times. An exact Nyquist differentiator can
therefore reconstruct $g'(0)$ as a convergent weighted expectation
over shifted executions. Each NSR branch is an ordinary experiment that shifts the source controls together along the tangent; a statically computed bandwidth certificate sets the shift distribution and cost.

The second realization is a stochastic \emph{parameter-shift rule}
(PSR). When the Hamiltonian terms touched by $\theta_\ell$ are
involutions, PSR
\cite{banchi2021measuring,leng2022differentiable} samples an insertion
time, interrupts the source evolution, and inserts one
parameter-independent operation. NSR and PSR compute the same
language-level derivative but spend different physical resources:
NSR preserves the source schedule while consuming waveform headroom
and sampling budget; PSR consumes interruption and inserted operations
while allowing branches associated with one generator to be reused
across parameters. They are two realizations of one language-level derivative, not separate end-to-end algorithms. Neither dominates structurally, so the compiler chooses between them.

Same-language closure turns derivative realization into a
compilation problem.
We instantiate DiffSimuQ on a zoned neutral-atom
architecture~\cite{bluvstein2024logical} with continuous Hamiltonian
evolution, movable atoms, and calibrated gates. DiffSimuQ normalizes
source and derivative programs into executable branches, extends SimuQ \cite{peng2024simuq} for
Hamiltonian realization, and represents derivative branches as
structured edits of the source realization. On this target, NSR
branches are usually waveform-coefficient edits and PSR branches schedule
insertions. The compiler then lowers either realization
through scheduling, atom transport, channel coordination, and pulse
generation to device-ready controls, while a static selector chooses
between the two strategies based on estimated execution cost. The lowering is architecture-specific; the language and AD
theory are device-independent.

On noisy hardware, the relevant optimization target is the objective
realized by the device rather than an inaccessible noiseless model.
Under shift-independent time-local Lindblad noise, NSR and PSR with
ideal insertions remain exact for the gradient of this device-realized
objective. This is not error mitigation: the derivative tracks the 
real-device landscape seen by the optimizer. Physical constraints
are a separate realization layer; finite waveform headroom can
truncate NSR shifts, while imperfect inserted operations can perturb a
PSR branch. We analyze and measure these realization errors separately
from the derivative semantics.

We evaluate DiffSimuQ from derivative accuracy through
compilation. In machine-level noisy emulation, NSR and PSR converge
with ordinary $N^{-1/2}$ sampling error until their modeled realization
floors, while a budget-adaptive finite difference reaches a larger
step/setpoint floor. In closed-loop inverse simulation, the derivative
programs reach the planted target more reliably than fixed
finite-difference steps and a derivative-free baseline. NSR and PSR
win in complementary structural regimes; the compiler's static
selector agrees with the measured winner on $88.3\%$ of the design
space and incurs at most $1.26\times$ overhead when wrong. Finally,
on structured programs up to 1000 qubits, differentiation reuses the
source compilation and adds at most $0.3\%$ per branch.

\vspace{1em}
In summary, DiffSimuQ makes three contributions:
\begin{itemize}[leftmargin=0.5cm]
    \setlength{\itemsep}{0pt}
    \setlength{\parskip}{0pt}
    \setlength{\parsep}{0pt}
    \setlength{\topsep}{0pt}
  \item \textbf{Compositional AD:} a device-independent language and
    transformation that maps complete open-loop experiments to
    same-language derivative programs.
  \item \textbf{Executable evolution derivatives:} a new Nyquist
    shift rule for general evolution and a
    parameter-shift specialization with complementary physical costs.
  \item \textbf{Derivative-aware compilation:} a shared compiler that
    reuses source realizations, lowers derivative branches to
    neutral-atom controls, and selects a realization strategy.
\end{itemize}
Together, these results make differentiability a reusable capability
of the quantum experiment stack rather than a bespoke derivation for
each workload and device.

\section{Quantum Experiment Background}
\label{sec:background}

\subsection{Analog Experiments as Programs}
\label{sec:analog-programs}

We describe the protocol of an experiment as an analog quantum program: prepare
$n$ qubits in an initial quantum state, switch on a
collection of physical effects, let the system evolve continuously
for a set duration, measure, and repeat. In the noiseless model, the
state of the $n$ qubits is a unit vector $\ket{\psi}\in\mathbb{C}^{2^n}$; the effects in
force at time $t$ are summarized by a Hermitian matrix $H(t)$, the
\emph{Hamiltonian}, under which the state obeys the Schr\"odinger
equation $\frac{d}{dt}\ket{\psi(t)}=-iH(t)\ket{\psi(t)}$; when $H$ is
constant over a duration $t$, the solution
is the unitary map
$\ket{\psi}\mapsto e^{-iHt}\ket{\psi}$. The resulting state at time $T$ is the \emph{evolved state} $\ket{\psi(T)}$.

On an actual machine, $H$ has a fixed structure,
\begin{equation}
  H(t)\;=\;\sum\nolimits_j u_j(t)\cdot A_j .
  \label{eq:H-anatomy}
\end{equation}
Each Hermitian $A_j$ (e.g., $X=
\begin{psmallmatrix}0&1\\1&0
\end{psmallmatrix}$ and $Z=
\begin{psmallmatrix}1&0\\0&-1
\end{psmallmatrix}$) is
one effect an instrument of the machine can produce---a laser drive on one
qubit, or an atom-atom interaction between a pair---and is fixed by the hardware. Each $u_j(t)$ is the
strength of the interaction, changing over time. 
The sum represents these effects acting simultaneously.

The goal is often to probe properties of an evolved state---a novel physics effect, or the answer to a
chemistry or optimization problem---since simulating candidates
classically is exponentially hard in $n$. The search therefore runs
on the hardware itself, with the controls as knobs.

In an analog quantum program, the strength $u_j(t)$ of \eq{H-anatomy} is determined by the
\emph{waveform} of a driving instrument $j$: for example, an amplitude envelope the
programmer stores into a waveform generator that drives a laser modulated on the atoms, or the distance between 
the atom pairs at time $t$. \emph{The waveforms are
the program text.} The continuous physical process
is the object being programmed~\cite{peng2024simuq}, in contrast with the more familiar gate
model, where a program is a sequence of discrete unitaries from a
fixed set. 

The smallest such program serves as the running example: drive one
qubit from $\ket{0}$ with one instrument. A laser shone on the qubit produces the bit-flip
effect $X$ ($A_j$ in \eq{H-anatomy}) whose
amplitude follows a programmed waveform $u(t)$, so
$H(t)=u(t)\,X$. Let the waveform be a linear ramp, \emph{parametrized} by $\theta_1$ and $\theta_2$:
$u(t)=\theta_1+\theta_2\,t/T$.
By the Schr\"odinger equation, starting from $\ket{0}$, the state at
time $T$ is
\begin{equation*}
  \ket{\psi(T)}
  =\cos\!\Big(\textstyle\int_0^T u(t)\,\d t\Big)\ket{0}
  -i\,\sin\!\Big(\textstyle\int_0^T u(t)\,\d t\Big)\ket{1},
\end{equation*}
where $\textstyle\int_0^T u(t)\,\d t=\big(\theta_1+\tfrac{\theta_2}{2}\big)\,T$.
Here, only the area under $u(t)$ enters because $X$ commutes with itself. 

The evolved state stays inside the quantum device, and a single
execution does not expose its state vector. \emph{Measurement}
probes the quantum system and returns classical outcomes; it is
stochastic and destructive: measuring $\ket{\psi}$ in the
computational basis returns a bit string $m\in\{0,1\}^n$ with
probability $|\braket{m}{\psi}|^2$ and destroys the state, so a
single run yields only $n$ classical bits. For the running example,
the evolved qubit reads $1$ with probability
$f(\theta_1,\theta_2)=\sin^2\!\big((\theta_1+\tfrac{\theta_2}{2})T\big)$
in a single run, and estimating $f$ requires repetition. The
experiment is therefore wrapped: repeat $N$ ``shots,'' average, and
incur statistical error of order $N^{-1/2}$. In general the wrapper fixes a \emph{readout}
$\lambda:\{0,1\}^n\to\mathbb R$ and reports the empirical mean of
$\lambda(m)=\lambda_m$, estimating
\begin{equation*}
  \mathbb{E}[\lambda]=\sum\nolimits_m \lambda_m\,|\braket{m}{\psi}|^2
  =\bra{\psi}M\ket{\psi},
  ~M=\sum\nolimits_m \lambda_m\ket{m}\bra{m}
  \label{eq:observable-expect}
\end{equation*}
with the diagonal \emph{observable} $M$ packaging the weights. 
A suitable choice of $\lambda$ extracts the physical properties of $\ket{\psi(T)}$: distance to a
desired state, an energy, or a correlation. 
To a programmer, the device is a
sampler, and every quantity of interest is a classical function of the parameters, estimated from samples.

The running example inspires a \emph{quantum control problem}: if the evolution starts from $\ket{0}$, how should we tune the laser parameters $\theta_1$ and $\theta_2$ so that a measurement at time $T$ has the highest probability of reading $1$?

\subsection{Derivative Estimation as Programs}
\label{sec:why-derivatives}

\begin{figure}[t]
  \centering
  \includegraphics[width=\linewidth]{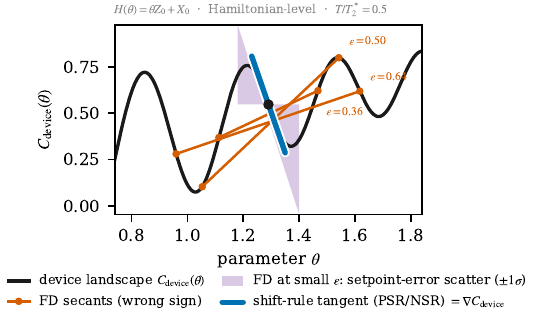}
  \caption{The finite-difference trap on a landscape narrower than
    the hardware's step resolution: the illustrated secants have the
    wrong sign, while the shaded band shows the noise cone near the
    step-resolution limit.
  }
  \label{fig:fd-trap}
\end{figure}

Real workloads---variational
eigensolvers~\cite{peruzzo2014variational}, combinatorial
optimization~\cite{farhi2014quantum},
Hamiltonian learning~\cite{wiebe2014hamiltonian}, inverse
quantum simulation~\cite{kokail2026inverse}---let the waveforms be arbitrary
expressions in $\theta\in\mathbb R^d$ and drive many mutually
non-commuting effects; then every parameter affects the
outcome. They leverage gradient descent
on the measured value from quantum experiment programs, which raises the core question of this paper: \textbf{given a program, can we estimate $\partial f/\partial\theta$?} Furthermore, \textbf{is unbiased estimation of $\partial f/\partial\theta$ itself a program?}

The obvious derivative estimator is the finite-difference secant:
run the experiment at $\theta_\ell\pm\epsilon/2$, subtract, divide:
$\partial f/\partial\theta_1\approx
\big[f(\theta_1+\tfrac{\epsilon}{2},\theta_2)
-f(\theta_1-\tfrac{\epsilon}{2},\theta_2)\big]/\epsilon$.
Its accuracy depends on the stepsize $\epsilon$. At any finite
$\epsilon$ the secant is generally biased, and increasing the shot
count does not remove this bias; shrinking $\epsilon$ inflates the
statistical error as $N^{-1/2}/\epsilon$ and amplifies finite
control-setpoint error. Thus $\epsilon$ couples two abstraction
levels---the landscape and the machine's control accuracy. A usable
step often exists (\app{fd}), but the programmer must find it.
\fig{fd-trap} shows a landscape where this usable window
collapses: large steps have the wrong sign, while small steps are
dominated by setpoint error.

Written as a program, finite difference is a transformation of
the source. Let $P(\theta_1,\theta_2)$ represent the running example program. Then, with formal notation defined in \sec{theory}, 
\begin{equation}
  \mathrm{FD}_\epsilon[P]
  =\tfrac{1}{\epsilon}\,P\big(\theta_1+\tfrac{\epsilon}{2},\theta_2\big)
  -\tfrac{1}{\epsilon}\,P\big(\theta_1-\tfrac{\epsilon}{2},\theta_2\big),
  \label{eq:fd-program}
\end{equation}
using two copies of the source, each with a weight and a shifted setting.
This paper shows that the same source program $P$ admits other compositional program transformations $\mathcal{D}[P]$ whose expected value equals the derivative.

\newcommand{\kw}[1]{\mathord{\text{\normalfont\sffamily\bfseries #1}}}
\newcommand{\kwev}{\kw{ev}}
\newcommand{\kwmeas}{\kw{meas}}
\newcommand{\kwlet}{\kw{let}}
\newcommand{\kwin}{\kw{in}}

\section{Programming Language Design}
\label{sec:theory}

A \emph{parameterized quantum experiment program} $P(\theta)$ packages the
experiment of \sec{analog-programs}, including evolutions and readouts. Moreover, it includes a \emph{classical sampler} over a parameter-independent classical distribution and a family of weighted programs. Few as they are, these constructs cover an analog experiment end to end. Here we formally define the syntax and semantics of such programs, and the blanket assumption for the differentiation theorems of \sec{ad}.

\subsection{Syntax}

\fig{language-syntax} presents the core syntax. A program $P$ denotes, for each valuation of the differentiation parameters
$\Theta=(\theta_1,\ldots,\theta_p)$ and of the sampled classical
variables in scope, an expected real value of an observable estimator
on a fixed $n$-qubit Hilbert space $\mathcal H$.

\begin{figure}[t]
  \centering
  \fbox{
    \begin{minipage}{0.94\columnwidth}
      \[
        \begin{array}{rcll}
          P,Q &::=& \kwlet\ \xi \sim \mu\ \kwin\ P & \textsc{sample} \\[0.25em]
          &\mid& \kwev(\tau_1,\tau_2,H);P & \textsc{evolve} \\[0.25em]
          &\mid& \kwmeas(m \mapsto f_m) & \textsc{measure} \\[0.25em]
          &\mid& \displaystyle\sum_{i\in I} w_i\, P_i &
          \textsc{family} \\[0.6em]
          H &::=& u \cdot A \ \mid\ H_1 + H_2 & \textsc{hamiltonian}
        \end{array}
      \]
      \vspace{-0.3em}
      \[
        \begin{array}{c}
          \xi, t, \tau_1,\tau_2\in\mathcal E_0,\qquad
          u,w\in\mathcal E_\Theta,\qquad
          \mathcal E_0\subseteq\mathcal E_\Theta\\[0.25em]
          (\partial/\partial\theta_\ell)\,\mathcal
          E_\Theta\subseteq\mathcal E_\Theta, \qquad|I|<\infty,\quad
          m\in\{0,1\}^n,\quad
          A\in\mathcal A .
        \end{array}
      \]
      \vspace{-0.3em}
    \end{minipage}
  }
  \caption{Core syntax of parameterized quantum experiment programs. Here $\mu$, and everything drawn from it, is parameter-free. The time variable $t$ in $H$ is bound by $\kwev$. $\mathcal A$ is the machine signature's finite set of Hermitian operators.}
  \label{fig:language-syntax}
\end{figure}

\paragraph{Expressions.}
Classical expressions form a language $\mathcal E_\Theta$ over the
differentiation parameters $\Theta$, the sampled variables in scope,
and (within an $\kwev$) the bound time variable $t$; the
subclass $\mathcal E_0$ contains the expressions independent of
$\Theta$; readout weights $f_m$ are expressions in
$\mathcal E_\Theta$ as well. In addition to supporting the
classical arithmetic used by the transformation, $\mathcal E_\Theta$
must satisfy \textbf{closure under partial differentiation},
$(\partial/\partial\theta_\ell)\,\mathcal E_\Theta
\subseteq \mathcal E_\Theta$: differentiating a classical expression must produce expressions the language can write. Any concrete
expression language with this property serves. The instantiation used
by our implementation is given in \app{expr}.

\paragraph{Hamiltonians as a description language.}
\label{par:H-normal}
The $\textsc{hamiltonian}$ production is~\eqref{eq:H-anatomy} in the
syntax, in the style of SimuQ~\cite{peng2024simuq}: the coefficients
are now expressions, $H\equiv\sum_j u_j\cdot A_j$, so waveforms can be differentiable functions of the parameters $\Theta$, classical samples $\xi$, and time variable $t$. 
On $n$ qubits, $\mathcal{A}$ includes operators like $X_a$ ($X$ on qubit $a$) and $Z_aZ_b$ (on qubits $a$ and $b$). A parameter $\theta_\ell$
\emph{touches} the Hamiltonian terms whose coefficients depend on it. 

\paragraph{Operational meaning of the constructs.}
$\kwev$ plays a pulse: the machine evolves under
$H$ from $\tau_1$ to $\tau_2$. $\kwmeas$ measures the system: it shines
readout light and records which detectors click, reporting the
weight $f_m$ of the observed bit string.
$\kwlet$ draws a classical sample from a distribution and stores it as a classical variable. A \textbf{family}
$\sum_i w_i\,P_i$ is a weighted batch of experiments: run each
branch on its own, estimate its expected outcome, and sum these
estimates with the weights $w_i\in\mathcal E_\Theta$, which may
depend on $\Theta$. Alternatively, a single run of a family may draw branches at random with probability $\propto|w_i|$
and reweight the outcomes with repetition. 

\paragraph{Derived forms.}
The $\textsc{family}$ constructor subsumes the additive and scalar
program forms:
\[
  \mathbf 0 \;:=\; \sum_{i\in\varnothing}w_i\,P_i,
  \quad
  a\,P \;:=\; \sum_{i\in\{\ast\}}a\,P,
  \quad
  P_1+P_2 \;:=\; \sum_{i\in\{1,2\}}1\cdot P_i .
\]


\paragraph{Initial states.}
The language has no initial-state construct: a program denotes a
function $\mathcal H\to\mathbb R$ and accepts any input state from $\mathcal{H}$, and
every theorem below is stated for arbitrary inputs. Preparing
$\ket{0}^{\otimes n}$ is an experimental convention of the target
machine, not a feature of the language.

\paragraph{The running example.} We may write
\begin{equation}
\label{eq:p-running}
P(\theta_1,\theta_2)\equiv \kwev\Big(0,T,\big(\theta_1+\theta_2\,\frac{t}{T}\big)\cdot X\Big);\
\kwmeas(m\mapsto m).
\end{equation}
If the input state is $\ket{0}$, the output denotes $f(\theta_1,\theta_2)$. 
In $P$, both parameters touch the single term $X$.
Its finite-difference program~\eqref{eq:fd-program} is a family of two
branches, the source $P$ with $\theta_1$ substituted:
\[
\begin{aligned}
  &\tfrac1\epsilon\,\big[\kwev\big(0,T,(\theta_1+\tfrac\epsilon2+\theta_2\,t/T)\cdot
  X\big);\kwmeas(m\mapsto m)\big]
  \;+\;\\
  &\big(-\tfrac1\epsilon\big)\big[\kwev\big(0,T,(\theta_1-\tfrac\epsilon2+\theta_2\,t/T)\cdot
  X\big);\kwmeas(m\mapsto m)\big].
\end{aligned}
\]

\subsection{Semantics and Differentiability Assumptions}

\fig{language-semantics} gives the
\textbf{denotational semantics}:
$\llbracket P\rrbracket:\mathcal H\to\mathbb R$ for each valuation of
$\Theta$ and of the free sampled variables (the valuation is
  suppressed in the notation; substitution $P[\xi\mapsto v]$ makes the
sampling rule explicit). \textbf{Well-formedness} req-
\newline uires finite evolution intervals with $\tau_1\le\tau_2$ and Hamiltonians that are bounded and piecewise continuous in time. Each $\mu$ in \textsc{sample} must be a probability measure and the body's denotational semantics must be Lebesgue $\mu$-integrable, so the integral in the sampling rule is defined.

\begin{figure}[t]
  \centering
  \fbox{
    \begin{minipage}{0.94\columnwidth}
      \[
        \begin{array}{rcl}
          \llbracket P\rrbracket &:& \mathcal H \to \mathbb R \\[0.5em]
          \llbracket \kwlet\ \xi\sim\mu\ \kwin\ P\rrbracket(\ket{\psi})
          &=& \displaystyle\int \llbracket P[\xi\mapsto
          v]\rrbracket(\ket{\psi})\,d\mu(v) \\[0.7em]
          \llbracket \kwev(\tau_1,\tau_2,H);P\rrbracket(\ket{\psi})
          &=& \llbracket
          P\rrbracket\big(U_H(\tau_2,\tau_1)\ket{\psi}\big) \\[0.5em]
          \llbracket \kwmeas(m\mapsto f_m)\rrbracket(\ket{\psi})
          &=& \displaystyle\sum_{m\in\{0,1\}^n}
          f_m\,|\braket{m}{\psi}|^2 \\[0.7em]
          \Big\llbracket \displaystyle\sum_{i\in
          I}w_i\,P_i\Big\rrbracket(\ket{\psi})
          &=& \displaystyle\sum_{i\in I} w_i\,\llbracket
          P_i\rrbracket(\ket{\psi})
        \end{array}
      \]
      \vspace{-0.3em}
    \end{minipage}
  }
  \caption{
    Denotational semantics: a program denotes an expected real value on an input state. Here, $U_H(b,a)=\mathcal T\exp\big(-i\int_a^b H(t)\,dt\big)$ is the time-ordered propagator of $H$ with its coefficient expressions evaluated in the current valuation.
    The family construct denotes the weighted sum of its branches, realized operationally by executing branches separately and combining estimates classically.
  }
  \label{fig:language-semantics}
\end{figure}

Differentiating a program (\sec{ad}) requires exchanging
$\partial/\partial\theta_\ell$ with the sampling integral, which is
not automatic. We state one blanket assumption, not per-theorem hypotheses:

\begin{assumption}[Differentiable programs]\label{as:diff}
  For every source program to be differentiated, the coefficient and
  readout expressions are smooth in $\Theta$ on an open neighborhood
  of the parameter point. The Hamiltonian and its parameter
  derivatives are bounded and piecewise continuous in time, with
  bounds locally uniform in $\Theta$. Each sampling construct
  admits, on such a neighborhood, a $\mu$-integrable bound
  dominating its body's derivative---so differentiation exchanges
  with sampling.
\end{assumption}
\noindent
For smooth bodies, finite-support $\mu$ satisfies this since the integral is a finite sum; \app{proofs} verifies it for this paper's programs. \sec{ad} follows it.

\section{Differentiation as Program Transformation}
\label{sec:ad}

\sec{why-derivatives} wrote finite differences as a program
transformation~\eqref{eq:fd-program}. This section supplies
automatic differentiation as program transformations: apply
$\mathcal D$ to the experiment's program, and the result is a family
of programs whose averaged outcome is the derivative. We define
soundness, give the rules, and prove them sound for every strategy
that fills the $\textsc{d-evolve}$ slot---NSR and PSR each do.

\subsection{Sound Differentiation Transformations}
\label{sec:ad-soundness}

\begin{figure}[t]
  \fbox{
    \begin{minipage}{0.98\columnwidth}
      \[
      \hspace{-0.18cm}
        \begin{array}{l@{~}l}
          \mathcal D^{\mathcal S}_\ell\big[\kwmeas(m\mapsto f_m)\big]
          \equiv \kwmeas\big(m\mapsto\frac{\partial f_m}{\partial\theta_\ell}\big)
          & \textsc{d-measure} \\[0.55em]
          \mathcal D^{\mathcal S}_\ell\big[\kwlet\ \xi\sim\mu\ \kwin\ P\big]
          \equiv \kwlet\ \xi\sim\mu\ \kwin\ \mathcal D^{\mathcal S}_\ell [P]
          & \textsc{d-sample} \\[0.55em]
          \mathcal D^{\mathcal S}_\ell\Big[\displaystyle\sum_{i\in
          I}w_i\,P_i\Big]
          \equiv \displaystyle\sum_{i\in I}\Big(\frac{\partial
          w_i}{\partial\theta_\ell}\Big)\,P_i
          +\sum_{i\in I}w_i\,\mathcal D^{\mathcal S}_\ell [P_i]
          & \textsc{d-family} \\[0.9em]
          \mathcal D^{\mathcal S}_\ell\big[\kwev(\tau_1,\tau_2,H);P\big]
          \equiv E;
          \mathcal D^{\mathcal S}_\ell [P]
          \;+\;\mathcal S^{E}_\ell[P]
          & \textsc{d-evolve}
        \end{array}
      \]
    \end{minipage}
  }
  \caption{The differentiation transformation
  $\mathcal D^{\mathcal S}_\ell$. In \textsc{d-evolve},
  $E=\kwev(\tau_1,\tau_2,H)$ is the head segment and
  $\mathcal S^{E}_\ell$ its evolution derivative under strategy
  $\mathcal S$ (\rdef{edp}), the only
  strategy-dependent rule, instantiated by NSR (\sec{shift})
  and PSR (\sec{kick}). No score term arises in
  \textsc{d-sample} ($\mu$ is parameter-free) and no boundary term in
  \textsc{d-evolve} (the $\tau_i$ are parameter-free); sums of
  families flatten to families.}
  \label{fig:diff-rules}
\end{figure}

A program's outcome is a real-valued function of $\Theta$;
differentiating it with respect to $\theta_\ell$ defines the target
$\ket{\psi}\mapsto
\tfrac{\partial}{\partial\theta_\ell}\llbracket P\rrbracket(\ket{\psi})$.
Automatic differentiation is then a property a program
transformation may or may not have:

\begin{definition}[Sound transformation]
  \label{def:ad}
  A program transformation $\mathcal T_\ell$ is a \emph{sound
  differentiation transformation} for $\theta_\ell$ if
  (i)~it maps every well-formed program to a well-formed program of
  the same language, and (ii)~for every well-formed $P$ and every
  input state,
  \begin{equation}
    \llbracket \mathcal T_\ell [P]\rrbracket(\ket{\psi})
    =\frac{\partial}{\partial\theta_\ell}
    \llbracket P\rrbracket(\ket{\psi}).
    \label{eq:soundness-obligation}
  \end{equation}
\end{definition}

Condition~(ii) is exactness of the derivative semantics with no
step-size parameter; condition~(i) keeps the derivative program
expressible in the same language. Throughout, \emph{exact} refers to
this semantic property of a transformation, and \emph{unbiased} to
the executed estimator whose sample mean converges to it
(\thm{endtoend}).

\subsection{A Strategy-Dependent Transformation}

\fig{diff-rules} gives our transformation
$\mathcal D^{\mathcal S}_\ell$, determined by a
\emph{differentiation strategy} specified by $\mathcal S$. To differentiate
$\llbracket P\rrbracket$ with respect to $\theta_\ell$, follow where
$\theta_\ell$ appears: in a readout weight, a family weight, or a
coefficient of a \emph{segment} $E=\kwev(\tau_1,\tau_2,H)$. The first
two are classical expressions, so their rules are ordinary calculus
and fixed. The segment is the physics: its derivative cannot be
computed from the state, which is never read out, so it must be a
transformation of the segment itself---and that transformation is
strategy-dependent.

To isolate the evolution part of the transformation, consider a segment $E$
followed by a continuation $P$. We perturb $E$'s Hamiltonian by an amount
$s$ in the parameter-tangent direction
$\dot H_\ell = \partial H/\partial\theta_\ell$, while holding $P$ fixed.
The derivative with respect to $s$ at $s=0$ of this one-dimensional executable family is exactly the contribution of $E$ to $\partial \llbracket E;P \rrbracket/\partial \theta_{\ell}$.
This gives the interface used below: any program transformation that realizes this local contribution can fill the \textsc{d-evolve} rule, while the rest of the derivative is assembled compositionally.

\begin{definition}[Evolution contribution]
\label{def:evcontrib}
Let $E = \kwev(\tau_1,\tau_2,H)$ be a segment continued by a program $P$ in $E; P$. The \emph{evolution contribution} of $E$ to $\partial\llbracket E; P\rrbracket/\partial \theta_\ell$ is
\[
\partial_\ell^{\mathrm{ev}}
\llbracket E;P\rrbracket(|\psi\rangle)
=
\left.
\frac{\d}{\d s}
\llbracket E_{\ell,s};P\rrbracket(|\psi\rangle)
\right|_{s=0}
\]
where $E_{\ell,s} := \kwev(\tau_1,\tau_2,H+s\dot H_{\ell}),$ $\dot H_{\ell}= \frac{\partial H}{\partial \theta_{\ell}}$, and $s\in\mathbb{R}$ is a fresh scalar perturbation parameter.
\end{definition}

The perturbation enters $E$ only; $P$ is untouched. Lemmas \ref{lem:herm} and \ref{lem:duhamel} identify
$\partial_\ell^{\mathrm{ev}}\llbracket E;P\rrbracket$ with the chain
rule's $E$-term.

\begin{definition}[Locally sound evolution derivative]
\label{def:evolution-derivative}
\label{def:edp}
\label{def:estimator}
An \emph{evolution derivative} $\mathcal S$ maps, for each
$\theta_\ell$ and segment $E$, a program $P$ to a program
$\mathcal S^E_\ell[P]$. It is \emph{locally sound} if
$\mathcal S^E_\ell[P]$ is well-formed whenever $E;P$ is, and for
every $E$, $P$, and input state $\ket{\psi}$,
\[
\llbracket \mathcal S^E_\ell[P]\rrbracket(\ket{\psi})
=\partial_\ell^{\mathrm{ev}}\llbracket E;P\rrbracket(\ket{\psi}).
\]
\end{definition}

No syntactic form is required, but every $S$ used in this paper has the shape
\[
\mathcal S^E_\ell[P]\;\equiv\;
\kwlet\ \bar\xi\sim\bar\mu\ \kwin\
\sum\nolimits_{i\in I} w_i\;\big[\,F_{i,1};\cdots;F_{i,k_i};\,P\,\big],
\]
where the samples $\bar\xi$, the weights $w_i$, and the segments
$F_{i,r}$ are built from $E$ and $\ell$ alone; $P$ enters as
the shared continuation.

\begin{theorem}[Soundness of $\mathcal D$]
  \label{thm:soundness}
  If $\mathcal S$ is a locally sound evolution derivative, then
  $\mathcal D^{\mathcal S}_\ell$ of Fig.\ref{fig:diff-rules} is a
  sound differentiation transformation.
\end{theorem}
\begin{proof}
  Structural induction on $P$. The $\textsc{d-measure}$,
  $\textsc{d-sample}$, and $\textsc{d-family}$ cases follow from definition:
  linearity, differentiation under the sampling integral, and the
  product rule. For $\textsc{d-evolve}$, differentiating
  $\bra{\psi}U^\dagger O_PU\ket{\psi}$ splits into the evolution
  contribution, matched by $\mathcal S^{E}_\ell[P]$ by local
  soundness, and the continuation's $\Theta$-dependence, matched by
  $E;\mathcal D^{\mathcal S}_\ell P$ by induction.
  Full proof in \app{proofs}.
\end{proof}

\begin{remark}[Closure]
  \label{rem:closure}
  Condition~(i) of \rdef{ad} is the well-formedness
  half of local soundness; it holds by inspection for both
  evolution derivatives in this paper---NSR's and PSR's---and is
  not revisited per strategy.
\end{remark}

\begin{remark}[Machines and regimes]
  \label{rem:machine-regimes}
  Soundness is machine-independent; cost is not. Each evolution
  derivative is realized in \sec{compiler} and priced in
  \sec{eval}. A bounded-amplitude device truncates the NSR
  series; the truncation bias is bounded at compile time
  (\lem{trunc-bias}) and measured in \sec{eval}.
\end{remark}

\begin{remark}[Noise]
  \label{rem:noise}
  The theorems above are stated for the ideal unitary semantics, but
  the derivative program runs on a real machine under noise. Under a
  realistic error model (\app{noise}), NSR remains sound for the
  landscape the device actually realizes, and PSR does so when its
  inserted segments execute
  ideally---Theorems~\ref{thm:noise-nsr} and~\ref{thm:noise-psr},
  same programs, same certificates. PSR's insertion bias stays
  bounded; FD's setpoint error is amplified as its step shrinks.
  \sec{eval} compares all three under that model.
\end{remark}

Each strategy now owes three things---its evolution derivative, its admissibility assumption, and local soundness.

\subsection{The NSR Evolution Derivative}
\label{sec:shift}

We first introduce a general strategy applicable to a broad class of analog
quantum experiments. It relies on a Nyquist shift rule (NSR), generalized from \cite{theis2023proper}. For a segment $E=\kwev(\tau_1,\tau_2,H)$ where $H=\sum_ju_jA_j$, its tangent Hamiltonian along
$\theta_\ell$ and angular-bandwidth bounds:
\begin{equation*}
  \dot H_\ell \;\equiv
  \sum_j
  \frac{\partial u_j}{\partial\theta_\ell}\cdot A_j ,
  \quad
  \Omega_\ell=\int_{\tau_1}^{\tau_2}
  \operatorname{diam}\big(\dot H_\ell(t)\big)\,\d t,
  \label{eq:bandwidth}
\end{equation*}
where $\operatorname{diam}(M)=\lambda_{\max}(M)-\lambda_{\min}(M)$ is a spectral diameter. 

\begin{assumption}[NSR condition]\label{as:bandwidth}
  Every $A_j$ has finite spectral diameter, and the compiler can produce a positive constant $\overline\Omega_\ell\ge\Omega_\ell$ valid on the parameter region of interest.
\end{assumption}

$\overline\Omega_\ell$ is computable classically in preprocessing,
e.g.\ as \newline $\int\sum_j|\partial u_j/\partial\theta_\ell|
\operatorname{diam}(A_j)\,dt$ (\app{nyquist}) or
tighter by grouping anticommuting terms
(\app{select}), and is
itself \emph{parameter-free}: any sufficiently large bound serves
and may be fixed at compile time---what soundness needs is its independence from $\Theta$, since the evolution derivative constructions are computed from it.

If $\dot H_\ell\equiv 0$ on the segment,
$\operatorname{NSR}^E_\ell[P]$ is the empty family. Otherwise, with shifts and sampling weights
\begin{equation*}
  h_\kappa=\frac{(2\kappa+1)\pi}{2\overline\Omega_\ell},
  \qquad
  \eta_\kappa=\frac{8}{\pi^2}\frac{1}{(2\kappa+1)^2},
  \qquad \kappa=0,1,2,\ldots,
\end{equation*}
we define the NSR evolution derivative:
\begin{align}
  \operatorname{NSR}^E_\ell[P]&\equiv{}
  \kwlet\ \kappa\sim\eta\ \kwin\
  \kwlet\ \sigma\sim\operatorname{Uniform}\{-1,+1\}\ \kwin
  \nonumber\\
  &\quad
  \overline\Omega_\ell\,(-1)^\kappa\sigma\;\Big[
    \kwev\big(\tau_1,\tau_2,\;
    H+\sigma h_\kappa\,\dot H_\ell\big);\,P
  \Big]
  \label{eq:shift}
\end{align}
The branch is an ordinary analog evolution, and the transformed program differs from the source only in its coefficient expressions $u_j+\sigma h_\kappa\,\partial u_j/\partial\theta_\ell\in\mathcal E_\Theta$.

The soundness of NSR comes from the following lemma, proven in this paper
(\app{nyquist}). It generalizes the Fourier analysis for a fixed Hamiltonian in~\cite{theis2023proper} to time-dependent, non-commuting controls, with a Paley–Wiener band-limitation argument and a Nyquist differentiator applied to evolution contributions. 

\begin{lemma}[Nyquist reconstruction for time-dependent evolutions]
  \label{lem:nyquist-core}
  Under \as{bandwidth}, for every $P$ and $\ket{\psi}$, let
  $g(h)=\llbracket E_{\ell,h};P\rrbracket(\ket{\psi})$.
  Then $g$ is band-limited in $h$ with angular bandwidth at most
  $\Omega_\ell$, and $g'(0)$ is exactly the evolution contribution.
  Consequently, for every positive
  $\overline\Omega_\ell\ge\Omega_\ell$,
  \begin{equation}
    g'(0)
    =\frac{4\overline\Omega_\ell}{\pi^2}
    \sum_{\kappa=0}^{\infty}\frac{(-1)^\kappa}{(2\kappa+1)^2}
    \big[g(h_\kappa)-g(-h_\kappa)\big].
    \label{eq:nyquist-recovery}
  \end{equation}
\end{lemma}

\begin{theorem}[Soundness of NSR]
  \label{thm:shift-local}
  \label{cor:shift-sound}
  Under \as{bandwidth}, $\operatorname{NSR}$ is a
  locally sound evolution derivative, and
  $\mathcal D^{\operatorname{NSR}}_\ell$ is a sound differentiation
  transformation.
\end{theorem}
\begin{proof}
  The $\kwlet$/\textbf{family} semantics of $\operatorname{NSR}^E_\ell[P]$ evaluates to
  the series~\eqref{eq:nyquist-recovery}; \lem{nyquist-core}
  and \thm{soundness} give the rest.
\end{proof}

Consider the running example $P(\theta_1, \theta_2)$ in \eq{p-running}, written as $E;M$ in short. Since $\dot H_1=X$, we may choose $\overline\Omega_1=2T$, so $h_\kappa=(2\kappa+1)\pi/(4T)$ and
\begin{align*}
    \operatorname{NSR}^E_1[M] \equiv \kwlet\ \kappa\sim\eta\ \kwin\
    \kwlet\ \sigma\sim\operatorname{Uniform}\{\pm1\}\ \kwin\ \\
    2T(-1)^\kappa\sigma\,\big[
      \kwev\big(0,T,(\theta_1+\sigma h_\kappa+\theta_2 t/T)\cdot
    X\big);\,M\,\big].
\end{align*}
It first draws a classical $\kappa$ from the distribution $\eta$, then draws a sign $\sigma$, and finally executes the corresponding shifted branch with the displayed weight. 

\subsection{The PSR Evolution Derivative}
\label{sec:kick}

When the Hamiltonian terms touched by $\theta_\ell$ are involutions, the evolution contribution admits another evolution derivative, given by the analog parameter shift rule (PSR)  \cite{leng2022differentiable, banchi2021measuring}. 

\begin{assumption}[PSR condition]\label{as:kick}
  Every term $u_j\cdot A_j$ of $H$ with
  $\partial u_j/\partial\theta_\ell\not\equiv 0$ satisfies $A_j^2=I$.
\end{assumption}

For example, if every $A_j$ above is a Pauli string, the condition is satisfied. This condition restricts the touched generators to involutions, whose spectral diameter is at most~2.

For a zero-duration segment, $\operatorname{PSR}^E_\ell[P]$ is the empty family. Otherwise, writing $\Delta\tau=\tau_2-\tau_1$ for the segment's duration, $\operatorname{PSR}$ cuts the segment at a sampled time
$\tau$ and inserts a \emph{parameter-free} segment
$\kwev(0,\tfrac{\pi}{4},s\cdot A_j)$ with $s=\pm1$, whose propagator is the unitary $e^{-is\pi A_j/4}$. $\operatorname{PSR}$ is defined by
\begin{align*}
  B_{j,s}(\tau)\;\equiv&\;
  \kwev(\tau_1,\tau,H);\,
  \kwev\big(0,\tfrac{\pi}{4},s\cdot A_j\big);\,
  \kwev(\tau,\tau_2,H);\,P, \\
  \operatorname{PSR}^E_\ell[P]\;\equiv&\;
  \kwlet\ \tau\sim\operatorname{Uniform}(\tau_1,\tau_2)\ \kwin\ \\
  &\sum\limits_{j,\,s} s\,\Delta\tau\,
  \frac{\partial u_j}{\partial\theta_\ell}(\tau)\;B_{j,s}(\tau).
\end{align*}

The soundness of PSR comes from a generalized version of the parameter-shift rule.

\begin{lemma}[{\cite[Theorem~3.2]{leng2022differentiable}}]
  \label{lem:psr-id}
  Under \as{kick}, for every $P$ and every input state,
  \begin{equation*}
   \partial_\ell^{\mathrm{ev}}
   \llbracket E;P\rrbracket(\ket{\psi})
    =\sum\limits_{j}\int_{\tau_1}^{\tau_2}
    \frac{\partial u_j}{\partial\theta_\ell}(\tau)
    \sum\limits_{s=\pm1} s\,\llbracket B_{j,s}(\tau)\rrbracket(\ket{\psi})\,\d\tau.
  \end{equation*}
  The sum is over the terms touched by $\theta_\ell$, and the segments of $B_{j,s}(\tau)$ carry $\theta_\ell$. A proof is presented in \app{proofs}.
\end{lemma}

\begin{theorem}[Soundness of PSR]
  \label{thm:psr-local}
  \label{cor:psr-sound}
  Under \as{kick}, $\operatorname{PSR}_\ell$ is a
  locally sound evolution derivative, and
  $\mathcal D^{\operatorname{PSR}}_\ell$ is a sound differentiation
  transformation.
\end{theorem}
\begin{proof}
  The zero-duration case is immediate. Otherwise, the Monte Carlo
  identity
  $\int_{\tau_1}^{\tau_2}f\d\tau
  =\Delta\tau\,\mathbb E_{\tau\sim\mathrm{Uniform}(\tau_1,\tau_2)}[f]$ identifies
  the semantics of the sampled form with the identity above;
  \lem{psr-id} and \thm{soundness} give the
  rest.
\end{proof}

Again, consider the running example $P(\theta_1, \theta_2)$ in \eq{p-running}, written as $E;M$ in short. We write
\begin{align*}
&\operatorname{PSR}^E_1[M]~\equiv~\kwlet\ \tau\sim\operatorname{Uniform}(0,T)\ \kwin\ \\
&\quad \sum_s s\,T\,[\kwev(0,\tau,u\cdot X);\,
  \kwev(0,\tfrac{\pi}{4},s\cdot X);\,
\kwev(\tau,T,u\cdot X);\,M\,]    
\end{align*}
Here $u=\theta_1+\theta_2\,t/T$. Each branch inserts one parameter-free segment into the source program. Since $B_{j,s}(\tau)$ does not distinguish $\ell$, parameters touching the same Hamiltonian term share branches and differ only in weights.
\begin{remark}[Gates are a realization choice]
  Realization and pricing of the segment belong to compilation; if an
  inserted segment admits a better realization---a calibrated gate,
  say---the compiler is free to exploit it.
\end{remark}
NSR and PSR also realize the same derivative semantics in two
physical ways; \sec{compiler} turns the freedom into a compiler decision.



\subsection{The Finite-Difference Fallback}
\label{sec:fd}

Parameter substitution stays in the language when both shifted
valuations lie in the source program's domain, so the
program~\eqref{eq:fd-program} of \sec{why-derivatives}
is, for any $\theta_\ell$, the derived transformation
$\operatorname{FD}^{\epsilon}_\ell(P)\equiv
\tfrac{1}{\epsilon}\,P[\theta_\ell\mapsto\theta_\ell+\tfrac{\epsilon}{2}]
+(-\tfrac{1}{\epsilon})\,
P[\theta_\ell\mapsto\theta_\ell-\tfrac{\epsilon}{2}]$.
It is not a sound differentiation transformation: its bias of order
$\epsilon^2$ violates condition~(ii) for all sufficiently small
$\epsilon>0$ wherever the third derivative is nonzero (\app{fd}),
and \sec{eval} measures it as the tuned numerical baseline.

\section{Compilation to Devices}
\label{sec:compiler}

The transformation of \sec{ad} produces another DiffSimuQ program,
while the device executes waveforms, transport, and gates. Because
derivative programs remain in the same language, source, NSR, PSR,
and FD share one compilation stack. As shown in \fig{arch}, programs
are normalized into executable branches, realized as machine-native
Hamiltonian segments, and lowered to device-ready controls. We reuse
the source realization across derivative branches: NSR primarily edits
control coefficients, while PSR inserts a parameter-independent
operation. 
Our compiler targets an emerging zoned neutral-atom
architecture~\cite{bluvstein2024logical} whose software/hardware boundary has not yet stabilized, and which runs both analog evolution and digital gates and therefore realizes both evolution derivatives.

\subsection{Execution Normal Form}
\label{sec:stdform}

An analog experiment program may interleave $\mathsf{let}$ binders, weights, and families with evolutions, while a device runs one
straight-line experiment at a time. We bridge the two with a common
normal form (proof in \app{execution}).

\begin{theorem}[Execution normal form]\label{thm:stdform}
  Every well-formed program is semantically equivalent to
  \begin{equation}
    \mathsf{let}\ \xi_1\sim\mu_1\ \mathsf{in}\ \cdots\
    \mathsf{let}\ \xi_k\sim\mu_k\ \mathsf{in}\
    \sum_{i\in I} w_i(\bar\xi)\,B_i(\bar\xi).
    \label{eq:stdform}
  \end{equation}
  Each $B_i$ is a sequence of evolutions followed by a
  measurement.
\end{theorem}

This is the shared interface for source, NSR, PSR, and FD programs (\rrem{closure}). The classical runtime draws the sampling prefix, selects and weights a branch, and accumulates its outcome; only the straight-line branch $B_i$ is sent downstream for machine realization. Normalized programs may have redundancies like repetitive branches, where batching and branch merging remove redundant executions.

Normal form also isolates the statistical cost of executing a program by the following theorem. 

\begin{theorem}[Execution guarantee]\label{thm:endtoend}
Suppose a normalized program $Q$ has readouts bounded in absolute
value by $R$ and total absolute branch weight bounded by $W$ almost
surely, and each branch is realized according to its program
semantics. Then its runtime execution is unbiased for
$\llbracket Q\rrbracket$ on any input state, and $O\!\left((WR)^2\epsilon_t^{-2}\log(1/\beta)\right)$
quantum executions suffice for accuracy $\epsilon_t$ with probability
$1-\beta$.
\end{theorem}

For $Q=\mathcal D^{\mathcal S}_\ell(P)$ with a sound strategy, its execution therefore is an unbiased estimator for
$(\partial/\partial \theta_\ell)\llbracket P\rrbracket$ by \thm{soundness}.
For $Q=\mathrm{FD}^{\epsilon}_\ell(P)$, it instead estimates the
finite-difference secant, which approaches the derivative only as
$\epsilon\to0$.

\subsection{Target Machine}
\label{sec:machine}%

Our target models on a cryogenic $^{133}$Cs tweezer array~\cite{jin2026extended} 
with the zoned neutral-atom layout of~\cite{bluvstein2024logical}: 
atoms occupy programmable optical
traps on a 2D plane, while a spatially-separated gate zone applies
two-qubit gates to pairs transported to it. The machine realizes
Hamiltonians of the form:
\begin{equation*}
  \sum_a\big(u^X_a(t)X_a+u^Y_a(t)Y_a+u^Z_a(t)Z_a\big)
  +u^{zz}(t)\sum_{a<b}J_{ab}Z_aZ_b ,
\end{equation*}
where $J_{ab}$ is fixed by the array geometry.

Two machine properties support the derivative realizations of
\sec{ad}. For NSR, the language coefficients $u_j(t)$ often map directly to drive waveforms, so a coefficient shift is a shift on the same physical channels; shifts of jointly synthesized interactions may instead require re-realization; the compiler separately checks whether it
fits within device headroom. For PSR, the $Z_aZ_b$ interaction can
be disabled during transport and restored afterward, allowing the
source evolution to halt, execute the inserted operation in the gate
zone, and resume. Device specs used by the
compiler and evaluation are specified in \app{machine}.

\paragraph{Device-realized objective.}
On hardware, a program denotes the expectation under the device's
noisy dynamics, not ideal evolution. Under the time-local Lindblad
model of \app{noise}, with shift-independent dissipators, NSR and PSR
remain unbiased for gradients of this device-realized objective.
Hardware-realization errors, including finite NSR headroom and
imperfect PSR insertions, are treated separately.

\subsection{Strategy-Aware Lowering and Optimization}
\label{sec:lower}

\begin{figure*}[t]
  \centering
  \includegraphics[width=\textwidth]{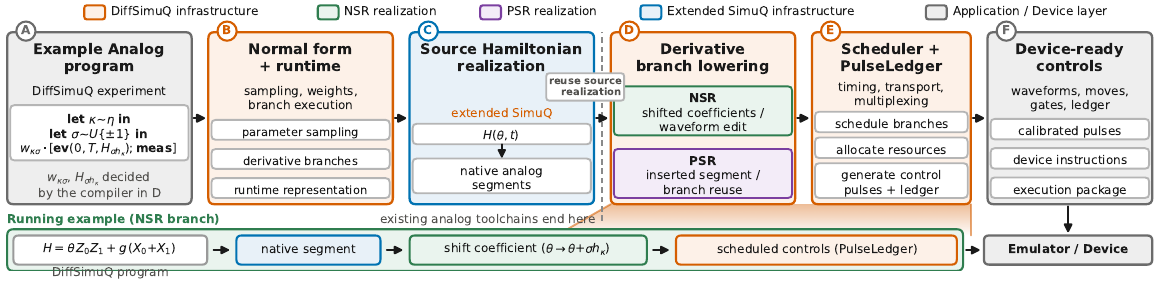}
  \caption{The lowering stack, with an NSR branch along the bottom (the PSR branch in
    \fig{branch}). Programs normalize to straight-line branches; extended SimuQ realizes the source Hamiltonian as machine-native segments. DiffSimuQ reuses the source to lower derivative branches as structured edits, then schedules them to controls and logs the result in the PulseLedger.}
  \label{fig:arch}
\end{figure*}

To realize the straight-line branches, we extend
SimuQ~\cite{peng2024simuq}, which maps a target Hamiltonian to
machine-native Hamiltonian segments. Our extension adds symbolic
Hamiltonian expressions for differentiating classical weights
(\app{expr}), a specialized solver for the target device
(\app{scaling}), and an abstract analog instruction set for the
tweezer array (\app{machine}). As shown in \fig{arch}, SimuQ realizes
the Hamiltonian fragments, while DiffSimuQ supplies the experiment
runtime and the remaining physical-control lowering.

More importantly, DiffSimuQ exploits the structure of derivative
programs rather than compiling every branch independently. Once the
source Hamiltonian is realized, derivative branches are represented
as structured edits of that realization, enabling strategy-specific
compile-time analyses and optimizations. NSR reuses the source
schedule and edits its coefficients; PSR reuses the source realization
around an inserted operation. These structures enable fusion, branch
reuse, hardware-feasibility checks, and strategy selection.

\paragraph{NSR lane.}
An NSR branch preserves the source Hamiltonian structure and replaces its
coefficient arrays by
$u_j+\sigma h_\kappa\,\partial u_j/\partial\theta_\ell$.
The compiler can therefore edit waveform parameters directly for most native controls; only shifts that change a synthesized interaction require re-realization. A static \emph{headroom} analysis determines
the largest playable shift index $M$. The resulting $\mathrm{NSR}^M$ may reject or truncate larger
shifts; Lemmas~\ref{lem:pfail} and~\ref{lem:trunc-bias} bound the
resulting realization error. 
When a parameter touches $k$ consecutive evolutions, \emph{fusion}
shares one $(\kappa,\sigma)$ draw and shifts them jointly, reducing
the branch inventory from $k$ to one (\rrem{fusion}).

\begin{figure}[t]
  \centering
  \includegraphics[width=\linewidth]{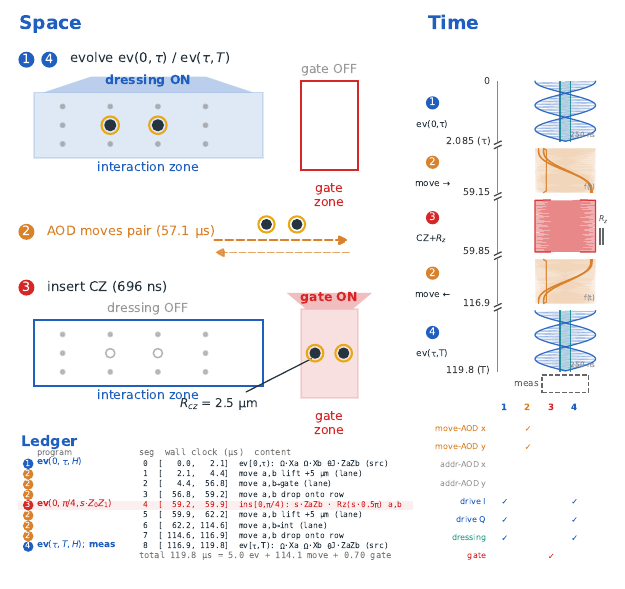}
  \caption{One device-ready compiled PSR branch $B_{j,s}(\tau)$ of the running
    instance, $A_j=Z_aZ_b$. Space: the pair is carried to the gate
    zone and back. Time: active channels per stage (axis not to
    scale). Ledger: the PulseLedger rows for the same branch, each
    group labelled with the program constructor it realizes. The
    inserted segment $\mathsf{ev}(0,\pi/4,s\cdot Z_aZ_b)$ is one CZ
    plus virtual $R_z$ rotations (\app{machine}).}
  \label{fig:branch}
\end{figure}

\paragraph{PSR lane.}
A PSR branch $B_{j,s}(\tau)$ cuts the source evolution at $\tau$ and
inserts the parameter-independent segment of \sec{kick}. The compiler
reuses the source realization on both sides of the insertion and
realizes the inserted operation once per generator class. On the
machine of \sec{machine}, a $Z_aZ_b$ insertion requires transport to
the gate zone and a calibrated gate, as shown in \fig{branch}.
When $p$ parameters touch the same Hamiltonian term, \emph{branch
reuse} shares the corresponding quantum executions and reweights
their outcomes, reducing the branch inventory by a factor of $p$.

\paragraph{Static strategy selection.}
The two optimized realizations expose different compile-time cost
structures. PSR cost follows from the touched terms and available
branch reuse, whereas NSR cost is estimated from the tangent-bandwidth
certificate $\overline\Omega_\ell$ and fusion opportunities. The
compiler combines these static bounds with a calibrated margin to select the cheaper realization
before execution. Both candidates implement the same derivative
semantics; hardware-specific realization errors are accounted for
separately. \app{select} gives the cost model, and
\sec{eval-select} evaluates the selector.

\subsection{Scheduling to Device-Ready Controls}
\label{sec:ledger}

A machine-native Hamiltonian segment specifies what the device should
do, but not yet when or on which physical channels it should happen.
The scheduler assigns each semantic action to wall-clock time and
lowers it to transport, gates, and per-channel waveforms, following
the real-time control model of~\cite{wu2026pulsedsl}.
Multiplexing maps these controls onto the fixed channel
complement: per-site drives share hardware by frequency, while the
gate zone is shared in time. \fig{branch} shows the resulting
device-ready control for one PSR branch, including transport to the
gate zone, the inserted operation, return transport, and resumed source
evolution.

On movable-atom hardware, position and frame are part of the machine
state, so waveforms alone do not fully identify the operation being
executed. DiffSimuQ therefore has a \emph{PulseLedger} alongside the
controls, recording each segment's program and wall-clock boundaries,
coefficient provenance, frame and transport state, and insertion
markers. The ledger makes the emitted schedule self-describing for
downstream analysis and calibrated error correction
(\app{compiler}).

\newcommand{\owed}[1]{{\color{red}\textbf{[owed: #1]}}}

\section{Evaluation}
\label{sec:eval}

The evaluation tests the compiler and system claims in this paper. 
We design the experiments to answer four questions: (1) do the exact derivative semantics survive device noise, and what
errors remain from physical realization; (2) does their accuracy matter to an optimizer; (3) do the two realizations have different costs; (4) does the transformed experiment compile at scale?

\paragraph{Device emulation.}
We execute the compiled machine nat-ive segments and normal-form orchestration on a QuTiP emulator~\cite{qutip} of the target machine in \sec{machine}. The emulator uses the device model of \app{noise}, including dephasing, insertion-gate infidelity, and control-setpoint errors.

\paragraph{Evaluation setup.}
We use transverse-field Ising models (TFIM) with different system scales for different questions. At two
qubits, where device-level emulation is tractable, we evaluate
gradient accuracy and closed-loop optimization.  At seven qubits, we
vary program structure to characterize the cost boundary between NSR
and PSR.  Finally, we scale structured TFIM programs from 10 to 1000
qubits to measure compilation cost; these large-system experiments
measure compilation only, not quantum execution or dynamics.

\subsection{Gradient Accuracy and Closed-Loop Optimization}
\label{sec:eval-machine}

\begin{figure}[t]
  \centering
  \includegraphics[width=\linewidth]{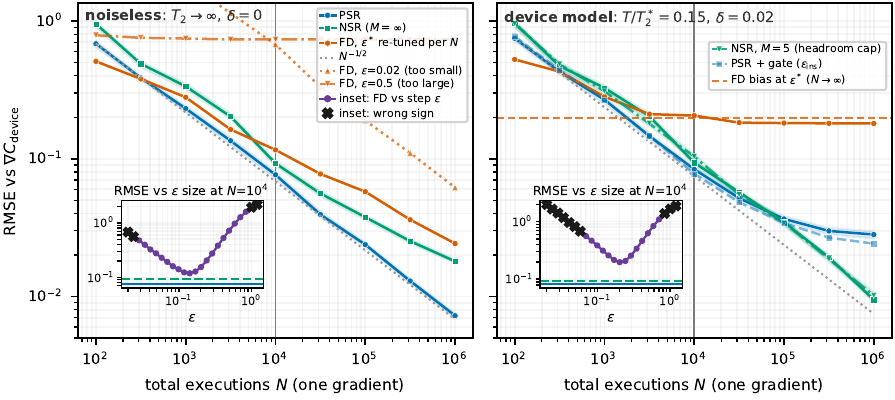}
\caption{Gradient error against execution budget. NSR and PSR
converge at $N^{-1/2}$. FD with $\epsilon^*$ re-tuned per $N$
follows $N^{-1/3}$ on the noiseless landscape (left) and reaches a bias floor under the device model (right, dashed); a
too-small $\epsilon$ amplifies shot noise and setpoint error, while a
too-large $\epsilon$ is truncation-limited. Inset: error against step
size at the budget marked by the vertical line.}
  \label{fig:F6}
\end{figure}

We first test estimator accuracy on a two-qubit TFIM, $H(\theta)=\theta Z_0Z_1+g(X_0+X_1),$
with $g$ fixed and $\theta$ the parameter. The measurement follows the observable $Z_0Z_1$, giving objective $C_{\mathrm{device}}$. We compare
each compiled estimator against $(\partial/\partial \theta) C_{\mathrm{device}}$
while varying the total execution budget $N$. As shown in
\fig{F6}, NSR and PSR converge at the expected $N^{-1/2}$ rate.
On the noiseless model (left), a budget-adaptive FD baseline,
whose step $\epsilon^*(N)$ is re-tuned at every execution budget,
is consistent and follows the expected numerical tradeoff as
$\epsilon^*(N)$ decreases. 
Under the device model (right), NSR and PSR continue to converge
until their separately modeled realization errors---finite NSR
headroom and imperfect PSR insertion---become visible. Adaptive FD,
however, reaches a substantially higher floor: decreasing
$\epsilon$ reduces its truncation error but amplifies sampling and
setpoint errors, while increasing $\epsilon$ restores truncation
error.

\begin{figure}[t]
  \centering
  \includegraphics[width=\linewidth]{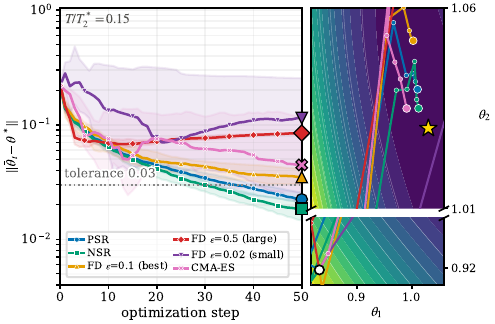}
  \caption{The two-qubit instance in closed loop: median and
    interquartile range of $\lVert\bar\theta_t-\theta^{\ast}\rVert$
    over 20 seeds, same optimizer and budget per step for every
    method. PSR and NSR reach the planted optimum and keep
    descending; every FD step and CMA-ES floor above them. Right:
    the valley on a broken $\theta_2$ axis, one seed per method,
    star $\theta^*$, circle the start.}
  \label{fig:loop}
\end{figure}

The remaining question is whether this gradient-error floor matters
to optimization.  We therefore run a two-parameter inverse-simulation
task with $H=\theta_1 Z_0Z_1+\theta_2(X_0+X_1),$
minimizing the residual between measured observables and their values
at a planted interior target $\boldsymbol{\theta}^{*}$.
Figure~\ref{fig:loop} reports 20 runs with the same gradient-descent 
optimizer and a 6000-shot cap for all gradient methods, together with
CMA-ES as a derivative-free baseline.  NSR and PSR reach the target
and continue descending, while every tested FD step and CMA-ES
plateau above them: a large FD step is truncation-limited, a small
one is unstable, and even the retrospectively best fixed step stalls
once its gradient error becomes significant relative to the local
curvature.

\subsection{Cost Tradeoffs and Strategy Selection}
\label{sec:eval-select}

NSR and PSR compute the same derivative but consume different
resources. NSR shifts all terms touched by one parameter together,
whereas PSR inserts one operation per touched Hamiltonian term and can
reuse that branch across parameters. We therefore expect their relative
cost to depend on two properties of the program: $p$, the number of
differentiated coefficients, and $q$, the number of device-alphabet
terms touched by each coefficient. Because quantum evolution time is a small portion of one execution (\app{window}), throughput is set by the number of executions.

\begin{figure}[t]
  \centering
  \includegraphics[width=0.81\linewidth]{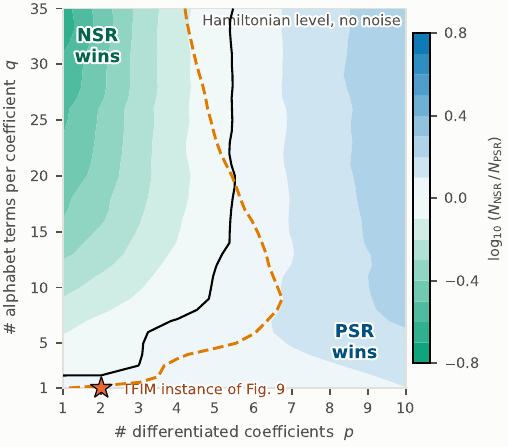}
\caption{NSR--PSR execution-cost tradeoff on seven qubits.
Each point varies $p$, the number of differentiated coefficients,
and $q$, the number of device-alphabet terms each touches.
Color shows $\log_{10}(N_{\mathrm{NSR}}/N_{\mathrm{PSR}})$:
blue favors PSR and green favors NSR. Black: measured crossing;
dashed orange: compiler selection boundary. The selector agrees
with the measured winner on 88.3\% of points and incurs at most
$1.26\times$ overhead when wrong. Star: the TFIM instance used in Sec. 6.1 (Fig. 9).}
  \label{fig:select}
\end{figure}

\fig{select} measures this tradeoff on a fixed seven-qubit
system. For each $(p,q)$, we generate programs whose tangents draw
$q$ terms from the 35-term device pool
$\{X_a,Z_a,Z_aZ_b\}$ and compare the execution budgets
$N_{\mathrm{NSR}}$ and $N_{\mathrm{PSR}}$ required for the same
gradient accuracy. The plane separates into two complementary regimes:
PSR wins when many parameters can reuse branches associated with the
same terms, while NSR wins when each parameter touches many terms,
because one combined Hamiltonian shift via fusion replaces multiple insertions.
NSR wins on 42\% of the measured plane. 

The compiler's static selector follows the same structural
information. As overlaid in \fig{select}, it agrees with
the measured winner on 88.3\% of the design space; when it chooses
the more expensive realization, the median overhead is $1.07\times$
and the maximum is $1.26\times$. The selector optimizes execution
cost: both candidates implement the same derivative semantics, while
their hardware-specific realization errors are characterized
separately. \app{select} gives the certificate construction,
calibration, and results on additional Hamiltonian families.

\begin{figure}[t]
  \centering
  \includegraphics[width=\linewidth]{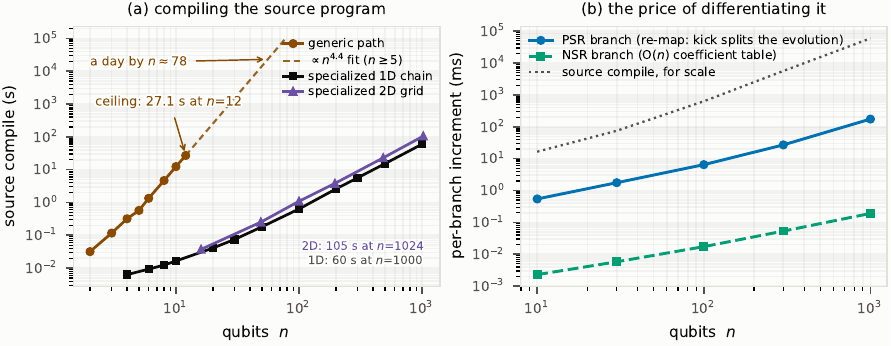}
  \caption{Compile cost against system size: (a) the source compile
    on the generic and the specialized paths; (b) the per-branch
    increment of differentiation, with the source compile for scale. At 1000 qubits, PSR adds 175 ms  and NSR 0.2 ms.}
  \label{fig:compile-curves}
\end{figure}

\subsection{Compilation at Scale}
\label{sec:eval-scale}

Our final question is whether differentiation creates a new
compilation bottleneck.  We therefore scale a structured TFIM family
from $n=10$ to $1000$ qubits and measure compilation only; the
corresponding quantum dynamics are not emulated at these sizes.
\fig{compile-curves}(a) shows that the dominant cost is
compiling the source Hamiltonian.  The generic SimuQ path becomes
impractical beyond roughly a dozen qubits, while a hardware-aware
specialization for our devices introduced in \sec{lower} scales to $1000$ qubits. \tab{strategies} summarizes the three realizations from derivative semantics through physical execution and per-branch compilation cost.

Once the source is compiled, derivative branches are comparatively
cheap.  At $n=1000$, the source compilation takes $59.6$\,s, while one
PSR branch adds $175$\,ms ($0.3\%$) and one NSR branch $0.191$\,ms
($3\times10^{-4}\%$).  NSR only edits the compiled coefficient table,
while PSR reuses the source realization on both sides of its inserted
operation.  Thus differentiation inherits the source program's
compilation bottleneck rather than introducing a new one. Across these experiments, the same language, derivative
transformation, and compilation stack apply from two to $1000$
qubits: the backend that compiles the experiment also compiles its derivative.

\begin{table}[t]
  \centering
  \small
  \setlength{\tabcolsep}{3.5pt}
  \caption{Derivative realizations: Machine-specific
  costs refer to the evaluated TFIM workload; * denotes
realization-preserving shifts; see full comparison in
  \app{scaling}.}
  \label{tab:strategies}
  \begin{tabular}{@{}lccc@{}}
    \toprule
      & PSR & NSR & FD \\
    \midrule
    branch
      & insertion & shift & re-dial \\
    step tuning
      & no & no & yes \\
    exact derivative
      & yes & yes & no \\
    schedule preserved
      & no & usually yes & usually yes \\
    device-level error source
      & insertion & headroom & step/setpoint \\
    compile/branch ($10^3$ q.)
      & 175\,ms & $\approx$~0.2\,ms* & $\approx$~0.2\,ms* \\
    \bottomrule
  \end{tabular}
\end{table}

\section{Related Work}
\label{sec:related}

\paragraph{Analytic gradients.}
Gate-level parameter-shift rules differentiate individual
gates~\cite{schuld2019evaluating}; Banchi and Crooks~\cite{banchi2021measuring} treat general
parameterized evolutions by sampled insertions;
and Leng et al. develop a pulse-level differentiable analog framework
with a stochastic evolution-gradient estimator and optimization
loop~\cite{leng2022differentiable}, whose local identity our PSR adapts. Theis~\cite{theis2023proper} derives exact Nyquist shift rules for
the time-independent perturbed evolution $e^{i(xA+B)}$, which we generalize in \lem{nyquist-core} to $H(t)+s\dot H(t)$ without commutativity
assumptions across controls or times. These works provide measurement-based estimators for individual evolutions. DiffSimuQ lifts such local estimators to full-program, compositional AD with same-language closure and denotational soundness, then lowers the resulting derivative programs through a shared compiler to device controls. Classical AD differentiates through an accessible trace~\cite{baydin2018automatic}, while a quantum device exposes no such trace.

\paragraph{Analog quantum programming and compilation.}
Pulse-level interfaces~\cite{alexander2020qiskit,openqasm3,
Silverio2022pulseropensource,Bloqade},
the analog compilers SimuQ~\cite{peng2024simuq} and
QTurbo~\cite{zhou2026qturbo}, the neutral-atom
compiler Atomique~\cite{wang2024atomique}, and circuit-layout
synthesis~\cite{tan2020optimal} provide complementary abstractions for
expressing or lowering quantum programs. DiffSimuQ extends this stack
with an explicit derivative-program transformation and
strategy-aware lowering.
Quantum circuit languages~\cite{Green2013,paykin2017qwire} and verified
compilers~\cite{hietala2021verified} provide formal semantics of gate-based quantum programs. 
DiffSimuQ instead formalizes analog experiment programs and program transformations whose derivatives are themselves programs, making derivative identities alternative realizations that the compiler can select and optimize.

\section{Conclusion}
\label{sec:conclusion}
DiffSimuQ shows that the gradient of an analog quantum experiment
can be obtained the way classical AD obtains gradients: by
transforming the program. A program written in our language is
mapped to a family of programs in the same language whose average
outcome is the derivative. Only the
rule for a continuous evolution depends on the strategy, and we give
two: a new Nyquist shift rule, which works for general time-dependent
control, and the parameter-shift rule, which applies when the
touched terms are involutions. Because the derivative is still a
program, the same compiler lowers it to neutral-atom controls and
picks a strategy based on estimated cost. DiffSimuQ makes differentiation 
a property of the experiment programming stack rather than a numerical 
procedure implemented outside it, enabling quantum control and variational 
systems to treat an analog quantum experiment as a differentiable component. 
More broadly, it provides a common interface on which new derivative rules, 
hardware backends, and hardware-aware optimization strategies can be developed 
without changing the experiment-level programming model.


\section{Acknowledgment}
\label{sec:Acknowledgment}
We thank Yikai Zheng for the initial discussions on semantics and program transformations, we also thank Yu-Hsuan Wu, Minnan Pei, Kazem Taram, Xiaodi Wu, and Jiaqi Leng for helpful discussions and their feedback. Y.~S. acknowledges support from the Princeton Quantum Initiative Graduate Fellowship and Purdue Quantum Theory Neighborhood program.

\bibliography{refs}

@article{openqasm3,
author = {Cross, Andrew and Javadi-Abhari, Ali and Alexander, Thomas and De Beaudrap, Niel and Bishop, Lev S. and Heidel, Steven and Ryan, Colm A. and Sivarajah, Prasahnt and Smolin, John and Gambetta, Jay M. and Johnson, Blake R.},
title = {OpenQASM 3: A Broader and Deeper Quantum Assembly Language},
year = {2022},
issue_date = {September 2022},
publisher = {Association for Computing Machinery},
address = {New York, NY, USA},
volume = {3},
number = {3},
issn = {2643-6809},
url = {https://doi.org/10.1145/3505636},
doi = {10.1145/3505636},
journal = {ACM Transactions on Quantum Computing},
month = {sep},
articleno = {12},
numpages = {50}
}

@article{Silverio2022pulseropensource,
  doi = {10.22331/q-2022-01-24-629},
  url = {https://doi.org/10.22331/q-2022-01-24-629},
  title = {Pulser: {A}n open-source package for the design of pulse sequences in programmable neutral-atom arrays},
  author = {Silv{\'{e}}rio, Henrique and Grijalva, Sebasti{\'{a}}n and Dalyac, Constantin and Leclerc, Lucas and Karalekas, Peter J. and Shammah, Nathan and Beji, Mourad and Henry, Louis-Paul and Henriet, Lo{\"{i}}c},
  journal = {{Quantum}},
  issn = {2521-327X},
  publisher = {{Verein zur F{\"{o}}rderung des Open Access Publizierens in den Quantenwissenschaften}},
  volume = {6},
  pages = {629},
  month = jan,
  year = {2022}
}

@misc{Bloqade,
  title={Bloqade: a Julia package for quantum computation and quantum dynamics based on neutral-atom architectures},
  author={QuEra},
  year={2022}, 
  url={https://queracomputing.github.io/Bloqade.jl/dev/}
}

@inproceedings{tan2020optimal,
  title={Optimal layout synthesis for quantum computing},
  author={Tan, Bochen and Cong, Jason},
  booktitle={Proceedings of the 39th International Conference on Computer-Aided Design},
  pages={1--9},
  year={2020}
}

@article{qutip,
title = {QuTiP: An open-source Python framework for the dynamics of open quantum systems},
journal = {Computer Physics Communications},
volume = {183},
number = {8},
pages = {1760-1772},
year = {2012},
issn = {0010-4655},
doi = {https://doi.org/10.1016/j.cpc.2012.02.021},
url = {https://www.sciencedirect.com/science/article/pii/S0010465512000835},
author = {J.R. Johansson and P.D. Nation and Franco Nori}
}

@article{alexander2020qiskit,
  title={Qiskit pulse: Programming quantum computers through the cloud with pulses},
  author={Alexander, Thomas and Kanazawa, Naoki and Egger, Daniel J and Capelluto, Lauren and Wood, Christopher J and Javadi-Abhari, Ali and McKay, David C},
  journal={Quantum Science and Technology},
  volume={5},
  number={4},
  pages={044006},
  year={2020},
  publisher={IOP Publishing}
}

@article{farhi2014quantum,
  title={A quantum approximate optimization algorithm},
  author={Farhi, Edward and Goldstone, Jeffrey and Gutmann, Sam},
  journal={arXiv preprint arXiv:1411.4028},
  year={2014}
}

@article{paykin2017qwire,
  title={QWIRE: a core language for quantum circuits},
  author={Paykin, Jennifer and Rand, Robert and Zdancewic, Steve},
  journal={ACM SIGPLAN Notices},
  volume={52},
  number={1},
  pages={846--858},
  year={2017},
  publisher={ACM New York, NY, USA}
}

@inproceedings{Green2013,
author = {Green, Alexander S. and Lumsdaine, Peter LeFanu and Ross, Neil J. and Selinger, Peter and Valiron, Beno\^{\i}t},
title = {Quipper: A Scalable Quantum Programming Language},
year = {2013},
isbn = {9781450320146},
publisher = {Association for Computing Machinery},
address = {New York, NY, USA},
url = {https://doi.org/10.1145/2491956.2462177},
doi = {10.1145/2491956.2462177},
booktitle = {Proceedings of the 34th ACM SIGPLAN Conference on Programming Language Design and Implementation},
pages = {333–342},
numpages = {10},
location = {Seattle, Washington, USA},
series = {PLDI '13}
}

@article{hietala2021verified,
author = {Hietala, Kesha and Rand, Robert and Hung, Shih-Han and Wu, Xiaodi and Hicks, Michael},
title = {A Verified Optimizer for Quantum Circuits},
year = {2021},
issue_date = {January 2021},
publisher = {Association for Computing Machinery},
address = {New York, NY, USA},
volume = {5},
number = {POPL},
url = {https://doi.org/10.1145/3434318},
doi = {10.1145/3434318},
journal = {Proc. ACM Program. Lang.},
month = {jan},
articleno = {37},
numpages = {29}
}

@article{theis2023proper,
  title={" Proper" Shift Rules for Derivatives of Perturbed-Parametric Quantum Evolutions},
  author={Theis, Dirk Oliver},
  journal={Quantum},
  volume={7},
  pages={1052},
  year={2023},
  publisher={Verein zur F{\"o}rderung des Open Access Publizierens in den Quantenwissenschaften}
}

@article{bluvstein2024logical,
  title={Logical quantum processor based on reconfigurable atom arrays},
  author={Bluvstein, Dolev and Evered, Simon J and Geim, Alexandra A and Li, Sophie H and Zhou, Hengyun and Manovitz, Tom and Ebadi, Sepehr and Cain, Madelyn and Kalinowski, Marcin and Hangleiter, Dominik and others},
  journal={Nature},
  volume={626},
  number={7997},
  pages={58--65},
  year={2024},
  publisher={Nature Publishing Group UK London}
}

@inproceedings{wang2024atomique,
  title={Atomique: A quantum compiler for reconfigurable neutral atom arrays},
  author={Wang, Hanrui and Liu, Pengyu and Tan, Daniel Bochen and Liu, Yilian and Gu, Jiaqi and Pan, David Z and Cong, Jason and Acar, Umut A and Han, Song},
  booktitle={2024 ACM/IEEE 51st Annual International Symposium on Computer Architecture (ISCA)},
  pages={293--309},
  year={2024},
  organization={IEEE}
}

@article{peng2024simuq,
  title={SimuQ: A framework for programming quantum Hamiltonian simulation with analog compilation},
  author={Peng, Yuxiang and Young, Jacob and Liu, Pengyu and Wu, Xiaodi},
  journal={Proceedings of the ACM on Programming Languages},
  volume={8},
  number={POPL},
  pages={2425--2455},
  year={2024},
  publisher={ACM New York, NY, USA}
}

@article{glaser2015training,
  title={Training Schr{\"o}dinger’s cat: Quantum optimal control: Strategic report on current status, visions and goals for research in Europe},
  author={Glaser, Steffen J and Boscain, Ugo and Calarco, Tommaso and Koch, Christiane P and K{\"o}ckenberger, Walter and Kosloff, Ronnie and Kuprov, Ilya and Luy, Burkhard and Schirmer, Sophie and Schulte-Herbr{\"u}ggen, Thomas and others},
  journal={The European Physical Journal D},
  volume={69},
  number={12},
  pages={279},
  year={2015},
  publisher={Springer}
}

@article{georgescu2014quantum,
  title={Quantum simulation},
  author={Georgescu, Iulia M and Ashhab, Sahel and Nori, Franco},
  journal={Reviews of Modern Physics},
  volume={86},
  number={1},
  pages={153--185},
  year={2014},
  publisher={APS}
}

@article{cerezo2021variational,
  title={Variational quantum algorithms},
  author={Cerezo, Marco and Arrasmith, Andrew and Babbush, Ryan and Benjamin, Simon C and Endo, Suguru and Fujii, Keisuke and McClean, Jarrod R and Mitarai, Kosuke and Yuan, Xiao and Cincio, Lukasz and others},
  journal={Nature Reviews Physics},
  volume={3},
  number={9},
  pages={625--644},
  year={2021},
  publisher={Nature Publishing Group UK London}
}

@article{baydin2018automatic,
  title={Automatic differentiation in machine learning: a survey},
  author={Baydin, Atilim Gunes and Pearlmutter, Barak A and Radul, Alexey Andreyevich and Siskind, Jeffrey Mark},
  journal={Journal of machine learning research},
  volume={18},
  number={153},
  pages={1--43},
  year={2018}
}

@article{schuld2019evaluating,
  title={Evaluating analytic gradients on quantum hardware},
  author={Schuld, Maria and Bergholm, Ville and Gogolin, Christian and Izaac, Josh and Killoran, Nathan},
  journal={Physical Review A},
  volume={99},
  number={3},
  pages={032331},
  year={2019},
  publisher={APS}
}

@article{banchi2021measuring,
  title={Measuring analytic gradients of general quantum evolution with the stochastic parameter shift rule},
  author={Banchi, Leonardo and Crooks, Gavin E},
  journal={Quantum},
  volume={5},
  pages={386},
  year={2021},
  publisher={Verein zur F{\"o}rderung des Open Access Publizierens in den Quantenwissenschaften}
}

@article{peruzzo2014variational,
  title={A variational eigenvalue solver on a photonic quantum processor},
  author={Peruzzo, Alberto and McClean, Jarrod and Shadbolt, Peter and Yung, Man-Hong and Zhou, Xiao-Qi and Love, Peter J and Aspuru-Guzik, Al{\'a}n and O’brien, Jeremy L},
  journal={Nature communications},
  volume={5},
  number={1},
  pages={4213},
  year={2014},
  publisher={Nature Publishing Group UK London}
}

@article{wiebe2014hamiltonian,
  title={Hamiltonian learning and certification using quantum resources},
  author={Wiebe, Nathan and Granade, Christopher and Ferrie, Christopher and Cory, David G},
  journal={Physical review letters},
  volume={112},
  number={19},
  pages={190501},
  year={2014},
  publisher={APS}
}

@article{kokail2026inverse,
  title={Inverse quantum simulation for quantum material design},
  author={Kokail, Christian and Dolgirev, Pavel E and van Bijnen, Rick and Gonzalez-Cuadra, Daniel and Lukin, Mikhail D and Zoller, Peter},
  journal={arXiv preprint arXiv:2601.12239},
  year={2026}
}

@article{leng2022differentiable,
  title={Differentiable analog quantum computing for optimization and control},
  author={Leng, Jiaqi and Peng, Yuxiang and Qiao, Yi-Ling and Lin, Ming and Wu, Xiaodi},
  journal={Advances in Neural Information Processing Systems},
  volume={35},
  pages={4707--4721},
  year={2022}
}

@inproceedings{zhou2026qturbo,
  title={QTurbo: A Robust and Efficient Compiler for Analog Quantum Simulation},
  author={Zhou, Junyu and Liu, Yuhao and Che, Shize and Mitra, Anupam and K{\"o}kc{\"u}, Efekan and Rrapaj, Ermal and Iancu, Costin and Li, Gushu},
  booktitle={Proceedings of the 31st ACM International Conference on Architectural Support for Programming Languages and Operating Systems, Volume 1},
  pages={203--218},
  year={2026}
}

@book{watrous2018theory,
  title={The Theory of Quantum Information},
  author={Watrous, John},
  year={2018},
  publisher={Cambridge University Press}
}

@article{evered2023high,
  title={High-fidelity parallel entangling gates on a neutral-atom quantum computer},
  author={Evered, Simon J. and Bluvstein, Dolev and Kalinowski, Marcin and Ebadi, Sepehr and Manovitz, Tom and Zhou, Hengyun and Li, Sophie H. and Geim, Alexandra A. and Wang, Tout T. and Maskara, Nishad and Levine, Harry and Semeghini, Giulia and Greiner, Markus and Vuleti{\'c}, Vladan and Lukin, Mikhail D.},
  journal={Nature},
  volume={622},
  number={7982},
  pages={268--272},
  year={2023}
}

@article{rydberg-review,
  title={Many-body physics with individually controlled {Rydberg} atoms},
  author={Browaeys, Antoine and Lahaye, Thierry},
  journal={Nature Physics},
  volume={16},
  pages={132--142},
  year={2020}
}

@article{rydberg-gate,
  title={Parallel implementation of high-fidelity multiqubit gates with neutral atoms},
  author={Levine, Harry and Keesling, Alexander and Semeghini, Giulia and Omran, Ahmed and Wang, Tout T. and Ebadi, Sepehr and Bernien, Hannes and Greiner, Markus and Vuleti{\'c}, Vladan and Pichler, Hannes and Lukin, Mikhail D.},
  journal={Physical Review Letters},
  volume={123},
  pages={170503},
  year={2019}
}

@article{jin2026extended,
  title={Extended Rydberg Lifetimes in a Cryogenic Atom Array},
  author={Jin, Junlan and Shi, Yue and Alaoui, Youssef Aziz and Deng, Jingxin and Lu, Yukai and Thompson, Jeff D and Bakr, Waseem S},
  journal={PRX Quantum},
  volume={7},
  number={3},
  pages={033009},
  year={2026},
  publisher={APS}
}

@article{manetsch2025tweezer,
  title={A tweezer array with 6,100 highly coherent atomic qubits},
  author={Manetsch, Hannah J and Nomura, Gyohei and Bataille, Elie and Lv, Xudong and Leung, Kon H and Endres, Manuel},
  journal={Nature},
  volume={647},
  number={8088},
  pages={60--67},
  year={2025},
  publisher={Nature Publishing Group UK London}
}

@article{li2019triply,
  title={Triply magic conditions for microwave transition of optically trapped alkali-metal atoms},
  author={Li, Gang and Tian, Yali and Wu, Wei and Li, Shaokang and Li, Xiangyan and Liu, Yanxin and Zhang, Pengfei and Zhang, Tiancai},
  journal={Physical Review Letters},
  volume={123},
  number={25},
  pages={253602},
  year={2019},
  publisher={APS}
}

@inproceedings{wu2026pulsedsl,
  title={A Pulse-Level {DSL} for Real-Time Quantum Control with Hardware Compilation and Emulation},
  author={Wu, Yu-Hsuan and Shi, Yue and Liu, Junyi and Peng, Yuxiang},
  booktitle={Workshop on Programming Languages for Quantum Computing (PLanQC), POPL},
  year={2026}
}

@misc{quantumtaiwan2026atoms,
  author={Chen, Ying-Cheng},
  title={Quantum Computing and Simulation with Neutral Atoms},
  howpublished={Quantum Taiwan, CADCH module},
  url={https://www.quantumtaiwan.org/en/modules/cadch/atoms},
  note={Accessed 7 September 2026}
}
\iftechrep
\clearpage
\appendix
\providecommand{\owed}[1]{{\color{red}\textbf{[owed: #1]}}}
\providecommand{\NSRctx}[2]{\operatorname{NSR}^{#1}_{\ell}[#2]}
\providecommand{\PSRctx}[2]{\operatorname{PSR}^{#1}_{\ell}[#2]}
\providecommand{\EDctx}[2]{\mathcal S^{#1}_{\ell}[#2]}

\section{The Expression Language}
\label{app:expr}

The theory of \sec{theory} requires only that the
classical expression language $\mathcal E_\Theta$ be closed under
partial differentiation. The instantiation used by our implementation
is generated by
\begin{align*}
  e ::= {}& c \mid \theta_\ell \mid \xi \mid t
  \mid -e \mid e_1+e_2 \mid e_1e_2 \mid e^{a}\\
  &\mid \exp(e) \mid \ln(e) \mid \sin(e) \mid \cos(e) \mid 1/e ,
\end{align*}
where $c$ ranges over real constants, $a$ over fixed real exponents,
$\xi$ over sampled variables in scope, and $t$ is admitted only inside
an $\mathsf{ev}$. Domains are restricted where the mathematical
functions require it ($\ln$, $1/e$, non-integer powers), and
well-formedness checks the domain conditions at each occurrence.
$\mathcal E_0$ is the sub-language not mentioning $\Theta$. Closure
under $\partial/\partial\theta_\ell$ holds by inspection of the generation
rules; differentiability is guaranteed on the checked domains.

\paragraph{How $\mathcal E_\Theta$ is realized.} The implementation
carries expressions symbolically through the frontend and
differentiates them there, before any compilation. Two kinds of
symbol coexist in a Hamiltonian term's coefficient: machine
variables, which the borrowed solver settles when it matches the
target's Pauli terms against instructions, and the program's
symbols $\theta_\ell$, $\xi$, and $t$ of the grammar above. Only
the second
kind is differentiated, and $\partial/\partial\theta_\ell$ on a
segment is term-wise: it returns the coefficient derivatives
$\partial u_j/\partial\theta_\ell$ that weight every branch of
Sec.~\ref{sec:stdform}'s normal form, with the operators $H_j$
untouched.

The two kinds are separated by binding, and this is what keeps the
change to the compiler small. Substituting values for
$\theta_\ell$ and for the sampled $\xi$ in scope turns a segment
back into an ordinary time-independent Hamiltonian over machine
symbols alone, so a branch---once its $\mathsf{let}$ prefix is
drawn and its shift chosen---is an ordinary compile target and the
solver never learns that differentiation happened. Time is the
exception the grammar already marks: $t$ is admitted only inside an
$\mathsf{ev}$, and it is not bound at compile time. A segment whose
coefficients still mention $t$ is factored as
$H(t)=\sum_kf_k(t)\,H_k$ with each $H_k$ time-independent, so each
$H_k$ compiles once through the ordinary path while its envelope
$f_k$ becomes a waveform on the machine's own knobs
(Appendix~\ref{app:machine}); $\partial f_k/\partial\theta_\ell$
remains a function of $t$ and is sampled with $\tau$. The
factorization is legal only when the groups act on disjoint sets of
sites in their multi-qubit terms, which is checked rather than
assumed: when two envelopes' interaction terms share a site, no
single knob realizes both, and the segment is Trotterized instead.

\section{Proofs}
\label{app:proofs}

Throughout, valuations of $\Theta$ and of free sampled variables are
suppressed; $H\equiv\sum_{j=1}^k u_j\cdot A_j$ is the normal form of
the Hamiltonian description; $U_H(b,a)$ is the time-ordered
propagator of Figure~\ref{fig:language-semantics}; units are
$\hbar=1$.

\subsection{Representation and Evolution Lemmas}

\begin{lemma}[Hermitian observable representation]\label{lem:herm}
  For every well-formed program $P$ there exists a Hermitian operator
  $O_P$ such that, for every input state $\ket{\psi}$,
  $\llbracket P\rrbracket(\ket{\psi})=\bra{\psi}O_P\ket{\psi}$.
\end{lemma}
\begin{proof}
  By structural induction on $P$.

  For $P=\mathsf{meas}(m\mapsto f_m)$, define
  $O_P=\sum_m f_m\ket{m}\bra{m}$; since every $f_m$ is real-valued,
  $O_P$ is Hermitian with the required semantics.

  For $P=\sum_{i\in I}w_iP_i$, with each $P_i$ represented by
  $O_{P_i}$, define $O_P=\sum_i w_iO_{P_i}$; real weights preserve
  Hermiticity, and the family semantics gives the equality.

  For $P=\mathsf{let}\ \xi\sim\mu\ \mathsf{in}\ Q$, with
  $Q[\xi\mapsto v]$ represented pointwise by $O_Q(v)$, define
  $O_P=\int O_Q(v)\,d\mu(v)$; the integral is Hermitian by linearity of
  Hermitian conjugation.

  For $P=\mathsf{ev}(\tau_1,\tau_2,H);Q$, with $Q$
  represented by $O_Q$ and $U=U_H(\tau_2,\tau_1)$,
  $\llbracket P\rrbracket(\ket{\psi})=\bra{\psi}U^\dagger O_QU\ket{\psi}$,
  so $O_P=U^\dagger O_QU$ is Hermitian.
\end{proof}

\begin{lemma}[Derivative of continuous-time evolution]\label{lem:duhamel}
  Let $H_\lambda(t)$ depend differentiably on a real parameter
  $\lambda$ and let $U_\lambda(b,a)$ be its time-ordered propagator.
  Then
  \begin{equation}
    \frac{\partial U_\lambda(b,a)}{\partial\lambda}
    =-i\int_a^b U_\lambda(b,\tau)\,
    \frac{\partial H_\lambda(\tau)}{\partial\lambda}\,
    U_\lambda(\tau,a)\,d\tau .
  \end{equation}
\end{lemma}
\begin{proof}
  The propagator satisfies
  $i\,\frac{\partial}{\partial b}U_\lambda(b,a)=H_\lambda(b)U_\lambda(b,a)$ with
  $U_\lambda(a,a)=I$. Differentiating in $\lambda$,
  \[
    \begin{gathered}
      i\,\frac{\partial}{\partial b}
      \Big(\frac{\partial U_\lambda}{\partial\lambda}\Big)
      =\frac{\partial H_\lambda}{\partial\lambda}\,U_\lambda
      +H_\lambda\,\frac{\partial U_\lambda}{\partial\lambda},\\
      \frac{\partial U_\lambda}{\partial\lambda}(a,a)=0,
    \end{gathered}
  \]
  and variation of constants yields the stated identity. Two instances
  are used below: $\lambda=\theta_\ell$, where by the normal form
  $\partial H/\partial\theta_\ell=\sum_j(\partial
  u_j/\partial\theta_\ell)A_j=\dot H_\ell$, the
  tangent Hamiltonian of the main text; and $\lambda=s$ for the shifted
  family $H+s\dot H_\ell$, where $\partial H/\partial s=\dot H_\ell$ as well.
\end{proof}

\subsection{Differentiability and Domination}

\begin{lemma}[Smoothness]\label{lem:smooth}
  Every expression of Appendix~\ref{app:expr} is $C^\infty$ in
  $\Theta$ on its checked domain, and for every program $P$ the map
  $\Theta\mapsto\llbracket P\rrbracket(\ket{\psi})$ is
  $C^1$ wherever the sampling exchanges of
  Assumption~\ref{as:diff} hold. Programs with no sampling
  construct---every source program of Sec.~\ref{sec:eval}---are
  $C^\infty$ by the propagator argument alone, which is the
  regularity the finite-difference expansions of
  Appendix~\ref{app:fd} consume.
\end{lemma}
\begin{proof}
  Induction on the expression grammar: the generators are
  $C^\infty$ on the checked domains and the formation rules preserve
  smoothness. For the propagator, $t\mapsto H(t)$ is bounded,
  piecewise continuous, and $C^\infty$ in $\Theta$, so smooth
  parameter dependence of linear ODEs in finite dimension makes
  $\Theta\mapsto U_H(\tau_2,\tau_1)$ smooth; the
  $\mathsf{meas}$ and family cases are finite linear combinations
  with smooth coefficients, and the $\mathsf{let}$ case is the
  exchanged integral, which Assumption~\ref{as:diff} makes $C^1$.
\end{proof}

\begin{lemma}[Domination certificates]\label{lem:dominate}
  At a construct $\mathsf{let}\ \xi\sim\mu\ \mathsf{in}\ P$,
  the exchange required by Assumption~\ref{as:diff} holds when:
  (i)~$\mu$ has finite support (unconditionally: the integral is a
  finite sum);
  (ii)~$\mu$ is countable with masses $p_v$ and
  \[
    \sum_v p_v\,\sup_{\theta\in N}
    \left|\frac{\partial}{\partial\theta_\ell}\llbracket
    P[\xi\mapsto v]\rrbracket(\ket{\psi})\right|<\infty
  \]
  for a neighborhood $N$ of the parameter point;
  (iii)~$\mu$ is continuous with compact support and
  \[
    (v,\theta)\mapsto\frac{\partial}{\partial\theta_\ell}
    \llbracket P[\xi\mapsto v]\rrbracket(\ket{\psi})
  \]
  is continuous, so the supremum over
  $\operatorname{supp}\mu\times\overline N$ is a finite constant
  dominator.
\end{lemma}
\begin{proof}
  (i) is immediate; (ii) and (iii) are dominated convergence with
  the displayed dominators.
\end{proof}

\paragraph{Verification for this paper.}
Every source program differentiated in Sec.~\ref{sec:eval} either
contains no sampling construct or draws from finite-support
distributions, so case~(i) applies and
Assumption~\ref{as:diff} holds unconditionally. The countable and
continuous samplers introduced by the transformation itself---$\eta$
over the shift index $\kappa$, the uniform insertion time---never
require the exchange: they occur only in \emph{output} programs,
and output programs are not differentiated again.

\subsection{Nyquist Analysis}
\label{app:nyquist}

Fix an evolution constructor $E=\mathsf{ev}(\tau_1,\tau_2,H)$ with
continuation $P$, so the segment is $E;P$; an input state
$\ket{\psi}$; and a parameter $\theta_\ell$ with tangent
$\dot H_\ell\not\equiv 0$ on the segment. Let $O_P$ represent the
continuation (Lemma~\ref{lem:herm}) and
$R=\|O_P\|_{\mathrm{op}}$. For $s\in\mathbb R$ let $U_s$ denote the
propagator of $H+s\dot H_\ell$ on
$[\tau_1,\tau_2]$ and define
\begin{equation}
  F(s)=\bra{\psi}\,U_s^\dagger\,O_P\,U_s\,\ket{\psi},
\end{equation}
so $F(0)$ is the unshifted segment value and $|F(s)|\le R$ by
unitarity. Write $\partial_\ell^{\mathrm{ev}}$ for the contribution
to $\partial\llbracket E;P\rrbracket(\ket{\psi})/\partial\theta_\ell$
obtained by differentiating the evolution while holding the
continuation fixed.

\begin{lemma}[Tangent correctness]\label{lem:tangent}
  $F'(0)=\partial_\ell^{\mathrm{ev}}$.
\end{lemma}
\begin{proof}
  Both derivatives are first-order variations of the same propagator
  along the same direction. By Lemma~\ref{lem:duhamel} with
  $\lambda=s$ at $s=0$, and with $\lambda=\theta_\ell$ at fixed
  continuation, respectively,
  \[
    \begin{aligned}
      \frac{\partial U_s}{\partial s}\Big|_{s=0}
      &=\frac{\partial U}{\partial\theta_\ell}\\
      &=-i\int_{\tau_1}^{\tau_2}
      U_0(\tau_2,\tau)\,\dot H_\ell(\tau)\,U_0(\tau,\tau_1)\,d\tau .
    \end{aligned}
  \]
  Combining each with its conjugate term gives, for both quantities,
  the identical commutator-insertion integral
  \[
    \int_{\tau_1}^{\tau_2}
    \bra{\phi_\tau}\,i[\dot H_\ell(\tau),O_\tau]\,\ket{\phi_\tau}\,d\tau,
  \]
  with $\ket{\phi_\tau}=U_0(\tau,\tau_1)\ket{\psi}$ and
  $O_\tau=U_0(\tau_2,\tau)^\dagger O_PU_0(\tau_2,\tau)$.
\end{proof}

\begin{lemma}[Entire extension of exponential type]\label{lem:type}
  $F$ extends to an entire function
  $\mathcal F:\mathbb C\to\mathbb C$ with
  \begin{equation}
    |\mathcal F(x+iy)|\le R\,e^{\Omega_\ell|y|}
    \qquad\text{for all }x,y\in\mathbb R,
  \end{equation}
  where
  $\Omega_\ell=\int_{\tau_1}^{\tau_2}
  \operatorname{diam}(\dot H_\ell(t))\,dt$. No commutativity between
  $H(t)$, $\dot H_\ell(t)$, or their values at distinct times is used.
\end{lemma}
\begin{proof}
  \emph{Centering.} Let $c(t)$ be the midpoint of the spectrum of
  $\dot H_\ell(t)$ and $\dot H^{c}_\ell=\dot H_\ell-c\,I$, so
  $\|\dot H^{c}_\ell(t)\|_{\mathrm{op}}
  =\tfrac12\operatorname{diam}(\dot H_\ell(t))$. Replacing
  $\dot H_\ell$ by $\dot H^{c}_\ell$ multiplies $U_s$ by the global
  phase $e^{+is\int c\,dt}$, which cancels in
  $U_s^\dagger(\cdot)U_s$; hence $F$ is unchanged and we may assume
  the tangent centered.

  \emph{Complex shift.} For $z\in\mathbb C$ define $U_z$ by
  $i\,\dot U_z=(H+z\dot H^{c}_\ell)U_z$,
  $U_z(\tau_1)=I$; the Dyson series converges locally uniformly
  in $z$, so $z\mapsto U_z(\tau_2)$ is entire. For any solution
  $\psi_t$ and $z=x+iy$,
  \[
    \begin{aligned}
      \frac{d}{dt}\|\psi_t\|^2
      &=2\,\mathrm{Im}\,\bra{\psi_t}(H+z\dot H^c_\ell)\ket{\psi_t}\\
      &=2y\,\bra{\psi_t}\dot H^c_\ell\ket{\psi_t}\\
      &\le2|y|\,\|\dot H^c_\ell(t)\|\,\|\psi_t\|^2,
    \end{aligned}
  \]
  so by Gr\"onwall
  $\|U_z(\tau_2)\|\le
  \exp\big(|y|\int\|\dot H^{c}_\ell\|\,dt\big)
  \le e^{\Omega_\ell|y|/2}$.

  \emph{Bilinear extension.} Define
  \[
    \mathcal F(z)=\bra{\psi}\,U_{\bar z}(\tau_2)^\dagger\,
    O_P\,U_z(\tau_2)\,\ket{\psi}.
  \]
  Since $z\mapsto U_z$ is
  entire, so is $z\mapsto U_{\bar z}^\dagger$ (the adjoint of an
  antiholomorphic family), hence $\mathcal F$ is entire; it agrees
  with $F$ on the real axis, and
  $|\mathcal F(x+iy)|\le
  R\,\|U_{\bar z}\|\,\|U_z\|\le R\,e^{\Omega_\ell|y|}$.
  Time ordering enters only through the defining ODE; no commutation
  relations are invoked.
\end{proof}

\begin{lemma}[Band limitation]\label{lem:band}
  A function bounded on $\mathbb R$ that extends to an entire function
  of exponential type $\Omega$ in the imaginary direction, as in
  Lemma~\ref{lem:type}, has distributional Fourier transform supported
  in $[-\Omega,\Omega]$.
\end{lemma}
\begin{proof}
  Boundedness on $\mathbb R$ makes $F$ a tempered distribution, and
  the growth bound makes $\mathcal F$ of exponential type $\Omega$;
  the Paley--Wiener--Schwartz theorem then localizes
  $\operatorname{supp}\widehat F$ to $[-\Omega,\Omega]$.
\end{proof}

\begin{lemma}[Exact Nyquist differentiator]\label{lem:diff}
  Let $f:\mathbb R\to\mathbb R$ be continuous and bounded, $|f|\le C$, with
  $\operatorname{supp}\widehat f\subseteq[-\Omega,\Omega]$, and let
  $\overline\Omega\ge\Omega$, $\overline\Omega>0$,
  $h_\kappa=(2\kappa+1)\pi/(2\overline\Omega)$. Then
  \begin{equation}
    f'(0)=\frac{4\overline\Omega}{\pi^2}
    \sum_{\kappa=0}^{\infty}\frac{(-1)^\kappa}{(2\kappa+1)^2}
    \big[f(h_\kappa)-f(-h_\kappa)\big],
    \label{eq:diff-series}
  \end{equation}
  the series converging absolutely
  ($\le 2C\sum(2\kappa+1)^{-2}<\infty$ termwise).
\end{lemma}
\begin{proof}
  \emph{Core case.} Suppose first
  $f(s)=\int e^{i\omega s}\,d\mu(\omega)$ for a finite complex measure
  $\mu$ on $[-\overline\Omega,\overline\Omega]$. Then
  $f(h_\kappa)-f(-h_\kappa)=\int 2i\sin(\omega h_\kappa)\,d\mu$, and by absolute
  convergence the sum and integral interchange:
  \[
    \begin{aligned}
      &\text{RHS of \eqref{eq:diff-series}}\\
      &\quad=\int\frac{8i\overline\Omega}{\pi^2}
      \sum_{\kappa\ge0}\frac{(-1)^\kappa}{(2\kappa+1)^2}
      \sin\!\Big(\frac{(2\kappa+1)\pi\omega}{2\overline\Omega}\Big)\,d\mu .
    \end{aligned}
  \]
  The classical triangle-wave Fourier series gives
  \[
    \sum_{\kappa\ge0}\frac{(-1)^\kappa\sin((2\kappa+1)u)}{(2\kappa+1)^2}
    =\frac{\pi}{4}u
  \]
  for
  $|u|\le\pi/2$ (the endpoint included by continuity); with
  $u=\pi\omega/(2\overline\Omega)$ and $|\omega|\le\overline\Omega$,
  the integrand becomes $i\omega$, so the right-hand side equals
  $\int i\omega\,d\mu=f'(0)$.

  \emph{Strict oversampling.} If $\Omega<\overline\Omega$, let
  $g_\nu(s)=f(s)\,\mathrm{sinc}(\nu s)$ with
  $\mathrm{sinc}(x)=\sin(x)/x$ and
  $0<\nu<\overline\Omega-\Omega$. Then $g_\nu\in
  L^2(\mathbb R)$ with $\widehat g_\nu\in L^2$ supported in
  $[-\Omega-\nu,\Omega+\nu]
  \subset[-\overline\Omega,\overline\Omega]$; compact support puts
  $\widehat g_\nu\in L^1$, so the core case applies to
  $g_\nu$ on the same grid. As $\nu\to0$:
  $g_\nu(\pm h_\kappa)\to f(\pm h_\kappa)$ termwise, dominated by
  $C\sum(2\kappa+1)^{-2}$; and
  $g_\nu'(0)=f'(0)\,\mathrm{sinc}(0)+f(0)\,\mathrm{sinc}'(0)
  =f'(0)$ since $\mathrm{sinc}$ is even with value $1$ at the origin.

  \emph{Endpoint.} If $\Omega=\overline\Omega$, apply the previous
  case to the dilation $f_\varrho(s)=f(\varrho s)$, $\varrho<1$,
  which has type $\varrho\overline\Omega<\overline\Omega$ on the
  same grid:
  \[
    \varrho f'(0)=\frac{4\overline\Omega}{\pi^2}
    \sum\frac{(-1)^\kappa}{(2\kappa+1)^2}
    [f(\varrho h_\kappa)-f(-\varrho h_\kappa)],
  \]
  and let
  $\varrho\to1$ using termwise convergence dominated by
  $2C\sum(2\kappa+1)^{-2}$.
\end{proof}

\begin{proof}[Proof of Lemma~\ref{lem:nyquist-core}]
  By the evolution semantics, the main text's $g$ coincides with $F$
  above. Lemmas~\ref{lem:type} and~\ref{lem:band} make $F$ bounded
  by $R$ and band-limited to $\Omega_\ell\le\overline\Omega_\ell$;
  Lemma~\ref{lem:diff} with $\overline\Omega=\overline\Omega_\ell$
  yields the series, and Lemma~\ref{lem:tangent} identifies
  $F'(0)=\partial_\ell^{\mathrm{ev}}$.
\end{proof}

\begin{remark}[Certificates]\label{rem:app-cert}
  Lemma~\ref{lem:diff} requires only
  $\overline\Omega_\ell\ge\Omega_\ell$: every upper bound is
  sound, so certification is a static analysis with one-sided error.
  The certificate enters cost alone: $C_{\mathrm{NSR}}
  =\overline\Omega_\ell R$ and $h_\kappa\propto
  \overline\Omega_\ell^{-1}$, so a loose bound inflates the shot
  budget quadratically ($N\propto C_{\mathrm{NSR}}^2$) and compresses
  the shifts, but never biases.
  Appendix~\ref{app:select} measures how much of the certificate
  the diameter reference recovers on the evaluated programs (the
    recovered fraction
    $\Omega_\ell/\overline\Omega^{\,\mathrm{AC}}_\ell$, from $1$ at
  small $q$ to $0.475$ at $q=35$).
\end{remark}

\begin{proof}[Proof of Theorem~\ref{thm:shift-local}]
  If $\dot H_\ell\equiv0$ on the segment, both sides are zero
  (the empty family, and $\partial_\ell^{\mathrm{ev}}=0$ by
  Lemma~\ref{lem:tangent} with $F$ constant). Otherwise each branch of
  $\NSRctx{E}{P}$ is a well-formed program---its
  coefficients $u_j+\sigma h_\kappa\,\partial
  u_j/\partial\theta_\ell$ and weight
  $\overline\Omega_\ell(-1)^\kappa\sigma$ lie in $\mathcal E_\Theta$---whose
  denotation on $\ket{\psi}$ is $F(\sigma h_\kappa)$. By the
  $\mathsf{let}$ semantics applied twice and the singleton family rule,
  \[
    \begin{aligned}
      &\llbracket\NSRctx{E}{P}\rrbracket(\ket{\psi})\\
      &\quad=\mathbb E_{\kappa\sim\eta}\,\mathbb E_\sigma
      [\overline\Omega_\ell(-1)^\kappa\sigma\,F(\sigma h_\kappa)],
    \end{aligned}
  \]
  the outer expectation converging absolutely since
  $|F|\le R$ and $\sum_\kappa\eta_\kappa\overline\Omega_\ell R<\infty$.
  Conditioning on $\kappa$,
  $\mathbb E_\sigma[\sigma F(\sigma h_\kappa)]
  =\tfrac12[F(h_\kappa)-F(-h_\kappa)]$, and with
  $\eta_\kappa=\tfrac{8}{\pi^2}(2\kappa+1)^{-2}$,
  \[
    \begin{aligned}
      &\llbracket\NSRctx{E}{P}\rrbracket(\ket{\psi})\\
      &\quad=\frac{4\overline\Omega_\ell}{\pi^2}
      \sum_{\kappa\ge0}\frac{(-1)^\kappa}{(2\kappa+1)^2}
      [F(h_\kappa)-F(-h_\kappa)]\\
      &\quad=F'(0),
    \end{aligned}
  \]
  the last step by Lemma~\ref{lem:diff}, applicable because $F$ is
  bounded by $R$ and band-limited to
  $\Omega_\ell\le\overline\Omega_\ell$
  (Lemmas~\ref{lem:type} and~\ref{lem:band}). Finally
  $F'(0)=\partial_\ell^{\mathrm{ev}}$ by Lemma~\ref{lem:tangent}.
\end{proof}

\subsection{The Parameter-Shift Identity}

\begin{lemma}[Signed parameter-shift identity]\label{lem:kickid}
  Let $A$ be Hermitian with $A^2=I$ and, for $s\in\{-1,+1\}$, let
  $V_{s}=e^{-is\pi A/4}$. Define the superoperator
  \begin{equation}
    \mathcal K_A[\,\cdot\,]:=\sum_{s=\pm1}s\,V_s\,(\cdot)\,V_s^\dagger .
  \end{equation}
  Then
  \begin{equation}
    \mathcal K_A[\rho]=-i[A,\rho]
    \qquad\text{for every operator }\rho,
    \label{eq:kickid-state}
  \end{equation}
  and equivalently, for every Hermitian $O$, in the adjoint
  (Heisenberg) direction,
  \begin{equation}
    \sum_{s=\pm1}s\,V_s^\dagger\,O\,V_s=i[A,O].
    \label{eq:kickid-obs}
  \end{equation}
\end{lemma}
\begin{proof}
  Since $A^2=I$, $V_s=\tfrac{1}{\sqrt2}(I-isA)$. Then
  \[
    \begin{aligned}
      V_s\,\rho\,V_s^\dagger
      &=\tfrac12(I-isA)\rho(I+isA)\\
      &=\tfrac12\big(\rho+A\rho A-is[A,\rho]\big),
    \end{aligned}
  \]
  and the signed sum over $s=\pm1$ cancels the first two terms, giving
  \eqref{eq:kickid-state}. Exchanging the roles of $V_s$ and
  $V_s^\dagger$ gives \eqref{eq:kickid-obs} by the same computation.
  The superoperator form \eqref{eq:kickid-state} is the one reused in
  the noise analysis, where $\mathcal K_A$ acts on density operators
  inside a Lindblad evolution; the ideal-semantics proofs below use the
  adjoint form \eqref{eq:kickid-obs}.
\end{proof}

\subsection{Soundness Cases}

\begin{theorem}[Measurement differentiation]\label{thm:app-meas}
  For every input state $\ket{\psi}$,
  \[
    \begin{aligned}
      &\llbracket\mathcal D^{\mathcal S}_\ell(\mathsf{meas}(m\mapsto
      f_m))\rrbracket
      \ket{\psi}\\
      &\quad=\frac{\partial}{\partial\theta_\ell}
      \llbracket\mathsf{meas}(m\mapsto f_m)\rrbracket(\ket{\psi}).
    \end{aligned}
  \]
\end{theorem}
\begin{proof}
  By the measurement semantics,
  \begin{align*}
    &\frac{\partial}{\partial\theta_\ell}\sum_m f_m|\braket{m}{\psi}|^2\\
    &\quad=\sum_m(\partial f_m/\partial\theta_\ell)|\braket{m}{\psi}|^2\\
    &\quad=\llbracket\mathsf{meas}(m\mapsto\partial
    f_m/\partial\theta_\ell)\rrbracket
    \ket{\psi}.\qedhere
  \end{align*}
\end{proof}

\begin{theorem}[Differentiation through classical sampling]
  \label{thm:app-let}
  Let $F(v)=\llbracket P[\xi\mapsto v]\rrbracket(\ket{\psi})$, regarded as a
  function of $v$ and $\Theta$. Assume the construct satisfies
  Assumption~\ref{as:diff}: in a
  neighborhood of the parameter point, $\partial F/\partial\theta_\ell$
  exists and admits a $\mu$-integrable dominating function $h$ with
  $|\partial F/\partial\theta_\ell(v)|\le h(v)$ for all $v$. If
  $\mathcal D^{\mathcal S}_\ell P$ computes the differential semantics of
  $P$, then $\mathsf{let}\ \xi\sim\mu\ \mathsf{in}\ \mathcal
  D^{\mathcal S}_\ell P$
  computes the differential semantics of
  $\mathsf{let}\ \xi\sim\mu\ \mathsf{in}\ P$.
\end{theorem}
\begin{proof}
  Because $\mu$ is parameter-free, dominated convergence permits
  differentiation under the integral sign:
  \[
    \begin{aligned}
      \frac{\partial}{\partial\theta_\ell}\!\int F(v)\,d\mu(v)
      &=\int\frac{\partial F}{\partial\theta_\ell}(v)\,d\mu(v)\\
      &=\int\llbracket\mathcal D^{\mathcal S}_\ell P[\xi\mapsto v]\rrbracket
      (\ket{\psi})\,d\mu(v),
    \end{aligned}
  \]
  which is the denotation of
  $\mathsf{let}\ \xi\sim\mu\ \mathsf{in}\ \mathcal D^{\mathcal S}_\ell P$.
\end{proof}

\begin{theorem}[Family differentiation]\label{thm:app-family}
  If $\mathcal D^{\mathcal S}_\ell P_i$ computes the differential semantics
  of $P_i$ for every $i\in I$, then
  $\mathcal D^{\mathcal S}_\ell\big(\sum_{i}w_iP_i\big)$ computes the
  differential semantics of $\sum_i w_iP_i$.
\end{theorem}
\begin{proof}
  By the product rule and the family semantics,
  \[
    \begin{aligned}
      &\frac{\partial}{\partial\theta_\ell}\sum_iw_i\llbracket
      P_i\rrbracket(\ket{\psi})\\
      &\quad=\sum_i(\partial w_i/\partial\theta_\ell)\llbracket
      P_i\rrbracket(\ket{\psi})\\
      &\qquad{}+\sum_iw_i\llbracket\mathcal D^{\mathcal S}_\ell
      P_i\rrbracket(\ket{\psi}),
    \end{aligned}
  \]
  which is the denotation of the two-family sum in the transformation
  rule; the sum of two finite families flattens to a finite family.
\end{proof}

\begin{proof}[Proof of Lemma~\ref{lem:psr-id}]
  Write $v_j=\partial u_j/\partial\theta_\ell$,
  $V_{j,s}=e^{-is\pi A_j/4}$ for the denotation of the inserted
  segment $\mathsf{ev}(0,\pi/4,s\cdot A_j)$, and
  \[
    \begin{aligned}
      B_{j,s}(\tau)={}&\mathsf{ev}(\tau_1,\tau,H);\\
      &\mathsf{ev}\big(0,\tfrac\pi4,s\cdot A_j\big);\,
      \mathsf{ev}(\tau,\tau_2,H);P
    \end{aligned}
  \]
  for the branch of the main text, with $O_P$, $\ket{\phi_\tau}$,
  and $O_\tau$ as in Appendix~\ref{app:nyquist}.

  \emph{Commutator insertion.} Lemma~\ref{lem:tangent}'s proof
  expresses $\partial_\ell^{\mathrm{ev}}$ as the
  commutator-insertion integral of $\dot H_\ell$; substituting the
  normal form $\dot H_\ell=\sum_jv_jA_j$ and using linearity,
  \begin{equation}
    \begin{aligned}
      &\partial_\ell^{\mathrm{ev}}\llbracket E;P\rrbracket(\ket{\psi})\\
      &\quad=\int_{\tau_1}^{\tau_2}\sum_{j:\,v_j\not\equiv0}
      v_j(\tau)\,\bra{\phi_\tau}i[A_j,O_\tau]\ket{\phi_\tau}\,d\tau,
    \end{aligned}
    \label{eq:psr-comm}
  \end{equation}
  the sum running over the terms touched by $\theta_\ell$.

  \emph{Kick identity.} For each touched $j$,
  Assumption~\ref{as:kick} gives $A_j^2=I$, and
  Lemma~\ref{lem:kickid}\,\eqref{eq:kickid-obs} yields
  \[
    \bra{\phi_\tau}i[A_j,O_\tau]\ket{\phi_\tau}
    =\sum_{s=\pm1}s\,
    \bra{\phi_\tau}V_{j,s}^\dagger O_\tau V_{j,s}\ket{\phi_\tau}.
  \]

  \emph{Branch identification.} Since the middle segment of
  $B_{j,s}(\tau)$ has propagator exactly $V_{j,s}$ and the outer
  segments reproduce $\ket{\phi_\tau}$ and $O_\tau$,
  $\bra{\phi_\tau}V_{j,s}^\dagger O_\tau V_{j,s}\ket{\phi_\tau}
  =\llbracket B_{j,s}(\tau)\rrbracket(\ket{\psi})$. Substituting
  into \eqref{eq:psr-comm} gives the lemma's display; the segments
  of $B_{j,s}(\tau)$ carry $\theta_\ell$ because $H$ does.
\end{proof}

\begin{proof}[Proof of Theorem~\ref{thm:psr-local}]
  Apply Lemma~\ref{lem:psr-id} and write the time integral as an
  expectation,
  \[
    \int_{\tau_1}^{\tau_2}g(\tau)\,d\tau
    =\Delta\tau\,\mathbb E_{\tau\sim\mathrm{Uniform}(\tau_1,\tau_2)}[g(\tau)]:
  \]
  the result is exactly the $\mathsf{let}$/family
  denotation of $\PSRctx{E}{P}$, whose branches are the
  $B_{j,s}(\tau)$ with weights $\Delta\tau\,v_j(\tau)\,s$.
\end{proof}

\begin{theorem}[Evolution case of the generic soundness theorem]
  \label{thm:app-ev}
  Let $\mathcal S_\ell$ be any locally sound evolution derivative
  (Definition~\ref{def:edp}), let $E=\mathsf{ev}(\tau_1,\tau_2,H)$,
  and suppose
  $\mathcal D^{\mathcal S}_\ell P$ computes the differential semantics of
  $P$. Then
  $E;\mathcal D^{\mathcal S}_\ell P
  +\EDctx{E}{P}$
  computes the differential semantics of $E;P$.
\end{theorem}
\begin{proof}
  With $O_P$ as in Lemma~\ref{lem:herm} and
  $U=U_H(\tau_2,\tau_1)$,
  \[
    \begin{aligned}
      &\frac{\partial}{\partial\theta_\ell}\bra{\psi}U^\dagger O_PU\ket{\psi}\\
      &\quad=\bra{\psi}(\partial U^\dagger/\partial\theta_\ell)O_PU\ket{\psi}\\
      &\qquad{}+\bra{\psi}U^\dagger(\partial
      O_P/\partial\theta_\ell)U\ket{\psi}\\
      &\qquad{}+\bra{\psi}U^\dagger O_P(\partial
      U/\partial\theta_\ell)\ket{\psi}.
    \end{aligned}
  \]
  The first and third terms are
  $\partial_\ell^{\mathrm{ev}}
  =\llbracket \EDctx{E}{P}\rrbracket(\ket{\psi})$ by local soundness. With
  $\ket{\phi}=U\ket{\psi}$, the induction hypothesis gives for the
  middle term
  \[
    \begin{aligned}
      \bra{\phi}(\partial O_P/\partial\theta_\ell)\ket{\phi}
      &=\tfrac{\partial}{\partial\theta_\ell}\llbracket
      P\rrbracket(\ket{\phi})\\
      &=\llbracket\mathcal D^{\mathcal S}_\ell P\rrbracket(\ket{\phi})\\
      &=\llbracket E;\mathcal D^{\mathcal S}_\ell P\rrbracket(\ket{\psi}).
    \end{aligned}
  \]
  The family semantics then gives exactly the denotation of the
  evolution rule.
\end{proof}

\begin{theorem}[Generic soundness; Theorem~\ref{thm:soundness} of
  the main text]
  For every locally sound evolution derivative $\mathcal S_\ell$ and
  under Assumption~\ref{as:diff},
  $\llbracket\mathcal D^{\mathcal S}_\ell P\rrbracket(\ket{\psi})
  =\tfrac{\partial}{\partial\theta_\ell}\llbracket P\rrbracket(\ket{\psi})$
  for every well-formed $P$ and input state.
\end{theorem}
\begin{proof}
  By structural induction on $P$. The $\mathsf{meas}$ case is
  Theorem~\ref{thm:app-meas}; the $\mathsf{let}$ case is
  Theorem~\ref{thm:app-let} with the induction hypothesis for the body;
  the family case is Theorem~\ref{thm:app-family} with the induction
  hypotheses for the branches; the $\mathsf{ev}$ case is
  Theorem~\ref{thm:app-ev} with the induction hypothesis for the
  continuation. These cases exhaust the syntax. Well-formedness of the
  output is condition~(i) of Definition~\ref{def:ad}, verified by
  construction in the proof sketch of Theorem~\ref{thm:soundness},
  and needs no separate proof. Instantiating $\mathcal S_\ell$ with the two
  locally sound evolution derivatives (Theorems~\ref{thm:shift-local}
  and~\ref{thm:psr-local}) yields the strategy-level soundness
  corollaries.
\end{proof}

\subsection{The Finite-Difference Transformation}
\label{app:fd}

Well-formedness of~\eqref{eq:fd-program}: the substitution
$\theta_\ell\mapsto\theta_\ell\pm\epsilon/2$ maps
$\mathcal E_\Theta$ into itself (the expression grammar contains
constants and addition), leaves $\mathcal E_0$, $\mu$, and the
parameter-free boundaries untouched, and the weights $\pm1/\epsilon$ are
constants; the result is a two-branch family of well-formed
programs. Its denotation is the central secant of
$f(\theta_\ell):=\llbracket P\rrbracket(\ket{\psi})$, and its error
has two sources, one in the program and one in the device.

\paragraph{Truncation.} For $f$ three times continuously
differentiable in $\theta_\ell$,
\begin{equation}
  \llbracket\operatorname{FD}^{\epsilon}_\ell(P)\rrbracket(\ket{\psi})
  =f'+\frac{\epsilon^2}{24}\,f'''+o(\epsilon^2),
  \label{eq:fd-trunc}
\end{equation}
by Taylor expansion of the two branches about the dialed point.
Where $f'''\ne0$ the bias is nonzero for all sufficiently small
$\epsilon>0$, so condition~(ii) of the soundness definition fails
in general. 

\paragraph{Setpoint.} The two programmed probe setpoints land at
$\theta\pm\epsilon/2+\delta_\pm$: each carries an offset
$\delta_\pm$, fixed over the shots taken there. Where the offsets
come from does not matter for what follows; what matters is that
FD divides by its step. Expanding each branch to first order in
its offset,
\begin{equation}
  \begin{aligned}
    &\llbracket\operatorname{FD}^{\epsilon}_\ell(P)\rrbracket(\ket{\psi})\\
    &\quad=f'+\frac{\epsilon^2}{24}\,f'''
    +\frac{\delta_+-\delta_-}{\epsilon}\,f'
    +\frac{\delta_++\delta_-}{2}\,f''\\
    &\qquad{}+o(\epsilon^2)+O\big(\epsilon\delta,\ \delta^2/\epsilon\big).
  \end{aligned}
  \label{eq:fd-setpoint}
\end{equation}
The fourth term displaces the evaluation point through $f''$. The
third is the trap: an offset at the scale of the
optimal $\epsilon$ is amplified to order one of the gradient, and
since $\delta_\pm$ are fixed over the estimate's shots (Table~\ref{tab:noise}: $\delta_\pm$ drawn independently,
zero mean, standard deviation $r$, per programmed setpoint), their
amplified contribution is a bias for each realized pair of probe
setpoints. Under the Sec.~\ref{sec:eval} model the realized offsets vary across
programmed setpoint draws, with mean effect only
$\mathbb E\llbracket\operatorname{FD}^{\epsilon}_\ell(P)\rrbracket
=f'+(\tfrac{\epsilon^2}{24}+\tfrac{r^2}{2})f'''+\cdots$ but
root-mean-square contribution
\begin{equation}
  S(\epsilon)=\frac{\sqrt2\,|f'|\,r}{\epsilon}.
  \label{eq:fd-fluct}
\end{equation}
For $f'\ne0$ and $f'''\ne0$, balancing $S$ against the leading
truncation bias in quadrature gives
\begin{equation}
  \begin{aligned}
    &\min_{\epsilon>0}\sqrt{\Big(\frac{\epsilon^2}{24}\,f'''\Big)^{\!2}
    +\frac{2f'^2r^2}{\epsilon^2}}\\
    &\quad=\frac{\sqrt3}{24^{1/3}}\,|f'''|^{1/3}\big(|f'|r\big)^{2/3}\\
    &\quad\approx0.60\,|f'''|^{1/3}\big(|f'|r\big)^{2/3},
  \end{aligned}
  \label{eq:fd-floor}
\end{equation}
attained at $\epsilon^\ast=(24\,|f'|\,r/|f'''|)^{1/3}$: the
minimum in the leading-order error model, independent of shot
count, and the analytic scaling behind the $\epsilon$ sweeps of
Figure~\ref{fig:F6}'s inset and Figure~\ref{fig:epssweep}. It is
a scaling, not a value: the truncation term is a Taylor estimate,
and at Figure~\ref{fig:F6}'s operating point, where
$\epsilon^\ast\overline\Omega_\ell\approx1.6$, it places
$\epsilon^\ast$ correctly ($0.217$ against the measured $0.21$)
while overestimating the floor itself ($0.23$ against $0.18$).

NSR and PSR program setpoints too, and carry offsets of the same
size---but they take no step, so nothing divides an offset. What
each does with it follows from when a draw is taken: under the
Sec.~\ref{sec:eval} model a draw occurs when a programmed value
changes and is held until it changes again. PSR's branches
program the source value, so they share one draw $\delta$: the
estimate is the exact derivative at the displaced point
$\theta+\delta$, an error $|f''|\,r$ at first order, never
amplified, and second order where $f''=0$ (mean $r^2f'''/2$,
root-mean-square $\tfrac{\sqrt3}{2}|f'''|r^2$). NSR re-dials
the coefficient on every execution, so every branch carries a
fresh draw: the draws average out with the shots, leaving only
the residual $r^2f'''/2$ of the smoothed landscape, which no
step amplifies. Figure~\ref{fig:epssweep} shows the three
exposures side by side on landscapes of different sharpness.

\begin{figure*}[t]
  \centering
  \includegraphics[width=\textwidth]{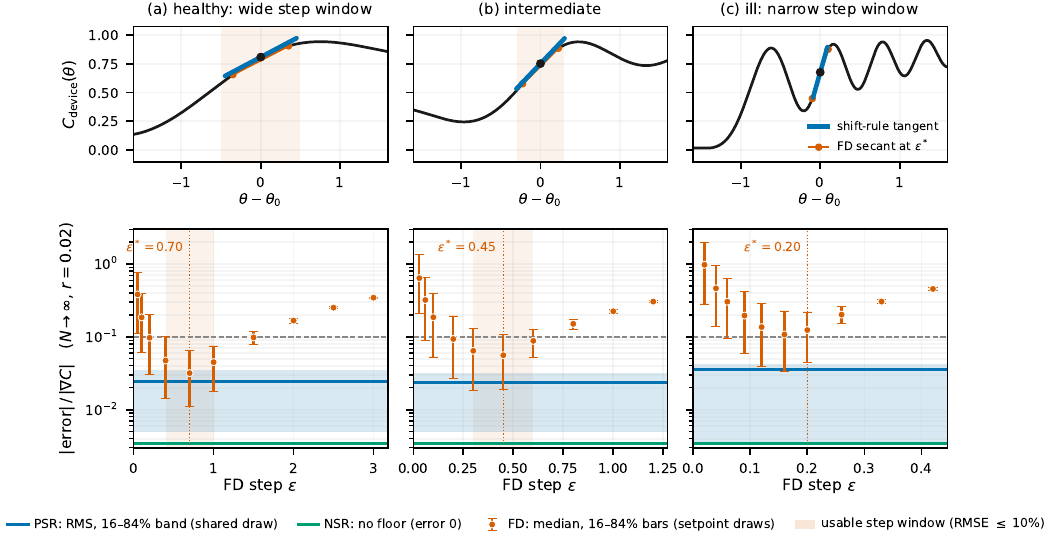}
  \caption{What a healthy and an ill landscape look like to a
    finite difference. Top: the device landscape
    $C_{\mathrm{device}}(\theta)$ around the operating point
    (black), the shift-rule tangent (blue), the finite-difference
    secant at its best step $\epsilon^\ast$ (orange), and the
    usable step window (shaded: RMSE at most $10\%$ of
    $|\nabla C|$). Bottom: the error of a single estimate as
    $N\to\infty$ with the setpoint error $r=0.02$---for FD the
    median and the 16--84\% range over setpoint draws at each
    step (orange bars); for PSR the root-mean-square and 16--84\%
    range of its one shared draw (blue); NSR has none. The three
    landscapes differ in sharpness ($\overline\Omega=2$, $4$,
    $14$); the operating points are generic, chosen so that PSR's
    displacement $|f''|\,r$ is the same ($2.5\%$ of
    $|\nabla C|$) in all three. Below $\epsilon\approx0.1$ the
    setpoint error is amplified by $1/\epsilon$ and the bars span
    a decade in every panel; above it the truncation bias takes
    over, slowly on the healthy landscape ($\epsilon=1$ is still
    at the $5\%$ level) and immediately on the ill one, where no
    step reaches $10\%$ and at the best step the setpoint draw
  alone decides whether the estimate is usable.}
  \label{fig:epssweep}
\end{figure*}

\begin{lemma}[Attenuation and displacement under dephasing]
  \label{lem:fd-noise}
  For the running example at constant drive, $H=uX$, under the
  dephasing dissipator $\Gamma(Z\rho Z-\rho)$ with
  $\Gamma<\omega_0:=2u$, write
  $\widetilde\omega=\sqrt{\omega_0^2-\Gamma^2}$ and
  $\varphi=\arctan(\Gamma/\widetilde\omega)$. The excited-state
  population from $\ket0$ is
  \begin{equation}
    \begin{aligned}
      P_1(T)&=\tfrac12\Big[1-e^{-\Gamma T}\big(\cos\widetilde\omega T
      +\tfrac\Gamma{\widetilde\omega}\sin\widetilde\omega T\big)\Big]\\
      &=\tfrac12\Big[1-\sqrt{1+\tfrac{\Gamma^2}{\widetilde\omega^2}}\,
      e^{-\Gamma T}\cos(\widetilde\omega T-\varphi)\Big].
    \end{aligned}
  \end{equation}
  The landscape's features attenuate with log-amplitude error
  $\Gamma T-\tfrac12\log(1+\Gamma^2/\widetilde\omega^2)$, growing
  with evolution time, while their positions displace by the phase
  $\varphi$; both $\widetilde\omega$ and $\varphi$ vary when $u$
  is scanned. Noise thus rescales and slightly translates the
  ideal landscape---closing the FD window from the
  $\epsilon_{\min}$ side earlier---but neither creates nor removes
  the trap mechanism, which is model-free.
\end{lemma}
\begin{proof}
  The Bloch equations for the stated generator reduce to a damped
  linear rotation in the $(y,z)$ plane; solving the $2\times2$
  system from $\ket{0}$ gives the first display, and the second is
  the harmonic-addition identity.
\end{proof}

\section{Noise: Channel Semantics}
\label{app:noise}

\subsection{Channel Semantics}

\begin{assumption}[Noise model]\label{as:noise}
  Each evolution constructor executes as the propagator
  $\Lambda_H(\tau_2,\tau_1)$ of the Lindblad master equation
  $\dot\rho=-i[H(t),\rho]+\mathcal D(t)\rho$: the Schr\"odinger
  equation plus a dissipation term $\mathcal D(t)$, which is
  bounded, measurable, and independent of $\Theta$, of the sampled
  variables, and in particular of the shifts. Segments compose as
  in Figure~\ref{fig:language-semantics}; $\mathsf{meas}$ is the
  Born rule---the measurement probabilities---on the evolved
  state. The channel denotation
  $\llbracket\cdot\rrbracket_{\mathrm{ch}}$ is the semantics with
  $U^\dagger(\cdot)U$ replaced by the adjoint propagator
  $\Lambda^\dagger$.
\end{assumption}
\noindent
The \emph{device landscape}
$C_{\mathrm{device}}:\Theta\mapsto\llbracket P\rrbracket_{\mathrm{ch}}
(\ket{\psi})$ is the objective the device realizes, and its
gradient $\nabla C_{\mathrm{device}}$ is what
Sec.~\ref{sec:eval} measures every estimator against;
unbiasedness below is with respect to it. ($\llbracket\cdot
  \rrbracket_{\mathrm{ch}}$ remains the semantics symbol;
  $\nabla C_{\mathrm{device}}$ is its gradient under a concrete
rate assignment, Appendix~\ref{app:noise-device}.)
Lemma~\ref{lem:herm} holds verbatim: the evolution case
sets $O_P=\Lambda^\dagger(O_Q)$, Hermitian with
$\|O_P\|\le\|O_Q\|$---the adjoint of a channel maps $I$ to $I$
and cannot increase operator norm---so every value and $R$-bound
carries over.

\begin{lemma}[Superoperator Duhamel]\label{lem:noise-duhamel}
  If $\mathcal L_\lambda(t)$ depends differentiably on $\lambda$
  with propagator $\Lambda_\lambda(b,a)$, then
  \[
    \begin{aligned}
      &\partial_\lambda\Lambda_\lambda(b,a)\\
      &\quad=\int_a^b\Lambda_\lambda(b,\tau)\,
      \partial_\lambda\mathcal L_\lambda(\tau)\,\Lambda_\lambda(\tau,a)\,d\tau.
    \end{aligned}
  \]
\end{lemma}
\begin{proof}
  The variation-of-constants computation of Lemma~\ref{lem:duhamel},
  at the superoperator level.
\end{proof}
\noindent
Tangent correctness transfers: with
$g(s)=\operatorname{Tr}[O_P\,\Lambda_s(\rho)]$ for the shifted
family $H+s\dot H_\ell$,
Lemma~\ref{lem:noise-duhamel} with
\[
  \partial_s\mathcal L=-i[\dot H_\ell,\cdot\,]=:\mathcal K(t)
\]
gives $g'(0)=\partial_\ell^{\mathrm{ev,ch}}$, exactly as in
Lemma~\ref{lem:tangent}; the dissipator contributes nothing because
it is independent of $s$ and $\Theta$.

\begin{lemma}[Analytic conjugation]\label{lem:noise-cplx}
  For Hermitian $A$ and $z\in\mathbb C$ let
  $\mathcal V_z(X)=e^{-izA}Xe^{+izA}$. Then $z\mapsto\mathcal V_z$
  is entire, $\mathcal V_s$ for real $s$ is unitary conjugation, and
  $\partial_z\mathcal V_z|_{z=0}=-i[A,\cdot\,]$.
\end{lemma}
\begin{proof}
  Both exponential factors are entire in $z$ and products of entire
  operator families are entire; the rest is the product rule.
\end{proof}

\begin{lemma}[Strip bound]\label{lem:noise-strip}
  If $A$ is centered, with spectrum in $[-d/2,d/2]$ and
  $d=\operatorname{diam}A$, then
  $\|\mathcal V_z\|_{1\to1}\le e^{|\operatorname{Im}z|\,d}$.
\end{lemma}
\begin{proof}
  Write $z=x+iy$: $\mathcal V_z=\mathcal V_x\circ
  \big(e^{yA}(\cdot)e^{-yA}\big)$; unitary conjugation is a
  trace-norm isometry and
  $\|e^{yA}Xe^{-yA}\|_1\le
  \|e^{yA}\|_{\mathrm{op}}\|e^{-yA}\|_{\mathrm{op}}\|X\|_1
  \le e^{2|y|d/2}\|X\|_1$.
\end{proof}

\begin{lemma}[Analytic propagator of the shifted generator]
  \label{lem:noise-trotter}
  Fix the segment and let
  \[
    \begin{gathered}
      \mathcal L_z(t)=\mathcal L_0(t)+z\,\mathcal K^c(t),\\
      \mathcal L_0=-i[H,\cdot\,]+\mathcal D,\qquad
      \mathcal K^c=-i[\dot H^c_\ell,\cdot\,],
    \end{gathered}
  \]
  with $\mathcal K^c$ the centered tangent. The propagator
  $\Lambda_z(\tau_2,\tau_1)$ of $\mathcal L_z$ is entire in $z$
  and satisfies
  \[
    \|\Lambda_z(\tau_2,\tau_1)\|_{1\to1}\le
    e^{\Omega_\ell|\operatorname{Im}z|}.
  \]
\end{lemma}
\begin{proof}
  All norms are $\|\cdot\|_{1\to1}$ (finite dimension;
  submultiplicative). \emph{Entirety.} On every compact
  $K\subset\mathbb C$,
  \[
    m_K=\sup_{t,\,z\in K}\|\mathcal L_z(t)\|<\infty,
  \]
  since $\mathcal D$ is bounded and measurable and $H$,
  $\dot H_\ell$ are bounded and piecewise continuous. The Dyson
  series
  \[
    \begin{aligned}
      &\Lambda_z(\tau_2,\tau_1)\\
      &\quad=\sum_{k\ge0}\int_{\tau_1<s_1<\cdots<s_k<\tau_2}
      \mathcal L_z(s_k)\cdots\mathcal L_z(s_1)\,ds
    \end{aligned}
  \]
  has $k$-th term a polynomial in $z$ of norm at most
  $[m_K(\tau_2-\tau_1)]^k/k!$, so it converges uniformly on $K$,
  and a uniform limit of polynomials is entire (Weierstrass).
  \emph{Strip bound.} Write
  $z=x+iy$. $\Lambda_x$ is the propagator of
  $\mathcal L_0+x\mathcal K^c$, a Lindblad generator with the
  Hermitian Hamiltonian $H+x\dot H^c_\ell$, hence a channel with
  $\|\Lambda_x\|=1$. Variation of constants gives
  \[
    \begin{aligned}
      \Lambda_z(t,\tau_1)={}&\Lambda_x(t,\tau_1)\\
      &{}+iy\int_{\tau_1}^t\Lambda_x(t,s)\,\mathcal K^c(s)\,
      \Lambda_z(s,\tau_1)\,ds,
    \end{aligned}
  \]
  and $\|\mathcal K^c(s)X\|_1\le2\|\dot H^c_\ell(s)\|_{\mathrm{op}}
  \|X\|_1=\operatorname{diam}(\dot H_\ell(s))\,\|X\|_1$, so
  \[
    \|\Lambda_z(t,\tau_1)\|\le1+|y|\int_{\tau_1}^{t}
    \operatorname{diam}(\dot H_\ell(s))\,\|\Lambda_z(s,\tau_1)\|\,ds,
  \]
  and Gr\"onwall's inequality yields
  \[
    \|\Lambda_z(\tau_2,\tau_1)\|\le
    \exp\Big(|y|\int_{\tau_1}^{\tau_2}
    \operatorname{diam}(\dot H_\ell)\,dt\Big)=e^{\Omega_\ell|y|}.
  \]
  No continuity of $\mathcal D$ beyond Assumption~\ref{as:noise}
  is required.
\end{proof}

\begin{lemma}[Band limitation, noisy]\label{lem:noise-band}
  $g$ extends to the entire function
  $g(z)=\operatorname{Tr}[O_P\,\Lambda_z(\rho)]$ with
  $|g(z)|\le R\,e^{\Omega_\ell|\operatorname{Im}z|}$ and
  $|g|\le R$ on $\mathbb R$; hence
  $\operatorname{supp}\widehat g\subseteq[-\Omega_\ell,
  \Omega_\ell]$ by Lemma~\ref{lem:band}.
\end{lemma}
\begin{proof}
  $|g(z)|\le\|O_P\|\,\|\Lambda_z(\rho)\|_1$ with
  Lemma~\ref{lem:noise-trotter}; on the real axis $\Lambda_s$
  preserves trace and positivity, so
  $\|\Lambda_s(\rho)\|_1=1$. Centering is free: conjugation by the scalar phase
  $e^{+is\int c\,dt}I$ is the identity superoperator, so replacing
  $\dot H_\ell$ by $\dot H^c_\ell$ does not change $\Lambda_s$
  at all in the channel picture.
\end{proof}

\subsection{Strategy Unbiasedness}

\begin{theorem}[NSR is exactly unbiased on the noisy device]
  \label{thm:noise-nsr}
  Under Assumptions~\ref{as:bandwidth} and~\ref{as:noise}, for
  every input state
  $\llbracket\NSRctx{E}{P}
  \rrbracket_{\mathrm{ch}}
  =\partial_\ell^{\mathrm{ev,ch}}
  \llbracket E;P\rrbracket_{\mathrm{ch}}$:
  the same program, the same certificate
  $\overline\Omega_\ell$, every branch executed under the
  device's own noise. Consequently
  $\mathcal D^{\operatorname{NSR}}_\ell$ is a sound
  differentiation transformation for
  $\nabla C_{\mathrm{device}}$.
\end{theorem}
\begin{proof}
  Branch denotations under
  $\llbracket\cdot\rrbracket_{\mathrm{ch}}$ are $g(\sigma
  h_\kappa)$: the branches are ordinary evolutions with shifted
  coefficients, and the noise, being shift-independent, acts on them
  identically. Lemma~\ref{lem:noise-band} bounds and band-limits
  $g$; Lemma~\ref{lem:diff} turns the sampled series into $g'(0)$;
  the superoperator Duhamel identifies $g'(0)$ with the noisy
  evolution contribution. The shared-rule case theorems are
  semantics-agnostic (linearity and dominated convergence) and
  assemble as in the generic soundness theorem.
\end{proof}

\begin{theorem}[PSR under noise, ideal insertion]
  \label{thm:noise-psr}
  Under Assumptions~\ref{as:kick} and~\ref{as:noise}, if each
  inserted segment executes as the ideal conjugation
  $V_{j,s}(\cdot)V_{j,s}^\dagger$, then
  $\llbracket\PSRctx{E}{P}
  \rrbracket_{\mathrm{ch}}
  =\partial_\ell^{\mathrm{ev,ch}}
  \llbracket E;P\rrbracket_{\mathrm{ch}}$.
\end{theorem}
\begin{proof}
  Time-local generators compose:
  \[
    \Lambda(\tau_2,\tau_1)=\Lambda(\tau_2,\tau)\Lambda(\tau,\tau_1)
  \] for every $\tau$, so divisibility is
  automatic and needs no hypothesis. The superoperator Duhamel
  inserts \[
    \partial_{\theta_\ell}\mathcal L(\tau)
    =\sum_j(\partial u_j/\partial\theta_\ell)(\tau)\,(-i[A_j,\cdot\,]);
  \] the kick identity in superoperator form,
  Lemma~\ref{lem:kickid}\,\eqref{eq:kickid-state}, replaces each
  $-i[A_j,\cdot\,]$ by the signed sum
  $\sum_s s\,V_{j,s}(\cdot)V_{j,s}^\dagger$; each term is then
  the channel denotation of the executable branch
  $B_{j,s}(\tau)$ with its ideal insertion at $\tau$, and the
  uniform-sampling identity finishes as in the proofs of
  Lemma~\ref{lem:psr-id} and Theorem~\ref{thm:psr-local}.
\end{proof}

\begin{lemma}[Insertion-error stability]\label{lem:ins-bias}
  Suppose each inserted segment executes within error
  $\epsilon_{\mathrm{ins}}$ of the ideal conjugation, in the
  sense that no experiment on the full register---the inserted
  pair together with anything entangled with it---can see a
  larger change in any outcome probability:
  \[
    \max_{j,s}\|\widetilde{\mathcal S}_{j,s}
    -V_{j,s}(\cdot)V_{j,s}^\dagger\|_\diamond\le
    \epsilon_{\mathrm{ins}}
  \]
  in the diamond
  norm~\cite{watrous2018theory}. Then the PSR estimator's bias
  obeys
  \begin{gather}
    \big|\mathbb E[\widetilde G]-
    \partial_\ell^{\mathrm{ev,ch}}\big|
    \le C_{\mathrm{PSR}}\,\epsilon_{\mathrm{ins}},
    \nonumber\\
    C_{\mathrm{PSR}}=2\Delta\tau\,
    \sup_\tau\textstyle\sum_j
    |\partial u_j/\partial\theta_\ell|(\tau)\,R,
  \end{gather}
  the constant of \thm{endtoend}. No factor in the
  bound grows as any experimental parameter is refined: nothing
  divides by a small step.
\end{lemma}
\begin{proof}
  Per branch the value is
  $\operatorname{Tr}[O_P\,\Lambda_{\mathrm{after}}\circ
  \mathcal S\circ\Lambda_{\mathrm{before}}(\rho)]$; replacing
  $\mathcal S=V_{j,s}(\cdot)V_{j,s}^\dagger$ by
  $\widetilde{\mathcal S}_{j,s}$ changes it by
  \[
    \begin{aligned}
      &\big|\operatorname{Tr}\big[O_P\,\Lambda_{\mathrm{after}}\circ
        (\widetilde{\mathcal S}_{j,s}-\mathcal S)\circ
      \Lambda_{\mathrm{before}}(\rho)\big]\big|\\
      &\quad\le\|O_P\|_{\mathrm{op}}\,
      \big\|\Lambda_{\mathrm{after}}\circ
      (\widetilde{\mathcal S}_{j,s}-\mathcal S)\circ
      \Lambda_{\mathrm{before}}\big\|_\diamond\\
      &\quad\le R\,\epsilon_{\mathrm{ins}},
    \end{aligned}
  \]
  using $|\operatorname{Tr}[O\,\Phi(\rho)]|\le
  \|O\|_{\mathrm{op}}\|\Phi(\rho)\|_1$, then
  $\|\Phi\|_{1\to1}\le\|\Phi\|_\diamond$, submultiplicativity of
  the diamond norm, and $\|\Lambda\|_\diamond=1$ for channels.
  The estimator averages branch values with total weight mass
  $W(\tau)\le C_{\mathrm{PSR}}/R$, giving the display.
\end{proof}

\paragraph{Scope.}
Assumption~\ref{as:noise} excludes amplitude-dependent noise---
dissipator rates tracking the instantaneous drive
(e.g.\ $\propto u_j$ or $u_j^2$)---because shifted branches would
then see shifted noise, breaking shift-independence. The
fixed-rate dephasing channel of the Sec.~\ref{sec:eval} noise
model satisfies the assumption. On the target machine the
interaction coefficient can in principle drag its own dephasing
rate (Appendix~\ref{app:machine},
Eq.~\eqref{eq:dress-tradeoff}); this paper holds the dissipator
fixed at the operating point and treats the drive's excursions as
a perturbation that leaves the rate approximately unchanged. The
regime where the rate tracks the drive materially is outside its
scope.

\subsection{The Device Error Model}
\label{app:noise-device}

Table~\ref{tab:noise} instantiates Assumption~\ref{as:noise} with
the model under which every Sec.~\ref{sec:eval} estimator
executes, stated as the emulator runs it. Every channel is
$\Theta$-, shift-, and draw-independent, as the assumption
requires. The setpoint row is not a channel at all---it charges
the act of dialing a setpoint, and is discussed below.

\begin{table}
  \caption{The device error model. Values are in the
    settings below; Appendix~\ref{app:machine} gives the machine's
  rate ranges they are justified against.}
  \label{tab:noise}
  \small
  \begin{tabular}{@{}>{\raggedright\arraybackslash}p{0.22\linewidth}>{\raggedright\arraybackslash}p{\dimexpr0.78\linewidth-2\tabcolsep\relax}@{}}
    \toprule
    effect & model \\
    \midrule
    dephasing &
    each qubit's phase randomizes: coherences decay as
    $e^{-t/T_2^*}$ (dissipator
      $\tfrac{\gamma_\phi}{2}(Z\rho Z-\rho)$,
    $\gamma_\phi=1/T_2^*$), on every segment except the inserted
    one \\
    inserted-gate error &
    the inserted gate is imperfect: within
    $\epsilon_{\mathrm{ins}}$ of ideal
    (Lemma~\ref{lem:ins-bias}), half coherent over-rotation, half
    random $Z$ \\
    setpoint error &
    each programmed value lands off by an independent zero-mean
    draw $\delta$ of standard deviation $r$, but the draw is fixed across repetitions of the same programmed point; a new draw is taken when the programmed value changes and held until it changes again.
    (FD: two draws per estimate; PSR: one shared draw; NSR: one
    per shift) \\
    \bottomrule
  \end{tabular}
\end{table}

\paragraph{The setpoint row is not a channel.} Dialing a
setpoint is an operation, and the row charges it an error: the
programmed value lands offset by $\delta$. This never enters the
channel semantics---it perturbs \emph{which program executes},
not how the executed program evolves---and every method programs
setpoints, so the question is what each estimator does with the
error, not who has it. A draw is taken when a programmed value
changes and held until it changes again. FD's definition takes
two probe setpoints and divides their difference by $\epsilon$:
the probes are two programmed values, hence two draws
$\delta_\pm$, which become the $(\delta_+-\delta_-)f'/\epsilon$
term of Eq.~\eqref{eq:fd-setpoint}, amplified to order one once the offset reaches the scale of
the optimal $\epsilon$, and frozen over the estimate's shots,
so no shot count removes the resulting bias.
(Eq.~\eqref{eq:fd-floor}; the FD floor of Figure~\ref{fig:F6}).
PSR's branches program the source value, so they share one draw:
the estimate is the exact derivative at the displaced point
$\theta+\delta$, an error $|f''|\,r$ with bounded weight---the
same exposure class as the insertion error of
Lemma~\ref{lem:ins-bias}---never divided by anything, and second
order where $f''=0$ (the floor of Figure~\ref{fig:F6}'s PSR
series; Appendix~\ref{app:fd}). NSR re-dials the coefficient on
every shift, so each branch carries a fresh draw and the
draws average out with the shots.

\paragraph{Experimental settings.} The three configurations
behind Sec.~\ref{sec:eval}, one per figure. Table~\ref{tab:noise}
applies throughout, with $r=0.02$, per-figure $T/T_2^*$ as given
below, and $\epsilon_{\mathrm{ins}}=10^{-3}$ where charged.

\emph{Figure~\ref{fig:fd-trap} (the trap).}
$H=\theta Z_0+X_0$ with readout $\langle Z_0\rangle$; segment
$T=12$~\textmu s, $T_2^*=24$~\textmu s. The anchor $\theta^\ast=1.290$
is a steep mid-flank point, not an optimum, selected by an
automated sweep for a non-marginal configuration:
$|\nabla C_{\mathrm{device}}|=4.35$ there, $61\%$ of the window
maximum, a quarter of the local ripple period from the nearest
extremum. Probes at $\epsilon\in\{0.36,0.50,0.64\}$ in the
$\theta\pm\epsilon/2$ convention, each secant carrying its own
setpoint draws.

\emph{Figure~\ref{fig:F6} (convergence).} Two-qubit TFIM,
$H(\theta)=\theta\,Z_0Z_1+g\,(X_0+X_1)$ with $g=1.0$ fixed; the
differentiated parameter is the coupling $\theta$, whose tangent
is the single Pauli $Z_0Z_1$, at $\theta_0=1.757$, where
$\nabla C_{\mathrm{device}}=-1.449$. The operating point is
chosen at the stress end of a class, not at a generic point:
$\theta_0$ is the stationary point of $\nabla C_{\mathrm{device}}$
($C''=0$) inside the coefficient window whose headroom cap gives
$M=5$. There every estimator's setpoint displacement is second
order, so the comparison isolates what each strategy does with
the draw---FD divides by $\epsilon$, PSR shares one draw, NSR
re-dials---and Figure~\ref{fig:epssweep} repeats the comparison
at generic points on three landscapes of different sharpness, so
the point is the stress end of a class rather than a singleton.
Segment $T=5.0$~\textmu s,
$T_2^*=33.3$~\textmu s---the dressed-regime budget of
Appendix~\ref{app:machine}; executions $N$ on the decade grid
$10^2$--$10^6$; the setpoint draw of Table~\ref{tab:noise} on
every estimator. The left panel repeats every estimator on the
noiseless landscape: $T_2\to\infty$, no setpoint draw, no gate
error, the same $\theta_0$ (where the ideal gradient is
$-1.728$). FD arms, in the $\theta\pm\epsilon/2$ convention: the
step re-tuned at every budget, i.e.\ at each $N$ the
RMSE-minimizing $\epsilon$ of the $22$-point grid over
$[0.02,1.2]$ ($60$ tuning repetitions on a separate seed), which
moves from $0.31$ at $N=10^2$ to $0.21$ from $N=10^3$ on under the
device model and to $0.064$ at $N=10^6$ without noise; and the
fixed steps $\epsilon=0.02$ and $0.5$ on the noiseless panel. The
dashed line of the device panel is FD's shot-free error at
$\epsilon^\ast=0.21$, $0.197$ from $600$ setpoint draws. NSR: certificate
$\overline\Omega_\ell=2T=10.0$ (the term-wise bound, exact for a
  single-Pauli tangent; not to be confused with the transverse
field $g$), cap $M=5$ with the rejecting sampler of
Appendix~\ref{app:execution}, excluding $3.4\%$ of draws. PSR:
the headline series runs ideal insertions; ``PSR$+$gate'' charges
$\epsilon_{\mathrm{ins}}=10^{-3}$. The inset sweeps FD's
root-mean-square error against $\epsilon$ on $24$ points of a
geometric grid over $[0.02,1.2]$ at fixed $N=10^4$, $100$ seeds
per point, with PSR and NSR at the same $N$ as flat references.

\emph{Figure~\ref{fig:loop} (closed loop).} Inverse-simulation
task on two qubits, $H(\theta)=\theta_1\,Z_0Z_1
+\theta_2\,(X_0+X_1)$, segment $T=2.5$~\textmu s,
$T_2^*=16.7$~\textmu s: minimize the residual
\[
  \begin{aligned}
    C(\theta)&=(\langle Z_0Z_1\rangle-a^\ast)^2\\
    &\quad+w\,(\langle Z_0+Z_1\rangle-b^\ast)^2
  \end{aligned}
\]
with $w=0.25$ and
$(a^\ast,b^\ast)$ the observable values at the target, so
$C(\theta^\ast)=0$ at the interior optimum
$\theta^\ast=(1.03,1.03)$ (condition number $\kappa=24$); start
$(0.832,0.919)$; parameter box $[0.2,1.4]^2$, which constrains the
iterate only (NSR's shifts and FD's probes may leave it). Fixed-step FD has a truncation-shifted fixed point, displaced from $\theta^\ast$ by $(\epsilon^2/24)H^{-1}C'''$ along
the soft direction. Gradient descent with a decaying step,
$\eta_t=\eta_0/(1+t/20)$ with
$\eta_0=1/\mu_{\mathrm{stiff}}=0.059$ set once from the measured
stiffness; the reported iterate is the tail average $\bar\theta_t$
of the second half of the trajectory so far; $50$ steps, $20$
seeds (median and interquartile band), and at most $6000$
executions per step for every method ($6000$ for FD, $4800$ for
  the shift rules, whose residual's two observables are charged
twice in the accounting); the setpoint draw of
Table~\ref{tab:noise} applies at every step to every estimator.
PSR's insertion-time integral is discretized on $32$
$\tau$-samples per generator; NSR draws from $30$ modes (series
truncation $2.0\%$ at the start, $0.7\%$ at $\theta^\ast$). FD
arms at $\epsilon\in\{0.1,0.5,0.02\}$ in the
$\theta\pm\epsilon/2$ convention, the best chosen retrospectively
from $\{0.1,0.15,0.2,0.3,0.5\}$. CMA-ES is the
$(\mu/\mu_w,\lambda)$ variant with $\lambda=6$, $\mu=3$, one
generation per step at $1000$ executions per candidate, each
candidate a programmed setpoint carrying the draw, and initial
step $\sigma_0=0.05$, the best of $\{0.03,0.05,0.1,0.2,0.3\}$
chosen retrospectively, the same privilege FD's best step has.
Median final distance at step $50$ (fraction of seeds within
tolerance $0.03$): NSR $0.018$ ($100\%$), PSR $0.022$ ($95\%$),
FD $\epsilon=0.1$: $0.035$ ($20\%$), CMA-ES $0.045$ ($30\%$), FD
$\epsilon=0.5$: $0.085$ ($0\%$), FD $\epsilon=0.02$: $0.114$
($35\%$). The right panel shows the cost landscape on a broken
$\theta_2$ axis, with one seed per method---the seed whose distance
curve is closest to the median curve---the start, and the target.

\emph{Figure~\ref{fig:epssweep} (the step window).} The
two-qubit TFIM of Figure~\ref{fig:F6} at three segment lengths,
$T\in\{1,2,7\}$~\textmu s, each at $T/T_2^*=0.15$, differing
only in the landscape's sharpness. On each, the operating point
is a generic steep point of typical sharpness, the one whose PSR
displacement $|f''|\,r/|\nabla C|$ is closest to $2.5\%$, with
a fixed seed; nothing is tuned against FD. The step window (RMSE
at most $10\%$ of $|\nabla C|$ as $N\to\infty$) is
$\epsilon\in[0.4,1.0]$ on the healthy landscape, best $4.6\%$
at $\epsilon^\ast$; $[0.3,0.6]$ on the intermediate one, best
$7.8\%$; and empty on the ill one, best $15.8\%$. The operating
point of Figure~\ref{fig:F6} is therefore not a singleton: on
every landscape the FD window is bounded below by $r$ and above
by the landscape's sharpness, and even its best step leaves
$5$--$16\%$ error, while the shift rules need no step and stay
at a few percent.

\paragraph{The three gradients.}
For each parameter $\theta_\ell$,
\begin{equation}
  \begin{gathered}
    \nabla_\ell C_{\mathrm{ideal}}
    =\frac{\partial}{\partial\theta_\ell}
    \llbracket P\rrbracket(\ket{\psi}),
    \\
    \nabla_\ell C_{\mathrm{device}}
    =\frac{\partial}{\partial\theta_\ell}
    \llbracket P\rrbracket_{\mathrm{ch}}(\ket{\psi}),\\
    \widehat{\nabla_\ell C}
    =\text{mean of }G\text{ (Def.~\ref{def:exec})},
  \end{gathered}
\end{equation}
where $\widehat{\nabla_\ell C}$ is the physically realized
estimate: the executed estimator under the full model---headroom
truncation at cap $M$ (Lemma~\ref{lem:trunc-bias}), insertion
error $\epsilon_{\mathrm{ins}}$ (Lemma~\ref{lem:ins-bias}), and
the setpoint draws $\delta$. Sec.~\ref{sec:eval} reports
error of $\widehat{\nabla C}$ against
$\nabla C_{\mathrm{device}}$ only;
$\nabla C_{\mathrm{ideal}}$ appears in no measurement.

\paragraph{Emulator scope.}
The emulator is QuTiP: it consumes the compiler's machine-native
Hamiltonian segments and the normal-form orchestration around
them, not the scheduled per-channel pulse arrays---which is why
the main text claims estimator soundness from
Figure~\ref{fig:F6}, and schedule/ledger correctness from
nothing in Sec.~\ref{sec:eval}. The inserted segment of a PSR
branch carries no dissipator; its imperfection enters only through
$\epsilon_{\mathrm{ins}}$, so evolution noise is shift-independent
exactly as Assumption~\ref{as:noise} requires.

\section{Normal Form and Sample Complexity}
\label{app:execution}



\subsection{Normalization}

\begin{lemma}[Hoisting commutations]\label{lem:hoist}
  For fresh $\xi$ (not free in $H$, the boundaries, the weights, or the
  other branches; always arrangeable by $\alpha$-renaming), the
  following equivalences preserve the denotational semantics:
  \begin{align}
    &\mathsf{ev}(\tau_1,\tau_2,H);\,
    (\mathsf{let}\ \xi\sim\mu\ \mathsf{in}\ P)\nonumber\\
    &\quad\equiv\mathsf{let}\ \xi\sim\mu\ \mathsf{in}\
    (\mathsf{ev}(\tau_1,\tau_2,H);P),\label{eq:hoist-ev}\\[0.4em]
    &\mathsf{ev}(\tau_1,\tau_2,H);\,\sum_{i\in I}w_i\,P_i\nonumber\\
    &\quad\equiv\sum_{i\in I}w_i\,\big(\mathsf{ev}(\tau_1,\tau_2,H);P_i\big),
    \label{eq:hoist-fam}\\[0.4em]
    &w_1(\mathsf{let}\ \xi\sim\mu\ \mathsf{in}\ P_1)+\sum_{i\ge2}w_i\,P_i\nonumber\\
    &\quad\equiv\mathsf{let}\ \xi\sim\mu\ \mathsf{in}\
    \Big(w_1P_1+\sum_{i\ge2}w_i\,P_i\Big).
    \label{eq:hoist-let}
  \end{align}
\end{lemma}
\begin{proof}
  \eqref{eq:hoist-ev}: both sides denote
  \[
    \int\llbracket P[\xi\mapsto v]\rrbracket
    (U_H(\tau_2,\tau_1)\ket{\psi})\,d\mu(v).
  \]
  \eqref{eq:hoist-fam}: both sides denote
  \[
    \sum_iw_i\llbracket P_i\rrbracket(U_H(\tau_2,\tau_1)\ket{\psi}).
  \]
  \eqref{eq:hoist-let}: the right-hand side denotes
  \[
    \begin{aligned}
      \int\Big[&w_1\llbracket P_1[\xi\mapsto v]\rrbracket(\ket{\psi})\\
      &{}+\sum_{i\ge2}w_i\llbracket P_i\rrbracket(\ket{\psi})\Big]d\mu(v);
    \end{aligned}
  \]
  the second group is constant in $v$ and $\mu$ is a
  \emph{probability} measure, so its integral contributes the group
  itself---this is where the well-formedness requirement that every
  $\mu$ have total mass one is used structurally. Nested families
  flatten by associativity of the family denotation.
\end{proof}

\begin{proof}[Proof of Theorem~\ref{thm:stdform}]
  Apply Lemma~\ref{lem:hoist} repeatedly, innermost first: every
  $\mathsf{let}$ migrates to the front (later measures may depend on
    earlier bound variables, so the prefix is a kernel chain, not a
  product), every family flattens, and every evolution distributes over
  families until each branch is a sequence of $\mathsf{ev}$
  constructors terminated by one $\mathsf{meas}$. Each step preserves
  the denotation by Lemma~\ref{lem:hoist}; the procedure is a
  structural recursion on the finite syntax tree---each application
  moves one construct past one constructor---so it terminates.
  Identical branches may then be merged,
  summing weights; merging preserves the denotation by the family
  semantics, and can strictly reduce the executed estimator's variance,
  which is governed by $\sum_i|w_i(\bar\xi)|$ through the bound of
  Definition~\ref{def:exec}.
\end{proof}

\begin{remark}[The $P_i^\pm$ of the main text]\label{rem:pm-branches}
  The local FD, NSR, and PSR constructions pair their branches by
  a sign: FD's \eqref{eq:fd-program} is the one-pair instance,
  $P^\pm=P[\theta_\ell\mapsto\theta_\ell\pm\epsilon/2]$ with
  weight $1/\epsilon$; an NSR output pairs the two signs $\sigma$
  of each drawn shift; a PSR output pairs the two kick signs $s$
  of each insertion. Writing the evolution contributions as
  $\sum_iw_i\,[P_i^+-P_i^-]$ names these pairs; derivatives of
  parameter-dependent readouts or family weights contribute
  additional terms without such pairs. Normalization does not
  disturb the pairs: Theorem~\ref{thm:stdform} hoists samplers,
  flattens families, and at most merges identical branches, so in
  normal form the $P_i^\pm$ appear among the branches $B_i$,
  each sign absorbed into its weight, and
  Definition~\ref{def:exec} executes them as such.
\end{remark}

\subsection{Sample Complexity}
\label{app:sample}

Fix the normal form
\[
  \begin{aligned}
    &\mathsf{let}\ \xi_1\sim\mu_1\ \mathsf{in}\cdots
    \mathsf{let}\ \xi_k\sim\mu_k\ \mathsf{in}\\
    &\qquad\sum_{i\in I}w_i(\bar\xi)\,B_i(\bar\xi)
  \end{aligned}
\]
with branch denotations
$\beta_i(\bar\xi)=\llbracket B_i(\bar\xi)\rrbracket(\ket{\psi})$,
readouts bounded by $R$ (so $|\beta_i|\le R$ and every single-shot
readout value is bounded by $R$), and weighted-$L^1$ bound
$W(\bar\xi)=\sum_i|w_i(\bar\xi)|\le\overline W$ almost surely.

\begin{definition}[Single-level executed estimator]\label{def:exec}
  Draw $\bar\xi$ from the kernel chain. If $W(\bar\xi)=0$, return
  $G=0$ without executing. Otherwise draw a branch index $i$ with
  probability $|w_i(\bar\xi)|/W(\bar\xi)$; execute $B_i(\bar\xi)$
  \emph{once}, observing outcome $m$; return
  \begin{equation}
    G \;=\; W(\bar\xi)\,\operatorname{sign}\!\big(w_i(\bar\xi)\big)\,
    f^{(i)}_m(\bar\xi).
  \end{equation}
  One copy of $G$ costs at most one quantum execution.
\end{definition}

\begin{proof}[Proof of \thm{endtoend}]
  \emph{Unbiasedness.} When $W(\bar\xi)=0$ every weight vanishes
  and $G=0$ equals the normal form's value at that draw.
  Otherwise, conditioning on $\bar\xi$ and the branch draw,
  $\mathbb E[G\mid\bar\xi,i]=W\operatorname{sign}(w_i)\beta_i$, so
    \[
      \mathbb E[G\mid\bar\xi]
      = \sum_i
        \frac{|w_i|}{W}
        W\,\operatorname{sign}(w_i)\beta_i
      = \sum_i w_i\beta_i .
    \]
    Taking the expectation over the sampling prefix gives
    \[
      \mathbb E[G]=\llbracket Q\rrbracket(\ket{\psi}).
    \]

  \emph{Convergence.} $|G|\le\overline W R=:C_X$ deterministically, so
  by Hoeffding's inequality the mean of
  $N=O\big(C_X^2\,\epsilon_t^{-2}\log(1/\beta)\big)$ i.i.d.\ copies is
  within $\epsilon_t$ of $\llbracket Q\rrbracket(\ket{\psi})$ with probability $1-\beta$;
  the total number of quantum executions is at most $N$.

    For $Q=D_\ell^{\mathcal S}[P]$ with a sound strategy,
    Theorem~4.4 gives
    \[
      \llbracket Q\rrbracket
      = \frac{\partial}{\partial\theta_\ell}\llbracket P\rrbracket,
    \]
    yielding the derivative-estimation specialization used in the
    main text.

  \emph{The constants.} For NSR, $W=\overline\Omega_\ell$ identically,
  so $C_{\mathrm{NSR}}=\overline\Omega_\ell R$: the certificate is the
  estimator scale, and the unboundedness of the sampled 
  \emph{shift}
  $h_\kappa$ never enters---the estimator's \emph{value} is bounded. For
  PSR, the two kick signs give \[
    \begin{aligned}
      W(\tau)&=\sum_{j,s}|s\,\Delta\tau\,\partial
      u_j/\partial\theta_\ell|(\tau)\\
      &=2\Delta\tau\sum_j|\partial u_j/\partial\theta_\ell|(\tau)
    \end{aligned}
  \] over the touched terms, so
  \[
    C_{\mathrm{PSR}}=2\Delta\tau\,
    \sup_\tau\sum_j|\partial u_j/\partial\theta_\ell|(\tau)\,R;
  \]
  for $J$
  touched terms with coefficients bounded by $\bar v$,
  $C_{\mathrm{PSR}}\le 2\Delta\tau\,J\bar v R$. Multiple
  differentiated evolutions contribute additively to $W$ through the
  flattened family.
\end{proof}

\paragraph{Execution scaling.}
For a parameter touching $J$ terms (the $q$ of
Appendix~\ref{app:select}) with $|v_j|\le\bar v$ over a segment of
length $T$, both estimator scales are at most $2J\bar vTR$: the
term-wise certificate bounds $C_{\mathrm{NSR}}$ by
$2\int\sum_j|v_j|\,dt\cdot R$, and
$C_{\mathrm{PSR}}\le2\Delta\tau\,J\bar vR$ with
$\Delta\tau=T$. The executions to reach target accuracy
$\epsilon_t$ therefore scale as
$N\propto(J\bar vTR/\epsilon_t)^2$---quadratic in the touched-term
count and in the segment length---and a global coefficient, one
$\theta$ multiplying every bond of an $n$-qubit chain, has
$J\propto n$. Appendix~\ref{app:select} is the account of how
much of that $J^2$ NSR actually pays: the diameter of a mixed
tangent concentrates below the sum of diameters, roughly as
$\sqrt J$, and that concentration is what moves the measured
boundary. At a fixed execution budget
(Figure~\ref{fig:loop}'s $6000$ per gradient) the per-gradient
\emph{compile} cost is instead the budget times the per-branch
increment of Appendix~\ref{app:scaling}.

\begin{remark}[Fused shifts]\label{rem:fusion}
  The lowering fuses runs of consecutive differentiated
  $\mathsf{ev}$ constructors: one $(\kappa,\sigma)$ draw serves
  the whole run, each segment's touched coefficients shifted by
  the same $\sigma h_\kappa v_j$. This is sound. On the
  concatenated interval the shifted family is the piecewise
  $H_i+s\dot H_{\ell,i}$, and
  Lemmas~\ref{lem:tangent}--\ref{lem:band} run verbatim on
  it---their proofs are ODE arguments indifferent to piecewise
  generators, and Lemma~\ref{lem:noise-trotter} also applies to
  bounded piecewise generators, so the noisy case transfers too. Hence $F$ is
  band-limited to $\sum_i\Omega_{\ell,i}$, the sum of per-segment
  certificates is a sound run certificate, and $F'(0)$ is the sum
  of the run's evolution contributions---exactly what the
  per-segment transformation delivers through the flattened
  family. The estimator scale is unchanged ($W=\sum_i
  \overline\Omega_{\ell,i}$ either way), so
  \thm{endtoend}'s shot count is identical; what
  fusion buys is the branch inventory: a run's many
  single-segment branches become one, which is what makes an NSR
  branch a single amplitude edit of the source schedule
  (Appendix~\ref{app:compiler}) even when the touched
  tangent spans several ledger segments, and what
  Appendix~\ref{app:scaling}'s per-branch compile increment is an
  increment of.
\end{remark}

\begin{remark}[Why not two loops]
  The nested procedure that first estimates every branch mean to
  accuracy $\alpha$ at each outer draw, then averages $N'$ outer
  draws, costs
  \[
    \widetilde O\big(n_b^3M^2\epsilon_t^{-4}\log^2(1/\beta)\big):
  \]
  each
  level pays its own $\epsilon_t^{-2}$, and the per-branch,
  per-draw guarantees force union bounds. Deferring \emph{all}
  averaging to the single level of Definition~\ref{def:exec} removes
  two powers of $\epsilon_t^{-1}$ and one power of the logarithm, because
  the only concentration argument is one Hoeffding bound on the bounded
  variable $G$. Batched deployment---fixing a drawn configuration and
  taking $B$ shots before redrawing, as hardware upload costs
  favor---changes the variance decomposition but not the
  $\epsilon_t^{-2}$ scaling, and preserves unbiasedness for any $B$.
\end{remark}

\subsection{Bounded-Device Realizability}

\begin{lemma}[Out-of-range mass]\label{lem:tail}
  If a waveform's headroom admits the shifts $h_\kappa$ only for
  $\kappa\le M$, the probability that one NSR draw exceeds it is
  \begin{equation}
    p_{\mathrm{out}}
    =\sum_{\kappa>M}\eta_\kappa
    \le\frac{4}{\pi^2(2M+1)} .
  \end{equation}
\end{lemma}
\begin{proof}
  $\eta_\kappa=\tfrac{8}{\pi^2}(2\kappa+1)^{-2}$ and, by the
  integral test,
  \[
    \sum_{\kappa\ge M+1}(2\kappa+1)^{-2}
    \le\int_M^\infty(2x+1)^{-2}dx=\frac{1}{2(2M+1)}.
  \]
\end{proof}

\begin{lemma}[Plan-level realizability]\label{lem:pfail}
  A gradient plan whose $i$-th differentiated site carries headroom
  cap $M_i$, with independent draws across sites, realizes all of
  its draws with probability at least
  $1-\sum_i p_{\mathrm{out},i}$; equivalently
  $p_{\mathrm{fail}}
  =1-\prod_i(1-p_{\mathrm{out},i})
  \le\sum_i p_{\mathrm{out},i}$.
\end{lemma}
\begin{proof}
  The Weierstrass product inequality (union bound).
\end{proof}

\begin{lemma}[Truncated-mode bias]\label{lem:trunc-bias}
  The deterministic truncation that keeps modes $\kappa\le M$ with
  their exact weights has bias
  \begin{equation}
    \begin{aligned}
      &\Big|\frac{4\overline\Omega_\ell}{\pi^2}
      \sum_{\kappa>M}\frac{(-1)^\kappa}{(2\kappa+1)^2}
      \big[F(h_\kappa)-F(-h_\kappa)\big]\Big|\\
      &\qquad\le\frac{4\overline\Omega_\ell R}{\pi^2(2M+1)} .
    \end{aligned}
  \end{equation}
\end{lemma}
\begin{proof}
  $|F|\le R$ termwise and the tail estimate of
  Lemma~\ref{lem:tail}.
\end{proof}

\paragraph{Realizing the cap.} The two capped variants of
Sec.~\ref{sec:lower} execute NSR under a headroom cap $M$, and the
ledger records which one ran. The \emph{rejecting} sampler draws
$(\kappa,\sigma)$ from $\eta$ as in the uncapped program; a draw
with $\kappa>M$ is caught at scheduling time, before pulse
synthesis, consumes no device execution, and is recorded as a
zero sample, with no renormalization. The \emph{truncated}
sampler draws $\kappa$ from $\eta$ conditioned on $\kappa\le M$
and scales the branch weight by the retained mass
$1-p_{\mathrm{out}}$. Both have expectation
\[
  \frac{4\overline\Omega_\ell}{\pi^2}
  \sum_{\kappa\le M}\frac{(-1)^\kappa}{(2\kappa+1)^2}
  \big[F(h_\kappa)-F(-h_\kappa)\big],
\]
the exact-weight truncation, so the bias bound of
Lemma~\ref{lem:trunc-bias} covers both; and both keep
$|G|\le\overline\Omega_\ell R$, so
\thm{endtoend}'s shot bound applies to both. They
differ in variance and execution cost: rejection spends a fraction
$p_{\mathrm{out}}$ of its draws on zero samples but never
synthesizes an unplayable pulse, while truncation executes every
draw at a weight smaller by $1-p_{\mathrm{out}}$. The $M=5$
series of Figure~\ref{fig:F6} is the rejecting sampler:
Sec.~\ref{sec:eval}'s excluded $3.4\%$ of draws are its zero
samples, $p_{\mathrm{out}}=0.034$ by Lemma~\ref{lem:tail}.

\section{Machine Notes}
\label{app:machine}

\paragraph{The machine.} A tweezer array holds single neutral
atoms in tightly focused laser beams, one atom per focus, on a
programmable two-dimensional grid of hundreds of sites; the qubit
is a pair of ground states of the atom's outermost electron,
coherent for seconds when left alone. Interactions are switched on
optically, by driving
the atoms toward Rydberg states---electronic orbits large enough
that two atoms micrometres apart shift each other's energies by
many megahertz~\cite{rydberg-review}. And because the traps are
light, they move: the same machine rearranges its geometry and
carries atoms between functional zones in the middle of a
program~\cite{bluvstein2024logical}. One apparatus therefore supplies a
programmable analog Hamiltonian, discrete gates, and transport
between the two---exactly the resources the two evolution
derivatives consume, and why this platform is the target. The
machine of this paper combines the cryogenic $^{133}$Cs array
of~\cite{jin2026extended} with the zoned layout
of~\cite{bluvstein2024logical}: an
analog (interaction) zone and a spatially separated gate zone
reached by movable tweezers. Its one unusual ingredient is a
$4$~K radiation shield suppressing blackbody-induced Rydberg
decay by roughly two orders of magnitude, which is what the gate
budget below is computed at. Nothing in this paper has been
executed on that apparatus. Figure~\ref{fig:apparatus} shows a
representative apparatus of this class.

\begin{figure}[t]
  \centering
  \includegraphics[width=\linewidth]{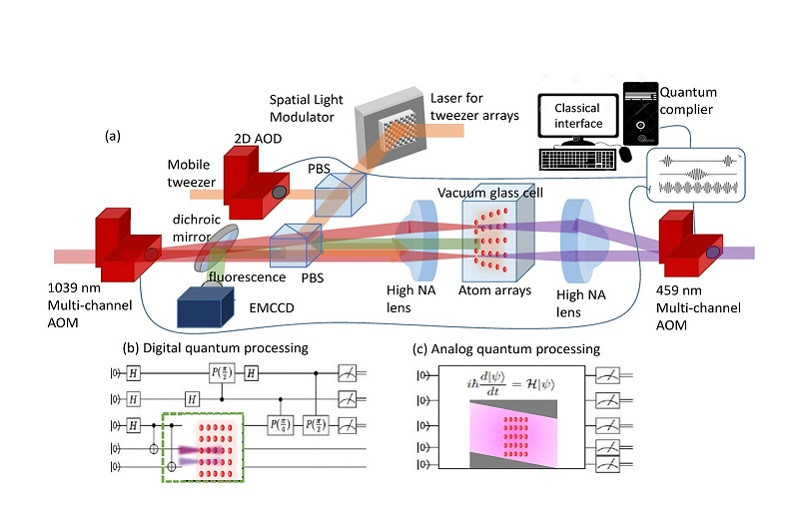}
  \caption{A representative tweezer-array apparatus, reproduced
    from~\cite{quantumtaiwan2026atoms} (CC~BY~4.0): one laser is shaped
    into the grid of traps---deflectors (AOD) and a spatial light
    modulator write the pattern---inside a vacuum cell, and
    cameras image the register through the same objective. Another
    laser propagates in the opposite direction and illuminates the
    atoms to implement single-qubit gates and interactions between
    qubits. The
    machine of~\cite{jin2026extended} shares this layout, with the
  addition of the cryogenic shield.}
  \label{fig:apparatus}
\end{figure}

\paragraph{The realizable Hamiltonian.} Start from what the
machine does digitally, which is the familiar menu: single-qubit
rotations about $X$, $Y$, or $Z$ on any site, and a CZ between a
chosen pair in the gate zone~\cite{rydberg-gate}. The analog
resource is the same menu left on: instead of firing a rotation
and stopping, the machine holds the generators on with
programmable, time-varying strengths---which is the machine
signature of Sec.~\ref{sec:machine},
\begin{equation}
  \begin{aligned}
    H(t)={}&\sum_a\big[u_a^x(t)X_a+u_a^y(t)Y_a+u_a^z(t)Z_a\big]\\
    &{}+\sum_{a<b}u^{zz}_{ab}(t)\,Z_aZ_b.
  \end{aligned}
  \label{eq:machine-H}
\end{equation}
Each coefficient denotes a realizable Hamiltonian control, not necessarily an independent physical channel. The single-qubit coefficients are driven by lasers only: $u^x_a$ and $u^y_a$ use one per-site drive laser with amplitude and phase lists before multiplexing, setting the strengths of the $X$ and $Y$ terms. $u^z_a$  is a per-site energy shift from an addressing beam—or costs nothing at all, since a $Z$ rotation can be book-kept in software. Two-qubit interaction terms require tuning both the lasers and the atom positions. Pair-specific $u^{zz}_{ab}$ terms are synthesized through the global dressing interaction together with the programmable geometry: atoms may be rearranged or moved out of the interaction zone to realize the requested pair structure, at a transport and scheduling cost.
The inserted $\pi/4$ kicks of
PSR are the digital menu again---$Z_a$ free in software, $X_a$ or
$Y_a$ a pulse of tens of nanoseconds~\cite{jin2026extended}, and $Z_aZ_b$
one calibrated CZ in the gate zone, Eq.~\eqref{eq:zz-lower}
below. Machines of this class can also couple atoms in a
flip-flop (XY-exchange) fashion~\cite{rydberg-review}; this one
does not, which is Appendix~\ref{app:select}'s exclusion of
$XX$/$YY$ tangents.

\paragraph{Two coherences.} The modeled qubit has two relevant coherence regimes. Dressing off, the atom is a ground-state qubit: trapped,
transportable, and coherent at the ten-second scale under
magic-wavelength trapping with dynamical
decoupling~\cite{li2019triply,manetsch2025tweezer}---idling and transport
are charged at this coherence, which is why the $114.1$~\textmu s
of transport in Figure~\ref{fig:branch} costs wall-clock time
and not error budget. Dressing on, the atom carries a small,
laser-controlled Rydberg
fraction $\varepsilon$ and inherits a scaled fraction of Rydberg
decoherence; the interaction scales as
$\varepsilon^2$ and the dressed decoherence rate as
$\varepsilon$, so at fixed detuning
\begin{equation}
  \theta\propto\varepsilon^2,
  \qquad
  \Gamma\propto\varepsilon\propto\sqrt\theta,
  \qquad
  T_2\propto\theta^{-1/2}.
  \label{eq:dress-tradeoff}
\end{equation}
The dressed $T_2$ is therefore an operating point, not a machine
constant: Sec.~\ref{sec:eval}'s $T/T_2^*=0.15$ corresponds, at
the $5.0$~\textmu s segment of Figure~\ref{fig:branch}, to a
dressed $T_2\approx33$~\textmu s, and deployed protocols on this
class of machine run interaction in short interleaved cycles, so
segments of this length are representative. The undressed
ground--Rydberg $T_2^*=6.2$~\textmu s of~\cite{jin2026extended} is
limited by atomic motion and is not the number
Sec.~\ref{sec:eval} divides by; the cryogenic Rydberg lifetimes,
extended roughly threefold
over room temperature by the shield~\cite{jin2026extended}, enter only
through the CZ budget. Because interaction strength and dressed
decoherence share one knob, the dissipator can in principle track
a differentiated coefficient; the paper's guarantees fix it at
the operating point (Assumption~\ref{as:noise}, scope in
Appendix~\ref{app:noise}). At operating density a single
blackbody-induced transition can avalanche through dipolar
coupling and void the array~\cite{jin2026extended}; such a shot is
detected as collective loss and discarded, a rejection fraction
suppressed by the shield and by $\varepsilon$.


\begin{table}[t]
  \caption{What the paper charges. Every row is consumed somewhere;
  the last column says where.}
  \label{tab:machine}
  \small
  \begin{tabular}{@{}>{\raggedright\arraybackslash}p{0.32\linewidth}>{\raggedright\arraybackslash}p{0.28\linewidth}>{\raggedright\arraybackslash}p{\dimexpr0.4\linewidth-4\tabcolsep\relax}@{}}
    \toprule
    quantity & value & charged at \\
    \midrule
    dressed interaction $T_2$ (operating point) &
    $\approx33$~\textmu s at the $5.0$~\textmu s segment &
    $T/T_2^*=0.15$: Table~\ref{tab:noise}, Sec.~\ref{sec:eval} \\
    ground-state coherence (idle, transport) &
    $\gtrsim10$~s, projected~\cite{li2019triply,manetsch2025tweezer} &
    transport charged zero error: Fig.~\ref{fig:branch} \\
    CZ duration &
    $696$~ns &
    the PSR branch: Fig.~\ref{fig:branch} \\
    transport, one relocation (lift, travel, drop) &
    $57.1$~\textmu s &
    the PSR branch ($119.8$~\textmu s total):
    Fig.~\ref{fig:branch} \\
    CZ error &
    budget $5\times10^{-4}$; charged $10^{-3}$ &
    $\epsilon_{\mathrm{ins}}$: Table~\ref{tab:noise} \\
    dressing amplitude range &
    supports shifts to $s_{\max}=1.94$: cap $M=5$ &
    NSR cap: Fig.~\ref{fig:F6}, App.~\ref{app:execution} \\
    \bottomrule
  \end{tabular}
\end{table}

The transport figure is the scheduled value of
Figure~\ref{fig:branch}'s ledger. Each relocation routes the
touched pair through a transit lane $5$~\textmu m off its row, so
the moving tweezers never sweep parked atoms: three minimum-jerk
legs at the $4$~\textmu m/\textmu s cap---lift ($2.34$~\textmu s),
travel over the $100$~\textmu m zone separation
($52.38$~\textmu s), drop ($2.34$~\textmu s)---$57.07$~\textmu s
per relocation, each leg a ledger row. Two relocations and the
gate make the $114.8$~\textmu s episode; together with the
interrupted $5.0$~\textmu s evolution the branch totals
$119.8$~\textmu s, all but $0.7$~\textmu s of it spent in the
ground-state regime.

\paragraph{Inserted operations by generator class.} A PSR branch
inserts $\mathsf{ev}(0,\tfrac{\pi}{4},s\cdot A_j)$ for a touched
generator $A_j$, with $s=\pm1$ (Sec.~\ref{sec:kick}). The
signature admits four classes:

\begin{itemize}\itemsep2pt
  \item $A_j=Z_a$: a $Z$ rotation, realized as a frame update in the
    control software. No pulse, no time.
  \item $A_j=X_a$ or $Y_a$: a $\pi/2$ pulse from the same drive
    laser---tens of nanoseconds at full power~\cite{jin2026extended}, below
    the scheduler's resolution.
  \item $A_j=Z_aZ_b$: one calibrated two-qubit gate plus frame
    updates, by~\eqref{eq:zz-lower} below. Physically an episode of
    transport legs and a gate: dressing off, relocation to the gate
    zone (lift, travel, drop), CZ, relocation back, dressing
    on---seven ledger rows of transport and gate where an NSR branch
    has a single segment (Appendix~\ref{app:compiler}).
  \item $A_j=X_aX_b$ or $Y_aY_b$: not in the signature. See
    Appendix~\ref{app:select}.
\end{itemize}

The two-qubit case is the only one that costs a gate:
\begin{equation}
  \begin{gathered}
    e^{-\mathrm{i}s\frac\pi4 Z_aZ_b}
    =e^{\mathrm{i}s\pi/4}\,\mathrm{CZ}_{ab}\,
    R_z^{(a)}\!\big(s\tfrac\pi2\big)\,
    R_z^{(b)}\!\big(s\tfrac\pi2\big),\\
    R_z(\phi)=e^{-\mathrm{i}\phi Z/2}.
  \end{gathered}
  \label{eq:zz-lower}
\end{equation}
Both signs use the \emph{same} $\mathrm{CZ}$: the sign enters only
through the virtual $Z$ frame and an unobservable global phase.
This is why the two PSR branches of Figure~\ref{fig:branch}
differ in the frame table alone, and why a first-order derivative
occupies the gate zone at most once per branch.
%

\paragraph{Gate error budget.} The infidelity
$\epsilon_{\mathrm{ins}}$ that Appendix~\ref{app:noise} charges to
a PSR insertion is a simulated error budget, not a measured gate
fidelity: a two-qubit gate has not been run on this apparatus. The
budget gives $5\times10^{-4}$ for a $\mathrm{CZ}$ at the cryogenic
Rydberg $T_1$ of~\cite{jin2026extended}, attainable at increased drive
power, where the leading error terms decrease with increasing
Rabi frequency~\cite{jin2026extended}. Table~\ref{tab:noise} charges the
conservative variant, $\epsilon_{\mathrm{ins}}=10^{-3}$, and that
value is worth stating plainly: it is the level two-qubit gates
have already been demonstrated at on room-temperature neutral-atom
hardware~\cite{evered2023high}. PSR's realization floor in
Sec.~\ref{sec:eval} therefore does not rest on a gate nobody has
built; it rests on one that has, with this machine's own budget a
factor of two below it. The paper claims device-ready control
throughout, and never hardware execution.

\paragraph{The instruction set, and its redundancy.} The compiler
does not see Eq.~\eqref{eq:machine-H} as a formula; it sees an
instruction set, and the one this machine is given has three
kinds of entry: a per-site detuning and a per-site drive,
realizing $u^z_a$ and the $u^x_a,u^y_a$ pair; a \emph{single
global} interaction instruction under one amplitude, realizing the
whole dressing sum $\sum_{a<b}C_6R_{ab}^{-6}\,n_an_b$ at the
geometry the compiler has chosen; and one derived $Z_aZ_b$ entry
per pair, which is what the inserted kick of
Assumption~\ref{as:kick} consumes. The global entry is the reason a
uniform coupling costs one instruction rather than one per pair,
and so is what makes the $O(n)$ machine of
Appendix~\ref{app:scaling} possible at all.

It also makes the set redundant, deliberately. With only two atoms in
the interaction zone, raising the global dressing amplitude
realizes their $Z_aZ_b$ coupling exactly as carrying them to the
gate zone for a single-qubit-plus-CZ sequence would: same
semantics, two realizations, and the compiler picks. It picks the
one that move fewer atoms, because moving them is what a
schedule pays for---$114.1$~\textmu s of the $119.8$~\textmu s
branch in Figure~\ref{fig:branch}. So the dressing entry carries
the couplings the program asks for, with its single-site residue
absorbed by the detunings, and the derived $Z_aZ_b$ entries are
left for the inserted kick. As with the choice of strategy in
Appendix~\ref{app:select}, both realizations are sound and the
choice is over cost. It is a choice at all only because this
machine has no always-on interaction: the coupling is present only
while its laser is on, which is also what
Assumption~\ref{as:kick}'s halt and resume needs.

\section{Compiler Notes}
\label{app:compiler}

Appendix~\ref{app:machine} said what the machine can do. This
appendix says what one run of it actually is, which slice of that
run the compiler owns, and what the compiler's output looks like
on the wires---the vocabulary Appendices~\ref{app:select}
and~\ref{app:scaling} price.

\subsection{One Execution, and the Window the Compiler Owns}
\label{app:window}

\begin{figure}[t]
  \centering
  \includegraphics[width=\linewidth]{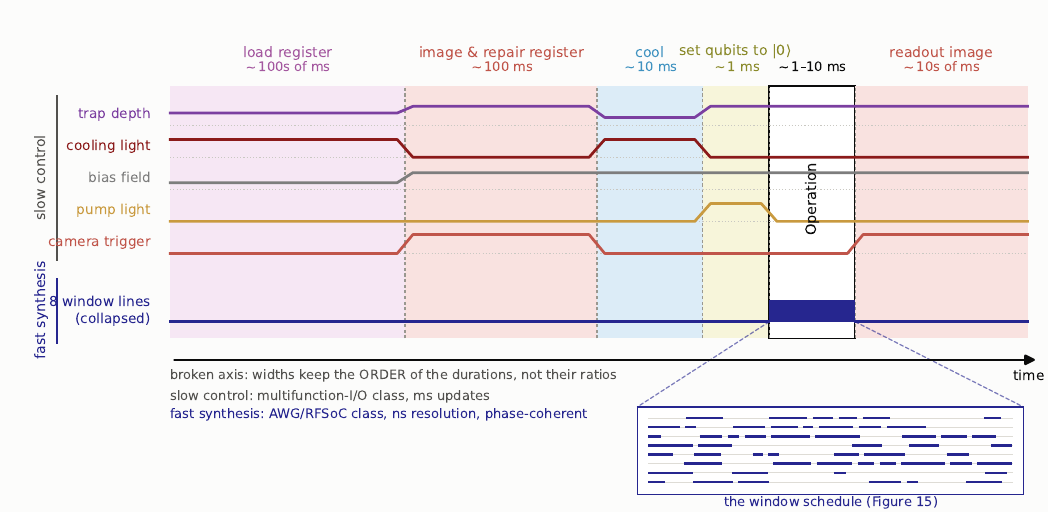}
  \caption{One execution cycle, generalized: redrawn
    after~\cite{manetsch2025tweezer} (CC~BY~4.0) with lab-specific
    labels replaced by their functions. The slow control group
    does all of the frame's work in millisecond envelopes and
    holds static values during the window; the fast synthesis
    group is silent through the entire frame and lights up only
    in the shaded sliver, where the compiled program---at most
  hundreds of microseconds---runs.}
  \label{fig:cycle}
\end{figure}

Everything the atoms feel is shaped light, and light is shaped by
electronics; compiling to this machine means, concretely,
computing the arrays its signal generators play. Two classes of
electronics run one execution (Figure~\ref{fig:cycle}). The
\emph{frame}---loading, register repair, cooling, setting the
qubits to $\ket0$, readout imaging---runs on cameras and slow
multifunction I/O: millisecond envelopes, no phase coherence
required. The \emph{window}---the programmed experiment---runs on
fast, phase-coherent synthesis channels feeding the light
modulators, at nanosecond resolution, on the eight lines of the
next subsection. The compiler's output lands only on the second
class.

The cycle reads off the figure. Atoms are loaded from a cold
cloud into the trap grid---arrival is random, so roughly half the
sites fill; a camera image finds the occupied sites and the traps
rearrange them into a defect-free register; the atoms are cooled;
a pumping pulse sets every qubit to $\ket0$; the window runs; a
final image reads the register out. The fast group is silent
through the whole frame; the slow group holds static values
through the whole window.

The frame is outside the compiler for three reasons, and they are
resource arguments, not merely scope. \emph{Resolution
economics:} every frame trace is a millisecond envelope that
cheap multifunction I/O drives; the window needs
nanosecond-resolution, phase-coherent synthesis, and those
channels are the scarce, expensive resource---which is why there
are eight of them, and why the inventory of the next subsection
is a budget rather than a list. \emph{Machine-dependence:} the
published sequence Figure~\ref{fig:cycle} is redrawn from is
lab-specific down to its laser names---that is why it had to be
generalized to be drawn at all. Only the window is portable, and
the compiler targets the Hamiltonian ability of
Sec.~\ref{sec:machine}, not any laboratory's frame.
\emph{Latency:} the only real-time, low-latency control in the
cycle lives inside the window, where the runtime follows the
real-time pulse DSL design of~\cite{wu2026pulsedsl} on RFSoC-class
hardware; the frame never touches that plane, and keeping it out
is what leaves the runtime extensible without dragging a
laboratory's frame along.
The language agrees with the boundary: Sec.~\ref{sec:theory} has no
initial-state construct---a program starts at a prepared register
and ends at $\mathsf{meas}$; preparing $\ket0^{\otimes n}$ is the
machine's convention, not the program's.

The compiler owns the window between ``qubits ready'' and the
readout trigger. Inside it: placing the atoms into the
\emph{programmed} geometry---which sites, which spacings; this is
where the pair couplings $J_{ab}$ of Sec.~\ref{sec:machine} are
set, so geometry is a compiled quantity, unlike the defect repair
of the frame---then the evolutions, any insertions with their
transport, and the measurement trigger. One consequence is worth
fixing before any number is read: the window is orders of
magnitude shorter than the cycle wrapped around it, so throughput
is set by the cycle and is identical across strategies. Cost in
this paper is therefore counted in \emph{executions}
(Appendices~\ref{app:execution} and~\ref{app:select}) and in
\emph{compile seconds} (Appendix~\ref{app:scaling}), never in
schedule microseconds---the wall-clock figures quoted for
branches here and in Figure~\ref{fig:branch} describe
structure, not cost.

\subsection{A Channel Is Not a Qubit}
\label{app:channels}

None of the window's channels is wired to one qubit. The drive
group's one physical line carries a sum of tones,
\begin{equation}
  s(t)=\sum_a A_a(t)\,\cos\!\big(\omega_a t+\varphi_a(t)\big),
  \label{eq:mux}
\end{equation}
one per site: the deflector maps frequency to position, so the
tone at $\omega_a$ lands on atom $a$ and nowhere else, and the
program's per-site coefficients \emph{are} the tone
envelopes---$u^x_a$ and $u^y_a$ of Eq.~\eqref{eq:machine-H} are
the two components of the complex envelope
$A_ae^{i\varphi_a}$. Synthesizing $s(t)$ from the envelopes,
sample by sample, is the compiler's job on this group, and two
constraints follow from the display rather than from any
datasheet. The line has one full scale, so amplitude range is
shared across tones: the composite $s(t)$ must fit it, and a
shifted envelope on any one site consumes headroom the whole
group feels---the headroom check of Appendix~\ref{app:execution}
therefore runs on the composite, not on a single waveform,
whenever the touched coefficients share a line (the running
  instance's shifts live on the single dressing line, where the
check is per-channel; Table~\ref{tab:machine}). And tones stay
independent only while their spacing exceeds the envelopes'
bandwidth, which bounds how fast per-site coefficients may change
on a shared line.

Locality, then, is manufactured, by three mechanisms.
\emph{Frequency} is Eq.~\eqref{eq:mux}. \emph{Position:} the
addressing AOD points light at chosen sites, and the movement AOD
makes locality by physically relocating atoms---into the
programmed geometry at the window's start, to the gate zone and
back for an insertion. \emph{Time:} the gate zone is one resource
shared serially, and contention never arises, because a
first-order derivative inserts at most one segment per branch---a
property of the transformation, not of scheduling luck. The
remaining channels are global by design and stay so: the dressing
field is one knob for the whole array, and pair structure comes
from geometry, which is compiled.

Concretely, the running instance's window uses eight lines in
five groups: the movement AOD ($x$ and $y$; executes the
transport plan), the addressing AOD ($x$ and $y$; site-selective
  energy shift realizing $Z_a$---optional, since $Z$ terms can
  instead be carried by the frame table---and site-selective
shelving in the analog zone), the single-qubit drive ($I$ and
$Q$, realizing $X_a$ and $Y_a$), the dressing field (the
$Z_aZ_b$ term), and the gate field of the digital zone (the CZ).
The frame table is software.

\subsection{Scheduling, and Why the Ledger Exists}
\label{app:ledger}

The scheduler assigns every semantic action to the wall clock
inside the window---segment boundaries, transport legs with their
settle time, frame-table updates---and runs the compile-time
analyses over the coefficient arrays: the certificate of
Assumption~\ref{as:bandwidth} and the headroom check that feeds
the capped variants of Appendix~\ref{app:execution}.

On a fixed-qubit machine, ``which qubit'' is a wire label and a
schedule is self-locating. Atom qubits carry one more coordinate:
\emph{position is state}. The same atom is an interaction-zone
qubit in one segment, cargo in the next, a gate operand in the
third, and the meaning of its couplings changes when it
moves---$J_{ab}$ is a function of where the atoms are. A bare
per-channel waveform dump therefore cannot answer the question
post-processing actually asks: where was atom $a$, and what did
$Z_aZ_b$ mean, during segment~3? The PulseLedger exists for that
question. Per segment it records the program- and wall-clock
boundaries (windows consume the latter only); the active
Hamiltonian terms under the geometry then in effect, with each
coefficient array and its provenance (source expression, or which
shift or insertion draw); the per-qubit frame-table state; the
transport plan; and a marker on inserted segments. Its consumers
are already in this paper: per-segment coherence charging reads
its rows (Appendix~\ref{app:machine}); the ``PSR$+$gate''
disclosure series of Figure~\ref{fig:F6} is computable from the
insertion markers alone; Appendix~\ref{app:scaling} counts its
entries; the shared draw of a fused run is recorded here
(Remark~\ref{rem:fusion}); and it is self-describing, so replay
and audit need the ledger alone, with no compiler state.

\subsection{A Branch Can be an Edit}
\label{app:branch-edit}

\begin{figure}[t]
  \centering
  \includegraphics[width=\linewidth]{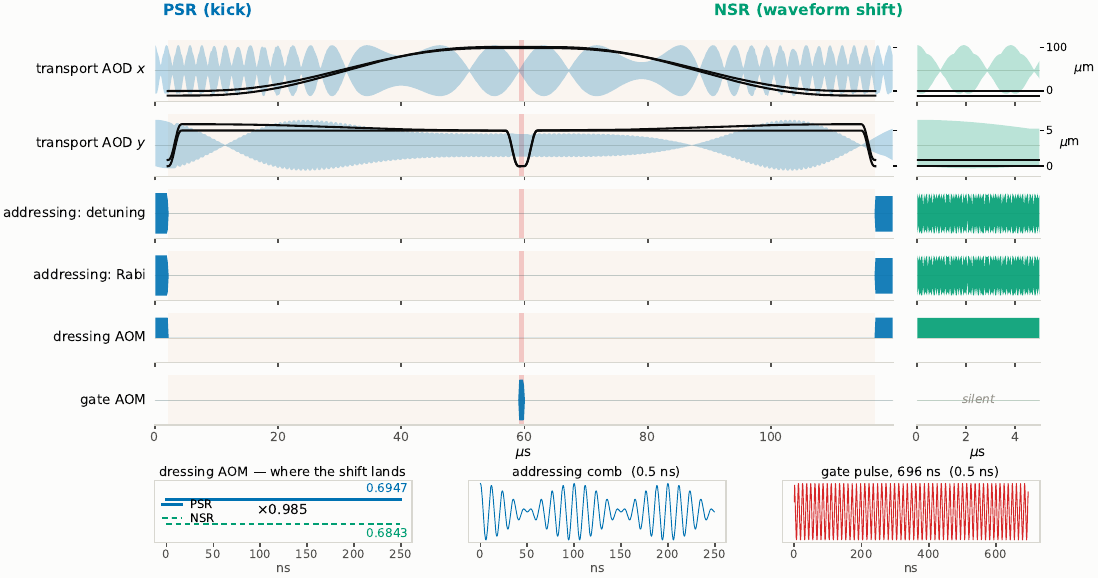}
  \caption{Both branch types, emitted from one compile of the
    running instance and drawn per channel with the PulseLedger's
    segment boundaries overlaid. The NSR branch is the source
    schedule with one amplitude edit ($5.000$~\textmu s; transport
    and gate lines silent). The PSR branch splices its
    transport-and-gate episode at the sampled $\tau$:
    $119.84$~\textmu s wall clock,
    of which $114.14$~\textmu s ($95.2\%$) is transport and gate
    around $5$~\textmu s of evolution---spent outside the dressed
    regime, at the idle coherence of
    Appendix~\ref{app:machine}, which is why the episode costs
    wall-clock time and $\epsilon_{\mathrm{ins}}$, not
    evolution-error budget. Figure~\ref{fig:branch}'s ledger lists
    the same nine segments of the same compile. Schedules are
    device-ready, not hardware-executed;
  no claim in Sec.~\ref{sec:eval} rests on their execution.}
  \label{fig:schedules}
\end{figure}

Where should a shift be realized? A general approach is to hand the shifted
Hamiltonian back to the solver. For derivative programs, however, this can
discard reusable source structure and pay an unnecessary price. A generic
re-solve at $H+\sigma h_\kappa \dot H_\ell$ can move the atoms and change
amplitudes by an order of magnitude, producing a different realization rather
than a small edit of the source. Pinning the solve to the source geometry can
instead become near-degenerate, with large dressing and detuning values
canceling to realize a small shift. When the source realization remains valid,
DiffSimuQ therefore intercepts the shift above the instruction-to-pulse
boundary: the touched terms share one coefficient rescaling, so the branch is
the source schedule with its amplitude arrays scaled---the scaled source lane reproduces the shifted
target to within $10^{-7}$, with residual equal to the source residual times $|\mathrm{scale}|$. Both rejected routes are measured and recorded in
the builder so they are not retried during branch compilation. The other two branch kinds
follow the same grammar: a PSR branch is the one structural
edit---the transport-and-gate episode of Figure~\ref{fig:schedules},
whose cost is what it makes the compiler emit and what
$\epsilon_{\mathrm{ins}}$ charges, not its microseconds---and an
FD branch is the same amplitude edit at a re-dialed setpoint,
with nothing new to emit. For NSR and FD the derivative program could need re-realization when a shift changes one or few unstructured term of a jointly synthesized interaction; for PSR the source realization can always be reused. This is the vocabulary the rest of the
paper prices: Appendix~\ref{app:select} counts branch
inventories, and Appendix~\ref{app:scaling} times per-branch
increments, of exactly these edits.

\section{Strategy Selection}
\label{app:select}

Both evolution derivatives are sound, so selection decides only
how many executions the same gradient costs. Throughout, $q$ is
the number of device-alphabet terms a parameter touches and $p$
the number of differentiated coefficients---the axes of
Figure~\ref{fig:select}---and $v_j=\partial u_j/\partial\theta_\ell$.
Every comparison below is certificate against certificate: the
reference $\Omega_\ell=\int_{\tau_1}^{\tau_2}
\operatorname{diam}(\dot H_\ell(t))\,dt$ is itself the tightest
bound Lemma~\ref{lem:type} supplies, not the exact spectral
content of $F$.

\subsection{The Comparison and What Makes It Blind}

\paragraph{How the plane is measured.} Each cell $(p,q)$ of
Figure~\ref{fig:select} holds six independently seeded programs of
$p$ tangents, every tangent drawing $q$ terms without replacement
from the device pool---$35$ terms on seven qubits, the $X_a$,
$Z_a$, and $Z_aZ_b$ of Sec.~\ref{sec:machine}, which is why $q$
tops out at $35$; the cell value is the mean over seeds of
$\log_{10}(N_{\mathrm{NSR}}/N_{\mathrm{PSR}})$, $11{,}550$
tangents in all. $N_{\mathrm{NSR}}$ is scored from the exact
spectral diameter of each tangent (an eigenvalue computation---the
reference oracle, never a compiler path), $N_{\mathrm{PSR}}$ from
the measured per-branch shot deviation at the operating point
(four high-entropy states, background drawn on the device pool,
readout $\sum_aZ_a$), with one branch per alphabet term serving
every coefficient that touches it; the certificate variants
replace the diameter by $\overline\Omega$ and the measured
deviation by its bound. No target accuracy needs quoting: $N$
scales as $C^2\epsilon_t^{-2}\log(1/\beta)$, the plane is drawn
in units where the target factor is one, and changing the target
leaves both the ratio colorbar and the boundary unchanged. Everything is at
the Hamiltonian level, noiseless: selection compares costs of two
estimators that Appendices~\ref{app:proofs}--\ref{app:noise}
already made sound.

The selection quantity is the ratio of full-gradient execution
budgets. It is assembled from the per-parameter estimator scales
of Appendix~\ref{app:execution},
\begin{equation}
  \begin{aligned}
    C_{\mathrm{NSR}}&=\overline\Omega_\ell R,\\
    C_{\mathrm{PSR}}&=2\Delta\tau\,\sup_\tau\sum_j|v_j|(\tau)\,R,
  \end{aligned}
  \label{eq:sel-constants}
\end{equation}
together with the branch counts each strategy needs for the whole
gradient: a PSR branch is per touched term and serves every
parameter touching it, while NSR draws one shift per differentiated
parameter. Those counts and the concentration factors enter both
sides symmetrically and cancel. The certificate does not.

On a Pauli alphabet the term-wise certificate of
Assumption~\ref{as:bandwidth} is
\begin{equation}
  \begin{aligned}
    \overline\Omega^{\,\mathrm{L1}}_\ell
    &=\int_{\tau_1}^{\tau_2}\sum_j|v_j|(t)\operatorname{diam}(A_j)\,dt\\
    &=2\int_{\tau_1}^{\tau_2}\sum_j|v_j|(t)\,dt,
  \end{aligned}
\end{equation}
the same $L^1$ mass $C_{\mathrm{PSR}}$ reads, so the predicted
ratio is \emph{independent of $q$}: touching more terms scales
both sides alike. The measured boundary is not, since one combined
shift replaces $q$ insertions and the diameter of a mixed
non-commuting tangent concentrates below the sum of diameters,
roughly as $\sqrt q$. The constants themselves are sound---at the
instance of Figure~\ref{fig:loop} ($p=2$, $q=1$), where the
certificate is exact, prediction and measurement agree to
$2.7\%$. The missing $q$ dependence alone is what puts the
compiler on PSR across the entire plane.

A tighter certificate is not free, because $\overline\Omega_\ell$
is not only the price NSR is quoted at but the quantity that sets
its shift grid, $h_\kappa=(2\kappa+1)\pi/(2\overline\Omega_\ell)$.
A smaller certificate enlarges every shift, so fewer of them fit
the device's amplitude range and the headroom cap $M$ falls.
Reducing the certificate by the $\sqrt2$ available on this
alphabet increases each shift by $\sqrt2$: $M=5$ becomes $M=3$,
the truncated tail $\sum_{\kappa>M}\eta_\kappa$ grows from
$0.034$ to $0.050$, and the squared estimator-range bound, which
scales as $\overline\Omega_\ell^{\,2}$, halves. The bound
$\overline\Omega_\ell R\,p_{\mathrm{out}}$ on the truncation
bias increases by about $1.04\times$. Halved variance against a
truncation-bias bound larger by $1.04\times$ is a trade whose
balance depends on the budget, not an improvement outright, and
it falls exactly in the large-$q$ region where grouping bites.
Proposition~\ref{prop:acsound} makes it a
compiler knob rather than a dilemma: every value in
$[\overline\Omega^{\,\mathrm{AC}}_\ell,
\overline\Omega^{\,\mathrm{L1}}_\ell]$ is sound, so the lowering
takes the smallest one the headroom check admits, and selection is
priced at the certificate the branch will actually run. At $q=1$
the interval is a point, which is why the NSR floor of
Figure~\ref{fig:F6} does not move.

\subsection{An Anticommutation-Grouped Certificate}

Inside a set of pairwise-anticommuting involutions the diameter is
not merely bounded but known.

\begin{lemma}[Anticommuting group diameter]\label{lem:acgroup}
  Let $\{A_j\}_{j\in g}$ be traceless Hermitian operators with
  $A_j^2=I$ and $A_jA_k=-A_kA_j$ for $j\ne k$, and let
  $v\in\mathbb R^{g}$. Then
  \[
    \Big(\sum_{j\in g}v_jA_j\Big)^2=\|v\|_2^2\,I,
  \]
  the spectrum is $\{\pm\|v\|_2\}$, and
  \[
    \operatorname{diam}\Big(\sum_{j\in g}v_jA_j\Big)=2\|v\|_2.
  \]
\end{lemma}
\begin{proof}
  Expanding the square, the cross terms cancel in pairs by
  anticommutation and the diagonal terms give $\sum_jv_j^2I$. An
  operator squaring to $\|v\|_2^2I$ has spectrum in
  $\{\pm\|v\|_2\}$, and tracelessness forces both signs to occur for
  $v\ne0$.
\end{proof}

\begin{definition}[Grouped certificate]\label{def:accert}
  Let $\mathcal G$ partition the terms touched by $\theta_\ell$ into
  pairwise-anticommuting groups. Set
  \begin{equation}
    \begin{aligned}
      \overline\Omega^{\,\mathrm{AC}}_\ell
      &=\int_{\tau_1}^{\tau_2}\sum_{g\in\mathcal G}
      \operatorname{diam}\Big(\sum_{j\in g}v_j(t)A_j\Big)\,dt\\
      &=2\int_{\tau_1}^{\tau_2}\sum_{g\in\mathcal G}\|v_g(t)\|_2\,dt,
    \end{aligned}
  \end{equation}
  the second equality on a Pauli alphabet by
  Lemma~\ref{lem:acgroup}.
\end{definition}

\begin{proposition}[Sound, and never worse]\label{prop:acsound}
  For every such partition,
  $\overline\Omega^{\,\mathrm{L1}}_\ell\ge
  \overline\Omega^{\,\mathrm{AC}}_\ell\ge\Omega_\ell$.
\end{proposition}
\begin{proof}
  The right inequality is subadditivity of $\operatorname{diam}$
  across the groups; the left is $\|\cdot\|_2\le\|\cdot\|_1$ inside
  each.
\end{proof}

So $\overline\Omega^{\,\mathrm{AC}}_\ell$ satisfies
Assumption~\ref{as:bandwidth} and replaces the term-wise bound
with no change to any soundness statement, and no partition is
ever the worse choice: partition quality is a tightness knob of
exactly the kind Remark~\ref{rem:app-cert} describes, one-sided
and paid in shots. Anticommutation of two Pauli strings is the
symplectic form on their bit-vectors, so the partition is a greedy
clique cover of a graph built in $O(q^2)$ bit operations; nothing
here touches the $2^n$ matrix, and the exact diagonalizations
quoted below serve as a reference oracle for this evaluation only,
never as a compiler path. Cover quality is insensitive to search
effort on this workload: over random restarts only $1.7\%$ of
tangents improve at all, by at most $1.21\times$.

One condition falls on the compiler rather than the mathematics.
Assumption~\ref{as:bandwidth} requires $\overline\Omega_\ell$ to
be parameter-free, since the shifts are computed from it, so the
partition must be fixed at compile time: group at a reference
point in the parameter region and take $\sup$ of $\|v_g\|_2$ over
that region. Any fixed partition is sound, so this costs tightness
and not correctness.

\paragraph{What the alphabet permits.} Pairwise-anticommuting sets
of Pauli strings on $n$ qubits reach size $2n+1$; the plane's pool
does not supply them. Its draw is $\{X_a,Z_a,Z_aZ_b\}$ only (a
unit test enforces that no $Y_a$ enters), and within that set
nothing anticommutes with both $X_a$ and $Z_a$, so every group has
size at most two and
$\overline\Omega^{\,\mathrm{AC}}_\ell\ge
\overline\Omega^{\,\mathrm{L1}}_\ell/\sqrt2$. The machine
signature itself also carries $Y_a$
(Eq.~\eqref{eq:machine-H}), which admits the size-three set
$\{X_a,Y_a,Z_a\}$ and a $\sqrt3$ bound; the plane does not
sample it, and Figure~\ref{fig:regimes}(c) deliberately steps
further outside to measure grouping on a richer alphabet. The
recovered fraction
$\Omega_\ell/\overline\Omega^{\,\mathrm{AC}}_\ell$ is $1$
up to $q\approx5$ and falls to $0.475$ at $q=35$. Grouping closes
part of the gap and provably cannot close all of it on this pool;
the ceiling is a property of the alphabet, not of the
certificate.

\subsection{The Residual Is Shape, and It Has One Parameter}

Let $\gamma$ be the measured execution ratio divided by the ratio
predicted from $\overline\Omega^{\,\mathrm{AC}}$. Across the plane
$\gamma$ is flat in $p$, spans $4.6\times$ in $q$, and follows
$\gamma_{\mathrm{med}}(q)\approx1.9\,q^{-1/2}$: the same $\sqrt q$
concentration the group bound recovers where the alphabet allows
it and misses where it does not. The compiler therefore applies a
margin to the comparison,
\begin{equation}
  \begin{gathered}
    \text{select NSR}\iff
    \gamma(q)\cdot\Big(\frac{C^{\mathrm{AC}}_{\mathrm{NSR}}}{C_{\mathrm{PSR}}}\Big)^2<1,\\
    \gamma(q)=\min\!\big(1,\ \gamma_0/\sqrt q\,\big),
  \end{gathered}
  \label{eq:margin}
\end{equation}
with a single calibrated constant $\gamma_0=1.86$.

Two things about \eqref{eq:margin} matter more than its accuracy.
First, the margin multiplies the \emph{comparison} and never the
certificate: the selected program executes with
$\overline\Omega^{\,\mathrm{AC}}_\ell$, which sets both the
estimator weight and the shift grid
$h_\kappa=(2\kappa+1)\pi/(2\overline\Omega_\ell)$, and which
Proposition~\ref{prop:acsound} keeps above $\Omega_\ell$.
Selection is a policy over two sound estimators; a wrong choice
costs executions and never bias. Second, $\gamma_0$ is a
calibration, not a bound---the worst case in
Table~\ref{tab:selector} is measured, not certified---and it is
admissible precisely because nothing downstream depends on it
being right.

Pinning the exponent rather than fitting it is what makes one
constant sufficient. Calibrated on $q\le8$ alone,
\eqref{eq:margin} extrapolates to $q\ge9$ at $90.0\%$ agreement
and $1.20\times$ worst case; a free-exponent fit extrapolates to
$78.9\%$ on $q\ge18$. The $\sqrt q$ comes from the diameter
argument rather than from the data, and it carries an
$n$-independent constant.

\begin{table}[t]
  \caption{Selectors against the measured winner on the plane of
    Figure~\ref{fig:select}. Forfeit is executions relative to the
    measured optimum, on the cells where the selector is wrong. Ties
  resolved to NSR.}
  \label{tab:selector}
  \small
  \centering
  \setlength{\tabcolsep}{3pt}
  \begin{tabular}{@{}lrrrr@{}}
    \toprule
    selector & agreement & \makecell[r]{NSR\\chosen} &
    \makecell[r]{median\\forfeit} & max \\
    \midrule
    $\overline\Omega^{\,\mathrm{L1}}$ (term-wise)
    & $57.7\%$ & $0\%$ & $1.49\times$ & $5.76\times$\\
    $\overline\Omega^{\,\mathrm{AC}}$
    & $71.4\%$ & $14.3\%$ & $1.35\times$ & $2.77\times$\\
    $\overline\Omega^{\,\mathrm{AC}}$, flat margin
    & $67.1\%$ & $58.6\%$ & $1.18\times$ & $1.84\times$\\
    $\overline\Omega^{\,\mathrm{AC}}$, \eqref{eq:margin}
    & $\mathbf{88.3\%}$ & $46.0\%$ & $1.07\times$ & $1.26\times$\\
    \midrule
    measured & --- & $42.3\%$ & --- & ---\\
    \bottomrule
  \end{tabular}
\end{table}

Table~\ref{tab:selector} is the sequence---its first row is the
always-PSR policy under another name ($0\%$ NSR chosen). Two rows
need reading
beyond their numbers. The flat margin is worse than no margin
because it overshoots at small $q$, where the certificate is
already exact, buying agreement in one region by losing it in
another---the residual is shape, so a constant cannot absorb it.
And \eqref{eq:margin}'s $88.3\%$ holds at $90.0\%$ on the
held-out $q\ge9$ cells, so the fit is not reading the region it is
scored on. Ties---$41$ of the $350$ cells, at floating-point
precision---are resolved to NSR throughout the table. The margin
makes the rule nearly irrelevant: after \eqref{eq:margin} the tie
set shrinks to six cells on this plane (five and eight on the
  families of Figure~\ref{fig:regimes}, which resolves them to
PSR), and either resolution moves agreement by less than
$1.7$ points.

\subsection{Other Hamiltonian Families}
\label{app:select-phase}

\begin{figure}[t]
  \centering
  \includegraphics[width=\linewidth]{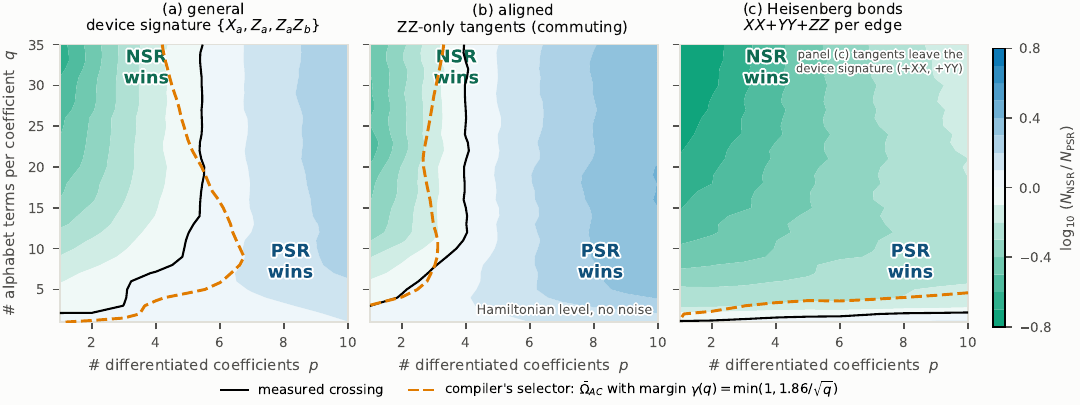}
  \caption{The plane of Figure~\ref{fig:select} repeated for three
    tangent families: (a) the general device pool, reproducing
    Figure~\ref{fig:select} exactly; (b) aligned $ZZ$-only
    tangents; (c) Heisenberg bonds. Color is the mean over seeds of
    $\log_{10}(N_{\mathrm{NSR}}/N_{\mathrm{PSR}})$ (blue: PSR
    wins; green: NSR wins); the thin black line is the measured
    crossing and the dashed orange line the selector of
    \eqref{eq:margin} with the same $\gamma_0=1.86$. Panel~(c)'s tangents extend the alphabet with
    $XX$/$YY$ terms; states, background, and readout stay on the
  device pool.}
  \label{fig:regimes}
\end{figure}

Figure~\ref{fig:regimes} repeats the plane on two families chosen
to sit either side of the general pool. \emph{$ZZ$-only tangents}
are aligned: every touched term commutes, so
Definition~\ref{def:accert}'s partition degenerates to singletons
and $\overline\Omega^{\,\mathrm{AC}}=
\overline\Omega^{\,\mathrm{L1}}$. Where the terms are also jointly
extremizable the term-wise bound is already tight, there is no gap
to close, and the NSR region shrinks. \emph{Heisenberg bonds} are
the opposite regime by outcome but not by mechanism: $X_aX_b$,
$Y_aY_b$ and $Z_aZ_b$ on one bond mutually \emph{commute}---they
are the components of the exchange operator---so grouping again
gains nothing on a single bond, and the contraction
($\operatorname{diam}=4$ against a term-wise $6$) is failure of
joint extremizability, not non-commutativity. Only across adjacent
bonds does anticommutation appear---$X_aX_b$ and $Y_bY_c$ differ
on one shared site---and the grouped certificate tighten: the
measured mean gain
$\overline\Omega^{\,\mathrm{AC}}/\overline\Omega^{\,\mathrm{L1}}$
is $1.000$ through $q=3$ (one bond) and $0.603$ at $q=35$, against
an identical $1.000$ at every $q$ on the aligned family and
$0.883$ on the general pool. Frustration is the same mechanism in the smallest
instance: $Z_1Z_2+Z_2Z_3+Z_3Z_1$ commutes term by term, but the
product of its three terms is the identity, so an odd number of
them can never be $-1$ at once; the spectrum is $\{3,-1\}$, the
diameter $4$, and both certificates give $6$.

Measured, with \eqref{eq:margin} carried over unchanged: agreement
is $88.3\%$ on the general family (the plane itself), $92.6\%$
on aligned, and $96.0\%$ on Heisenberg, with worst-case forfeits
$1.26\times$, $1.39\times$, and $2.17\times$ against measured
NSR shares of $42.3\%$, $27.7\%$, and $94.3\%$. The exposed
case was the aligned family, where the certificate is inert at
singleton grouping and $\gamma$ carries the whole correction while
$\gamma_0$ was calibrated on the general pool---and it closes on
the conservative side: the margin raises agreement from $72.3\%$
to $92.6\%$ while choosing NSR on only $20.9\%$ of cells against
the $27.7\%$ measured share. It under-claims NSR rather than
over-claiming it, so indexing the margin on the grouping gain
$\overline\Omega^{\,\mathrm{AC}}/\overline\Omega^{\,\mathrm{L1}}$
instead of $q$---the repair held in reserve---is unnecessary, and
is not implemented.

\paragraph{Why the plane has no $XX$/$YY$ column.} Two-local $XX$
and $YY$ tangents are absent from Figure~\ref{fig:select} because
they are not in the device signature of Sec.~\ref{sec:machine}
(Appendix~\ref{app:machine}). Figure~\ref{fig:regimes}(c)
supplies exactly that column at the Hamiltonian level, where
selection is well-posed because \eqref{eq:sel-constants} and
Definition~\ref{def:accert} read only the tangent's
alphabet-level structure---and it is NSR territory at a $94.3\%$
measured share. On the machine the case is stronger still: a PSR
insertion for an $XX$ term would need a gate realization other
than the $ZZ$ identity~\eqref{eq:zz-lower}, so the strategy that
wins on cost is also the one that lowers without new hardware.

\paragraph{The non-involutive regime.} Generators failing
$A_j^2=I$---a number operator, or a spin-$S$ generator with
$S\ge1$---fall outside Assumption~\ref{as:kick}, so NSR is the
only sound evolution derivative there and selection is vacuous.
Lemma~\ref{lem:acgroup} does not apply either, and the certificate
reverts to the term-wise bound. On an all-Pauli alphabet every
term is involutive, so the regime cannot be instantiated on the
target machine; it is positioned, not measured.

\section{Compiler Scaling Breakdown}
\label{app:scaling}

\begin{table*}[t]
  \caption{The full comparison behind Table~1. Every cell names
    its source; the two evolution derivatives are sound estimators
    of the same gradient, FD is the baseline the language admits
  without a soundness theorem.}
  \label{tab:full}
  \small
  \begin{tabular}{@{}>{\raggedright\arraybackslash}p{0.21\linewidth}*{3}{>{\raggedright\arraybackslash}p{\dimexpr0.26333\linewidth-2\tabcolsep\relax}}@{}}
    \toprule
    & NSR & PSR & FD \\
    \midrule
    branch mechanism &
    amplitude edit of the source schedule at a drawn shift
    (App.~\ref{app:compiler}) &
    gate-zone episode spliced at a drawn time
    (App.~\ref{app:machine}) &
    the source re-dialed at $\theta\pm\epsilon/2$ \\
    sound for the ideal landscape &
    exact in expectation (Thm~\ref{thm:shift-local}) &
    exact in expectation; needs $A_j^2=I$
    (Thm~\ref{thm:psr-local}) &
    no: bias $\epsilon^2f'''/24$ \eqref{eq:fd-trunc} \\
    under the device model (Table~\ref{tab:noise}) &
    unbiased for $\nabla C_{\mathrm{device}}$
    (Thm~\ref{thm:noise-nsr}) &
    unbiased up to insertion error
    (Thm~\ref{thm:noise-psr}, Lemma~\ref{lem:ins-bias}) &
    setpoint term amplified by $1/\epsilon$
    \eqref{eq:fd-setpoint} \\
    setpoint exposure (App.~\ref{app:fd}) &
    none in the limit: a fresh draw per execution averages out;
    residual $r^2f'''/2$ &
    one shared draw: $|f''|\,r$, second order
    ($\tfrac{\sqrt3}{2}|f'''|r^2$) where $f''=0$ &
    two draws, amplified by the step:
    $\sqrt2\,r|f'|/\epsilon$ \eqref{eq:fd-fluct} \\
    residual, as realized &
    truncation $\le4\overline\Omega R/\pi^2(2M+1)$
    (Lemma~\ref{lem:trunc-bias}) &
    $\le C_{\mathrm{PSR}}\,\epsilon_{\mathrm{ins}}$
    (Lemma~\ref{lem:ins-bias}) &
    RMSE floor
    $\approx0.60\,|f'''|^{1/3}(|f'|r)^{2/3}$
    \eqref{eq:fd-floor} \\
    branches per gradient &
    one draw $(\kappa,\sigma)$ per parameter &
    two per touched term, reused across parameters
    (App.~\ref{app:select}) &
    two per parameter \\
    estimator scale $C_S$ &
    $\overline\Omega_\ell R$ &
    $2\Delta\tau\,\sup_\tau\sum_j|v_j|\,R$ &
    $2R/\epsilon$ \\
    per-branch compile ($n{=}1000$) &
    $\approx 0.2$~ms* (Table~\ref{tab:scaling}) &
    $175$~ms (Table~\ref{tab:scaling}) &
    $\approx 0.2$~ms* (Table~\ref{tab:scaling}) \\
    per-branch device time (running instance) &
    $5.0$~\textmu s &
    $119.8$~\textmu s, $95.2\%$ transport$+$gate
    (App.~\ref{app:compiler}) &
    $5.0$~\textmu s \\
    \bottomrule
  \end{tabular}
\end{table*}

\begin{table}[H]
  \caption{Compile cost on the specialized path (1D TFIM chain;
    medians). Fitted exponents over $n=10$--$1000$:
    source $\sim n^{1.8}$, one PSR branch $\sim n^{1.2}$
  ($n^{1.4}$ over $n\ge100$), one NSR branch $\sim n^{1.0}$.}
  \label{tab:scaling}
  \small
  \centering
  \setlength{\tabcolsep}{4pt}
  \begin{tabular}{@{}rrrrr@{}}
    \toprule
    $n$ & \makecell[r]{source\\(s)} & \makecell[r]{+PSR branch\\(ms)} &
    \makecell[r]{+NSR branch\\(ms)} & \makecell[r]{ledger\\entries} \\
    \midrule
    10   & 0.017 & 0.54  & 0.0023 & 46 \\
    50   & 0.18  & 2.9   & 0.0090 & 206 \\
    100  & 0.63  & 6.5   & 0.017  & 406 \\
    500  & 15.2  & 57.5  & 0.093  & 2\,006 \\
    1000 & 59.6  & 175   & 0.191  & 4\,006 \\
    \bottomrule
  \end{tabular}
\end{table}

\noindent
Figure~\ref{fig:compile-curves} (Sec.~\ref{sec:eval-scale}) plots
Table~\ref{tab:scaling} with fitted slopes, the measured generic
path ($n=2$--$12$), and its extrapolation past the measured
ceiling; the generic-path exponent depends on the fit window
($4.39$ over $n\ge5$, $3.79$ over all eight points). Wall clock on
macOS-26.5-arm64; schedule ops and pulse ledger only, no
pulse-shape synthesis; on the 2D grid the compiled
nearest-neighbour model is exact and the diagonal $J/8$ tail
($\approx14\%$ of the $L^1$ mass) is declared dropped.

\paragraph{Generic path and specialization.}
The generic all-pairs SimuQ solve is measured from $n=2$
($0.032$\,s) to its practical ceiling at $n=12$
($27.1$\,s); the asymptotic-window slope is $n^{4.4}$
(Figure~\ref{fig:compile-curves}), reaching a day of compiling
near $n\approx78$. On the specialized path, re-solving from scratch at a shifted parameter
costs essentially a full source solve---$99.4\%$ of one,
measured at $n=300$. On the TFIM path measured here,
the lowering reuses the source solve for every derivative branch;
no measured branch pays the generic all-pairs solve. For paths
where a shift changes one or few unstructured terms of a jointly synthesized interaction,
NSR and FD may instead require a new source solve, whereas PSR
always reuses the source realization (\app{branch-edit}). The specialization is
target-aware, not a new machine description: chain locality
reduces the global solve to per-bond solves with an analytic warm
start, which is what the table times. No claim is made that
arbitrary-Hamiltonian SimuQ compilation scales.

\paragraph{Per-gradient compile cost.}
The runtime compiles one branch per execution
(Definition~\ref{def:exec}), so per-gradient compile cost is the
execution budget times the per-branch increment. At the
6000-execution budget of Figure~\ref{fig:loop}'s closed loop and
the $n=1000$ rates: PSR $\approx17$\,min, NSR and FD
$\approx1$\,s, and the source solve runs once, amortized across
all branches. At a fixed accuracy target the budget itself grows
with the touched-term count as in Appendix~\ref{app:execution}'s
execution-scaling paragraph.

\fi

\end{document}